%% file: spiral_paper.tex
\documentclass{aastex631}
\shorttitle{Effect of Environment on Spirals, Bars, Concentration, and Quenching}
\shortauthors{Smith et. al.}
\graphicspath{{./}{figures/}}

\begin{document}

\title{The Effect of Environment on Galaxy Spiral Arms, Bars, Concentration, and Quenching}

\author[0000-0002-8521-5240]{Beverly J. Smith}
\affiliation{East Tennessee State University \\
Department of Physics and Astronomy, Box 70652\\
Johnson City TN 37614, USA}

\author{Mark L. Giroux}
\affiliation{East Tennessee State University \\
Department of Physics and Astronomy, Box 70652\\
Johnson City TN 37614, USA}

\author[0000-0002-6490-2156]{Curtis Struck}
\affiliation{Iowa State University \\
Department of Physics and Astronomy, Ames IA 50011}



\begin{abstract}

For a sample of 4378 nearby spiral and S0 galaxies, 
\citet{2020ApJ...900..150Y}
used Fourier analysis of Sloan Digital Sky Survey images to show that the strengths of the spiral arms and the pitch angles of the arms are inversely correlated with central concentration. In the current study, we search for trends in the 
\citet{2020ApJ...900..150Y}
spiral arm parameters with environment and specific star 
formation rate (sSFR).
When comparing galaxies with similar concentrations, we do not find a significant difference in the arm strengths or pitch angles
of spiral galaxies in clusters compared to field galaxies.  
When differences in concentration are taken into account,
we also find no significant difference in the 
parameter f3 for cluster spirals compared to field spirals, 
where f3 is 
the normalized m = 3 Fourier amplitude.
When concentration is held fixed, both arm strength and pitch angle are correlated with sSFR, but f3 is not.  These relations support the suggestion by 
\citet{2015ApJ...802L..13D}
of a `fundamental plane' of spiral structure involving pitch angle, bulge stellar mass, and gas surface density.  We discuss these results in terms of theories of spiral arm production and quenching in galaxies.
To aid comparison with earlier studies based on Galaxy Zoo, we explore how
the \citet{2020ApJ...900..150Y} parameters 
relate to similar parameters measured 
by Galaxy Zoo (i.e., f3 vs.\ number of arms,
pitch angle vs.\ winding parameter, and concentration vs.\ bulge class).

\end{abstract}



\section{Introduction} \label{sec:intro}

Most large galaxies come in one of two basic flavors: 
disk-dominated galaxies with on-going
star formation and spiral patterns, or spheroidal-dominated galaxies 
with mostly older stars.  The evolutionary relationship between these
two types of galaxies remains
a matter of debate.   
How the spiral patterns in disk galaxies form and evolve with
time, and the effect of a central bar
on spiral patterns, are also uncertain.  Another open
question is why there is variation from galaxy to galaxy 
in the number of spiral
arms, the length and prominence of these arms,
and the pitch angle of the arms.
To better understand these issues, 
studying the environmental dependence of different 
types of galaxies is useful.

\subsection{Morphology, Quenching, and Environment} \label{sec:intro_morph}

Dense clusters of galaxies tend to host more 
elliptical and S0 galaxies,
while galaxies in the field are more likely to be spirals
\citep{1974ApJ...194....1O,
1980ApJ...236..351D}.
Cluster galaxies tend to have lower specific star formation rates (sSFRs)
than field
galaxies
\citep{2004MNRAS.353..713K,
2012MNRAS.424..232W},
where the 
sSFR is defined as the star formation rate (SFR) per stellar mass (M*).
A larger fraction of the spirals
in clusters have early-type morphologies compared to 
field spirals 
\citep{2003MNRAS.346..601G}.
In a cluster, the fraction of 
galaxies that are 
late-type drops with decreasing clustercentric radius
while the fraction of 
early-type spirals rises, 
at least until about 0.3 virial radii, when it 
drops in the inner core 
\citep{2003MNRAS.346..601G}.
The fraction of blue (star-forming) spirals in clusters
drops gradually with decreasing clustercentric radius,
while
the fraction of red (quiescent) spirals 
increases to a maximum at about 0.4$\times$ the virial
radius, then decreases in the inner cluster 
\citep{2009MNRAS.393.1324B}.
The fraction of
red spirals peaks in intermediate environments,
inside the infall region of clusters
but not in their cores 
\citep{2009MNRAS.399..966S,
2009MNRAS.393.1302W,
2010MNRAS.405..783M}.
Passive spirals tend to have larger bulges, higher
central concentrations, and larger central stellar mass densities
than blue spirals 
\citep{2006MNRAS.367.1394K,
2010ApJ...719.1969B,
2013ApJ...776...63F,
2014MNRAS.441..599B}.
These observations suggest that, as spirals become quenched,
their bulge-to-disk ratios tend to grow, although 
there is a lot of scatter in this relation.

A number of mechanisms
have been suggested that may enhance the
relative brightness
of the spheroidal component of a galaxy relative to its disk.
These include 1) minor mergers 
\citep{1998ApJ...502L.133B,
2001A&A...367..428A,
2005A&A...437...69B,
2009MNRAS.396.1972P,
2010ApJ...715..202H},
2) gas-rich major mergers 
\citep{2005ApJ...622L...9S,
2006ApJ...645..986R,
2009ApJ...691.1168H,
2016ApJ...821...90A,
2017MNRAS.470.3946S},
3) central star formation triggered by inflowing gas driven
by galaxy interactions 
\citep{2000MNRAS.317..667M, 2007AJ....133..791S,
2008MNRAS.385.1903L},
bars 
\citep{1990ApJ...363..391P},
and/or spiral arms
\citep{2014MNRAS.440..208K, 2022arXiv220206932Y},
4) fading of the disk due to quenching 
\citep{2009MNRAS.394.1213W,
2016ApJ...818..180C},
5) tidal heating of a disk in the gravitational field of a cluster
\citep{2003ApJ...582..141G,
2020MNRAS.496.2673J},
or 
6) tidal stripping of stars from the outer disks of galaxies in clusters
by fast
interactions with multiple neighbors (`harrassment', 
\citealp{1996Natur.379..613M,
1998ApJ...495..139M,
1999MNRAS.304..465M})
or from the tidal effect of the cluster as a whole 
\citep{1990ApJ...350...89B,
2003ApJ...582..141G,
2004AJ....127.1344A,
2020A&A...638A.133L}.

Possible quenching mechanisms 
include:
1) ram pressure stripping of cold disk gas by a hot intracluster
medium (ICM) 
\citep{2017MNRAS.466.1275B,
2021arXiv210913614B},
2) starvation (also called strangulation),
defined as the removal of hot circumgalactic gas by 
an interaction with the ICM 
\citep{1980ApJ...237..692L,
2000MNRAS.318..703B,
2000ApJ...540..113B,
2008MNRAS.383..593M,
2009MNRAS.399.2221B}.
Hot gas removal also involves ram pressure stripping,
as well as processes such as viscuous stripping and thermal
evaporation 
\citep{1982MNRAS.198.1007N,
1977Natur.266..501C,
2021PASA...38...35C};
starvation is often distinguished from ram pressure stripping 
of the cold gas because it quenches much more slowly
(timescale of a few Gyrs) by
preventing later infall and fueling of star formation
\citep{2021PASA...38...35C}.
3) tidal stripping of gas from the disks of spirals by
interactions 
\citep{2009MNRAS.400.1962K},
4) disk gas being driven into the center of a galaxy
by a bar or an interaction, depleting the outer disk
of gas while triggering central star formation
and building up the stellar bulge 
\citep{2000MNRAS.317..667M, 
2007AJ....133..791S,
2008MNRAS.385.1903L},
and 
5) 
heating of the interstellar gas
by feedback from active galactic nuclei (AGN) 
\citep{2014MNRAS.441..599B,
2016MNRAS.462.2559B,
2020MNRAS.499..230B,
2022arXiv220107814B}.
6) Increasing the central spheroidal component
of a galaxy could potentially
stabilize a gaseous disk, slowing star formation,
a process known as morphological quenching
\citep{2009ApJ...707..250M,
2013MNRAS.432.1914M,
2020MNRAS.495..199G}.
7) Tidal heating of disks by the gravitational field of the
cluster could reduce gravitational instabilities
and therefore suppress star formation 
\citep{2003ApJ...582..141G}.
8) Another possible factor 
is 
assembly bias, 
in which the assembly history of galaxies plays a role in how
susceptible a galaxy is to environmental effects 
\citep{2021arXiv211111252S}.
In this picture, spirals with more
gas delay quenching longer, and galaxies with less gas become red
more quickly.

The relative importance of the above processes depends upon environment.
Ram pressure stripping of cold gas is most effective
close to the core of a massive cluster
\citep{1999MNRAS.308..947A,
2021arXiv210913614B},
though
it may operate further out 
\citep{2020MNRAS.498.4327X}
and in 
galaxy groups 
\citep{2012MNRAS.427.2841F,
2013MNRAS.436...34C,
2017MNRAS.466.1275B}.
Strangulation may
operate out to 
three virial radii in clusters or beyond
\citep{2013MNRAS.430.3017B,
2018MNRAS.475.3654Z,
2021MNRAS.502.1051A},
and
may also 
occur in small groups if the velocity dispersion is high
enough 
\citep{2009MNRAS.399.2221B,
2008ApJ...672L.103K}.
Historically, galaxy mergers were thought to be unimportant in clusters
because of the large relative velocities of cluster galaxies 
\citep{1980ComAp...8..177O,
1984ApJ...276...26M}.
However, galaxy interactions and mergers are sometimes observed
in the outskirts of nearby clusters 
\citep{2006MNRAS.373..167M},
and
mergers can occur 
in groups infalling into clusters
\citep{2013MNRAS.435.2713V,
2020MNRAS.498.3852B}.
Supporting this idea of cluster galaxies being `pre-processed' in groups,
larger proportions of 
quenched galaxies
relative to field galaxies 
are seen
out to 3 virial radii from the
center of a cluster 
\citep{2012MNRAS.424..232W,
2012MNRAS.420..126L,
2014MNRAS.439.3564C,
2017MNRAS.467.3268R},
and
the fraction of red spirals is higher than the field
out to 3 $-$ 5 virial radii 
\citep{2009MNRAS.393.1324B}.
Pre-processing in groups may be driven
in part
by galaxy interactions and mergers
\citep{2004cgpc.symp..277M}.
Galaxy interactions
and mergers may be enhanced when a group falls into a cluster, due to
the effect of the dark matter halo of the cluster on the group
\citep{2003ApJ...582..141G,
2006asup.book..115S,
2008MNRAS.385L..38M}.
Pre-processing may also involve strangulation/starvation 
in groups 
\citep{2004PASJ...56...29F,
2013MNRAS.435.2713V}.

\subsection{Spiral Patterns and Bars}

It is still uncertain
how the spiral patterns in disk galaxies are affected by 
concentration, 
environment, and quenching.
Traditionally, spiral galaxies are classified into
Hubble types based on both
their bulge-to-disk (B/D) ratio and the tightness of their spiral arms
\citep{1926ApJ....64..321H,
1961hag..book.....S,
1991rc3..book.....D}.
Spiral galaxies with later Hubble types 
tend to have larger pitch angles 
although there is 
a large amount of scatter 
\citep{1981AJ.....86.1847K,
1999A&A...350...31M,
2002A&A...388..389M,
2018ApJ...862...13Y,
2020MNRAS.493..390S,
2019ApJ...871..194Y,
2020ApJ...900..150Y},
and 
not all studies find a correlation \citep{1998MNRAS.299..685S}.
In some studies,
central concentration
was found to be anti-correlated with pitch angle, although
with considerable
scatter 
\citep{2013MNRAS.436.1074S,
2019ApJ...871..194Y};
in other
studies, little or no correlation was found 
\citep{2015MNRAS.446.4155K,
2019A&A...631A..94D}.
Galaxy Zoo citizen scientist estimates of arm tightness and bulge prominence 
do not show a correlation
\citep{2017MNRAS.472.2263H,
2019MNRAS.487.1808M,
2021MNRAS.504.3364L,
2022MNRAS.509.3966W},
however, Galaxy Zoo
arm tightness estimates weakly correlate
with 
Hubble type classifications made by experts
\citep{2013MNRAS.435.2835W}.
An anti-correlation between pitch angle and rotation curve
maximum velocity V$_{max}$ has also been seen
\citep{1981AJ.....86.1847K,
1982ApJ...253..101K,
2017MNRAS.471.2187D,
2019ApJ...877...64D}.
In some studies, 
spiral arm pitch angles were found to correlate
with the shear rate in the disks of galaxies 
\citep{2005MNRAS.359.1065S,
2006ApJ...645.1012S,
2014ApJ...795...90S,
2019ApJ...871..194Y},
but other studies did not see a correlation \citep{2019A&A...631A..94D}.
Pitch angle also correlates 
with the galaxy central velocity
dispersion and therefore the total bulge mass 
\citep{2008ApJ...678L..93S,
2013ApJ...769..132B}.
Since central velocity dispersion correlates with the mass of the
central black hole 
\citep{2000ApJ...539L...9F,
2013ApJ...764..184M,
2016ApJ...818...47S},
pitch angle also correlates with the black hole mass
\citep{2008ApJ...678L..93S,
2013ApJ...769..132B}.

Spiral galaxies are classified into 
`arm classes' based on the appearance of their
spiral arms, following the Elmegreen scheme 
\citep{1981ApJS...47..229E,
1982MNRAS.201.1021E,
1987ApJ...314....3E,
2011ApJ...737...32E}.
In this system, galaxies are divided into `grand design', `multi-arm',
and `flocculent' galaxies based on the appearance of their spiral arms.
Flocculent galaxies are defined as having multiple fragmented irregular arms,
and 
grand design galaxies have two long continuous arms. Multi-armed galaxies have
less fragmented and irregular arms than flocculent galaxies, but more
than two arms.
For a given Hubble type, grand design galaxies tend to have stronger
arms 
\citep{2011ApJ...737...32E}.
Arm class is somewhat correlated
with Hubble type, with grand design galaxies more likely to be early-type
spirals and flocculent galaxies late-type, but with a lot of scatter 
\citep{1982MNRAS.201.1021E,
2017MNRAS.471.1070B,
2019A&A...631A..94D}.
Galaxies with strong two-armed
patterns have larger B/D ratios than other spirals 
\citep{2017MNRAS.471.1070B}.
In Galaxy Zoo,
two-armed spirals
were found to be redder than galaxies with more arms 
\citep{2016MNRAS.461.3663H}.
This is consistent with the trend of arm class with Hubble type, 
since earlier
type spirals tend to be redder than later types 
\citep{1994ARA&A..32..115R}.

How spiral patterns in galaxy disks are formed 
and maintained is still uncertain.  
Four basic
models of spiral formation and maintenance 
have been suggested: classical spiral
density waves, 
with a fixed spiral wave 
traveling through the disk
\citep{1964ApJ...140..646L},
spiral patterns 
produced by gravitational instabilities in a differentially-rotating
disk enhanced by swing amplification 
\citep{1965MNRAS.130..125G,
1966ApJ...146..810J,
2012ApJ...751...44S,
2013ApJ...763...46B},
spiral patterns induced
by interactions 
\citep{1972ApJ...178..623T,
1992AJ....103.1089B,
2010MNRAS.403..625D,
2011MNRAS.414.2498S},
and spirals
driven by bars
\citep{1976ApJ...209...53S,
1979ApJ...233..539K,
1980A&A....88..184A}.
The observed anti-correlation between 
pitch angle and concentration is consistent 
with classical density wave theory 
\citep{1964ApJ...140..646L,
1975ApJ...196..381R}.
Models of transient spiral arms
are able to produce
a relation between pitch angle and shear rate,
and therefore mass distribution
\citep{2013A&A...553A..77G,
2014ApJ...787..174M,
2016ApJ...821...35M,
2014PASA...31...35D}.
Simulations of
spiral arm formation by 
gravitational instabilities predict that arms 
wind up with time
\citep{2019MNRAS.490.1470P},
as do
models of spiral production by interactions
\citep{2008ApJ...683...94O,
2010MNRAS.403..625D,
2011MNRAS.414.2498S}.
The variation in pitch angle for a given bulge prominence has been
cited as evidence for spiral arm winding 
\citep{2019MNRAS.487.1808M,
2021MNRAS.504.3364L}.

In models of spiral production via gravitational instabilities, 
the number of arms depends upon the mass distribution of the 
galaxy 
\citep{2001A&A...368..107F,
2013ApJ...766...34D,
2015ApJ...808L...8D,
2016ApJ...821...35M}.
These kinds of models
typically
do not produce two-armed patterns 
\citep{2015ApJ...808L...8D},
meaning that
two-armed galaxies need another explanation.
Since two-armed spirals can be produced in interactions
according to simulations
\citep{2008ApJ...683...94O,
2010MNRAS.403..625D,
2011MNRAS.414.2498S},
interactions are often invoked as the cause
of two-armed spiral patterns in galaxies
\citep{2014PASA...31...35D,
2016MNRAS.461.3663H}.
\citet{2017MNRAS.471.1070B}
and 
\citet{2020ApJ...900..150Y}
explained
the observed trend of increasing numbers of arms with decreasing bulge
dominance 
by classical spiral density wave theory
in galaxies with large bulges, and 
by gravitational instabilities 
in galaxies with small bulges. 
According to density wave theory, 
a high Toomre Q parameter 
\citep{1969ApJ...158..899T}
is produced by 
a large bulge; this
reflects incoming waves, creating
a long-lived two-armed spiral pattern 
\citep{1985IAUS..106..513L,
1989ApJ...338..104B,
2016ApJ...826L..21S}.
In contrast,
a galaxy with a weak bulge lacks a high Q center, and therefore does not
produce a stable density wave pattern.
\citet{2020ApJ...900..150Y}
suggest 
that in galaxies with small bulges, spiral patterns are instead 
produced by random gravitational instabilities in the disk.
Models of spiral structure produced
by self-gravity in isolated differentially-rotating disks are able to
reproduce the multiple short arm fragments seen in flocculent galaxies,
supporting this interpretation 
\citep{2011MNRAS.410.1637S,
2012MNRAS.421.1529G,
2013ApJ...766...34D,
2018MNRAS.478.3793D,
2021arXiv211005615S}.

Several studies have
investigated how the number
of spiral arms in galaxies and the arm class varies with environment.
\citet{1982MNRAS.201.1035E}
found
similar fractions of grand design patterns 
among cluster and non-cluster galaxies
when resolution
effects were taken into account. 
However, 
\citet{2011JKAS...44..161C}
found proportionally
more grand design spirals
in clusters.  
Using Galaxy Zoo results,
\citet{2016MNRAS.461.3663H}
found 
a larger fraction of
two-armed spirals in higher density regions. 
Some studies found that the 
fraction of grand design galaxies increases with local density or
size of host group/cluster
\citep{1987ApJ...314....3E,
2011JKAS...44..161C,
2014JKAS...47....1A}.
\citet{2020MNRAS.493..390S}
found that non-isolated spirals
are more likely to have two arms compared to isolated spirals.
Simulations show that a gravitational 
interaction with the 
potential of a cluster can induce a two-armed spiral pattern 
in a disk galaxy
\citep{1990ApJ...350...89B,
1993ApJ...408...57V,
2017ApJ...834....7S}.
Spiral patterns can also be induced or modified by ram pressure
stripping in clusters, possibly producing a flocculent pattern
\citep{2001MNRAS.328..185S,
2021MNRAS.500.1285B}.

The relationship between a central bar and the spiral pattern
is also not settled.
Bar strength and arm strength tend to be correlated 
\citep{2004AJ....128..183B,
2005AJ....130..506B,
2020ApJ...900..150Y}.
This correlation
has been used to argue that bars drive spirals 
\citep{1976ApJ...209...53S,
1979ApJ...233..539K}
or, alternatively, that the conditions in disks that 
favor bar production
also favor strong arms 
\citep{2010ApJ...715L..56S,
2019A&A...631A..94D}.
Using Galaxy Zoo data, 
\citet{2019MNRAS.487.1808M}
find
that for a given bulge prominence barred galaxies have more open spiral
arms on average, while 
\citet{2021MNRAS.504.3364L}
do not see
a correlation between bar strength and pitch angle.
Some studies see a correlation between bar strength and central
concentration or earlier Hubble type 
\citep{2007MNRAS.381..401L,
2011arXiv1111.1532G},
but others see the reverse
\citep{2008ApJ...675.1194B,
2009A&A...495..491A}.

Theory suggests that 
bar strengths may be enhanced in clusters by
tidal forces from the cluster as a whole 
\citep{1990ApJ...350...89B,
2016ApJ...826..227L,
2020A&A...638A.133L}.
Consistent with this idea,
\citet{2012MNRAS.423.1485S}
conclude
based on Galaxy Zoo data
that
barred galaxies tend to be in denser environments than unbarred galaxies.
\citet{1981ApJ...244L..43T}
also found higher fractions of barred galaxies
in clusters.
Other studies, however, did not find a difference
in bar fraction with environment
\citep{1986MNRAS.223..139K,
2009A&A...497..713B,
2009A&A...495..491A,
2010ApJ...711L..61M,
2011arXiv1111.1532G,
2021arXiv211111252S}.

In the current paper, we search for 
environmental variations in the spiral
arm parameters of cluster galaxies compared to galaxies in the field.
We use the 
\citet{2020ApJ...900..150Y}
set of spiral
arm parameters. 
In Section \ref{sec:sampledata}, we describe
the sample and the data.  In Section \ref{sec:clusters}, 
we compare cluster and field
galaxies.  
In Section \ref{sec:sSFR}, 
we compare with sSFR.
To put our results into context with earlier studies of galaxy
structure and morphology based on Galaxy Zoo citizen scientist 
measurements, 
in Section \ref{sec:galzoo}
we compare the 
\citet{2020ApJ...900..150Y}
spiral parameters with
parameters from Galaxy Zoo.  
We discuss our results in terms of
theories of galaxy evolution, spiral arm production, and quenching
in Section \ref{sec:discussion}.  
Conclusions are given in Section 
\ref{sec:summary}.

\section{Sample and Data} \label{sec:sampledata}

\subsection{Galaxy Data}

\begin{figure}[ht!]
\plotone{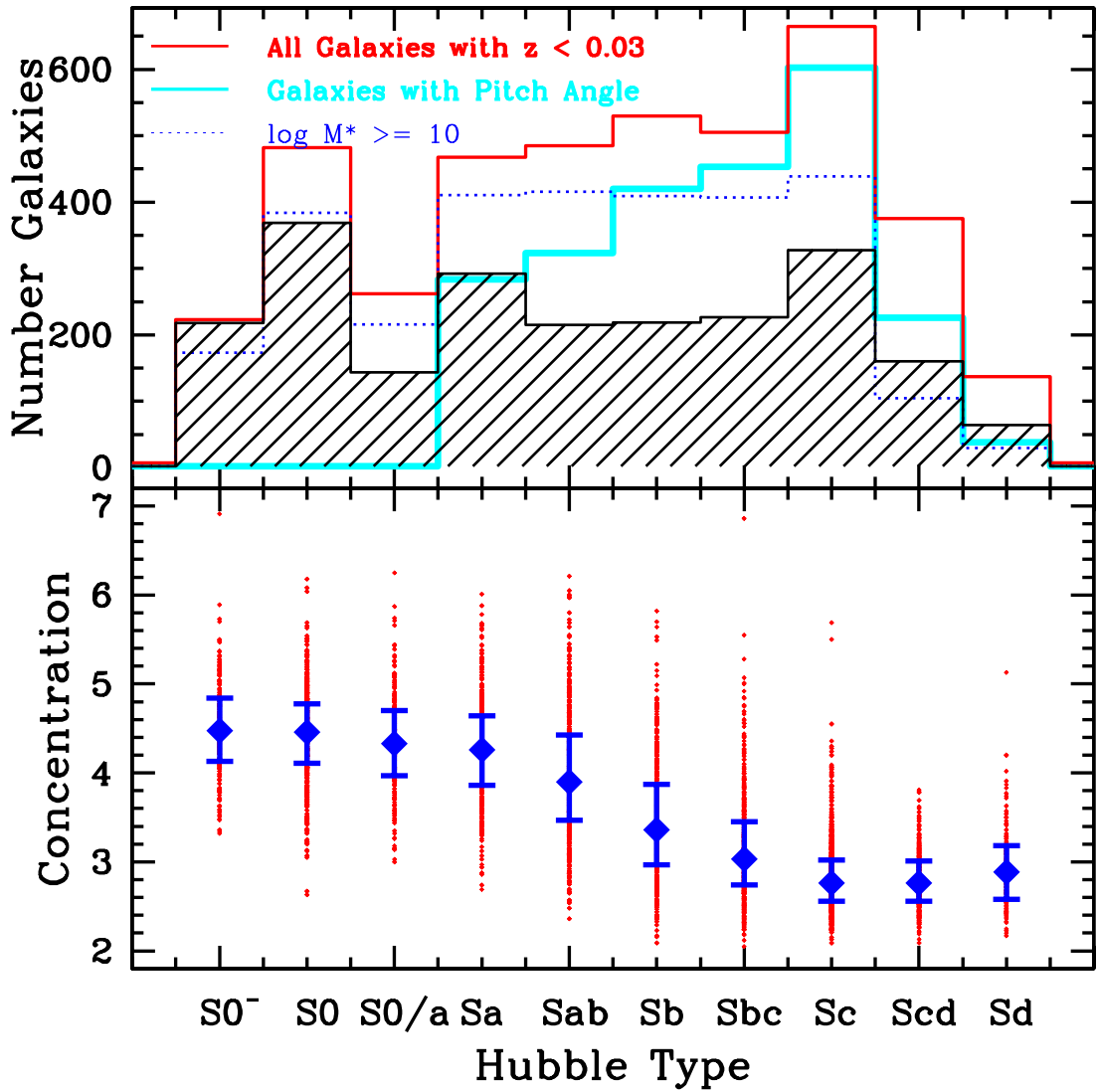}
\caption{
Top panel: 
Histograms of Hubble types for the galaxies 
in the 
\citet{2020ApJ...900..150Y}
sample with z $<$ 0.03.
Red histogram: all galaxies. 
Hatched black histogram: unbarred galaxies.
Cyan histogram: galaxies with measured pitch angles.
Dotted blue histogram: galaxies with log M* $\ge$ 10.
Bottom panel:
Concentration vs.\ Hubble Type.
Individual galaxies are plotted as red dots.  The blue filled diamonds
give the median value for each Hubble type.  The blue errorbars
represent the range from first quartile to third quartile.
\label{fig:C_vs_TType}}
\end{figure}

We started with the 
4378 spiral and S0 galaxies studied by 
\citet{2020ApJ...900..150Y}.
Their primary sample was 
selected 
from the NASA Sloan Atlas (NSA), version 0.1.2 
\citep{2011AJ....142...31B}
to have a redshift range of 0.005 $\le$ z $<$ 0.03,
magnitudes in r $<$ 14.5, and to
be relatively face-on (ellipticity $<$ 0.5).
\citet{2020ApJ...900..150Y}
classified the galaxies by eye into standard Hubble types
using Sloan Digital Sky Survey (SDSS) images, 
and excluded ellipticals from their final
sample but retained the S0 galaxies along with the spirals.
They visually inspected their sample
to exclude ongoing mergers, thus galaxies
undergoing strong interactions with massive companions are 
under-represented in their sample, however, some close pairs are included.
As an extension to their sample, 
\citet{2020ApJ...900..150Y}
added galaxies with 21 cm
HI measurements from the xGASS survey 
\citep{2008ApJ...685L..13C},
which expanded their redshift
range to z $<$ 0.05.
In the current study, we restrict our sample to the 
4062 galaxies with z $<$ 0.03.
The distribution of Hubble types of these 4062 galaxies is given
in the top panel of 
Figure \ref{fig:C_vs_TType}.

\citet{2020ApJ...900..150Y}
conducted Fourier analyses of the SDSS r images of their
galaxies after deprojecting the images to face-on.  They
measured the following parameters: 
1) pitch angle $\phi$ of the spiral arms, 
2) spiral arm strength s,
3) f3, defined as the fractional contribution of the m=3 Fourier amplitude 
to the sum of the 1 $>$ m $\ge$ 4 Fourier amplitudes,
and 4) bar strength.
To better see faint spiral patterns, they constructed structure maps 
(unsharp masked images) by subtracting from the deprojected image a smooth axisymmetric model
derived from azimuthally averaged isophotes.
Using these 
structure maps, they confirmed that many of the galaxies classified as S0 galaxies have faint spiral patterns.
We will refer to the galaxies in this sample as `spirals' because spiral patterns were
discerned in the disks, although some were classically determined to be S0 galaxies.
Out of the 4062 galaxies with z $<$ 0.03, 
bar strengths were measured for 
1846 (45\%). 
In Figure 
\ref{fig:C_vs_TType} (top panel), the hatched region represents
unbarred galaxies.   
In this sample, 
later morphological types 
are more likely to be barred compared to earlier types.
The blue dotted histogram 
in the top panel of 
Figure \ref{fig:C_vs_TType}
indicates galaxies
with stellar mass M* $\ge$ 10$^{10}$ M$_{\sun}$.
Following 
\citet{2020ApJ...900..150Y}, we use
the stellar masses provided by the NSA in the following analysis.

For 1729 of the 4062 galaxies (43\%),
\citet{2020ApJ...900..150Y}
were not able to measure a pitch angle,
but estimates of f3 and arm strength were possible.
Pitch angle measurements are available for 
68\% of the barred galaxies and 
49\% of unbarred galaxies.
The cyan histogram in 
the top panel of 
Figure \ref{fig:C_vs_TType}
shows galaxies with pitch angle measurements.
None of the 946 S0-, S0, SB0, S0/a, and SB0/a galaxies have pitch
angle measurements.   Of the 3116 galaxies with
Hubble types between Sa and Sd,
only 783 lack pitch angle measurements.
This shows that measuring the full set of spiral parameters 
is more difficult in galaxies with large bulges.   
In the following analysis, we investigate how this observational bias
affects various correlations.

To search for AGN activity in these galaxies, we cross-correlated
the list with the Max Planck Institute for
Astronomy/Johns Hopkins University (MPA/JHU)
catalog 
\citep{2004astro.ph..6220B},
which provides SDSS optical spectral classes of the
galaxies.   A total of 489 of the 
\citet{2020ApJ...900..150Y}
galaxies with 
z $<$ 0.03 are
classified as Baldwin-Phillips Terlevich (BPT) class 4 in the 
MPA/JHU catalog, indicating an AGN that
is not a LINER (low ionization nuclear emission region) galaxy
(i.e., it is a Seyfert galaxy). 
We also searched the
\citet{2010A&A...518A..10V}
catalog of active galaxies, and
found 54 additional 
sample
galaxies 
that were classified as
Seyfert 1 or Seyfert 2.   
In total, there are 543 optically-selected AGN in this sample.
We also searched for radio-bright AGN in the sample using
the compilation of 
\citet{2012MNRAS.421.1569B}, and found only eight with z $<$ 0.03;
only one of these was not previously identified as an AGN.
The AGNs are discussed further in Section \ref{sec:sSFR}.

\citet{2020ApJ...900..150Y}
also measured the ``concentration index" C,
which they define as C = 5 log(R80/R20), where
R80 and R20 are the radii enclosing 80\% and 20\%, respectively,
of the total SDSS r flux.
In the bottom panel of Figure \ref{fig:C_vs_TType},
we plot these concentrations against the Hubble Type.
A general trend of decreasing concentration
along the Hubble sequence is present, however, this trend flattens
at both ends. There is little distinction
in concentration
between Sc, Scd, and Sd galaxies in this sample, or between S0-, S0,
and S0/a galaxies. 

\subsection{Environmental Parameters}

We cross-correlated the 
\citet{2020ApJ...900..150Y}
list
of galaxies with the 
\citet{2017MNRAS.470.2982L}
SDSS luminosity-based catalog of galaxy groups and clusters
derived using only SDSS redshifts.
The 
\citet{2017MNRAS.470.2982L}
cluster/group catalog is an update of
the well-cited 
\citet{2005MNRAS.356.1293Y}
cluster/group catalog.
We define cluster/group members
as galaxies that lie
within 3R$_{\rm 200}$ and 3$\sigma$$_{\rm v}$ of the
central position and velocity of the 
\citet{2017MNRAS.470.2982L}
cluster/group, where R$_{\rm 200}$ is the
cluster radius at which the density drops to 200 times the critical
density\footnote{R$_{\rm 200}$ is approximately 0.7 times the 
virial radius for typical clusters at z = 0 
\citep{2021MNRAS.501.5073O}.}
and 
$\sigma$$_{\rm v}$ is the velocity
dispersion of the cluster.  
These are generous criteria for defining cluster membership,
and may contain infalling galaxies as well as
some interlopers.
We define massive clusters
as those with halo masses greater than 10$^{14}$ M$_{\sun}$
with at least 50 catalogued members.   
Our \citet{2020ApJ...900..150Y}
sample has 138 galaxies 
that meet this
criteria for massive cluster membership.
Moderate clusters are identified as those with 
halo masses between 10$^{13}$ and 10$^{14}$ M$_{\sun}$, with at least 20 known
members.   
There are 118 
\citet{2020ApJ...900..150Y}
galaxies
with z $<$ 0.03 in moderate clusters.
We label as `groups' structures in the 
\citet{2017MNRAS.470.2982L}
list with halo masses less than 10$^{13}$ M$_{\sun}$ and at least four known
members.  There are 399 galaxies 
with z $<$ 0.03 that meet these
criteria.

We define field galaxies using the relatively restrictive
definition given by 
\citet{2011MNRAS.416.2882M}:
galaxies 
that are not within 10 Mpc from the centers of any group/cluster
in the 
\citet{2017MNRAS.470.2982L}
group/cluster catalog with
more than four members and a halo mass 
greater than 10$^{12.5}$ M$_{\sun}$.
Our
\citet{2020ApJ...900..150Y}
sample has 2265 galaxies that meet
our definition of field galaxies.

\begin{figure}[ht!]
\plotone{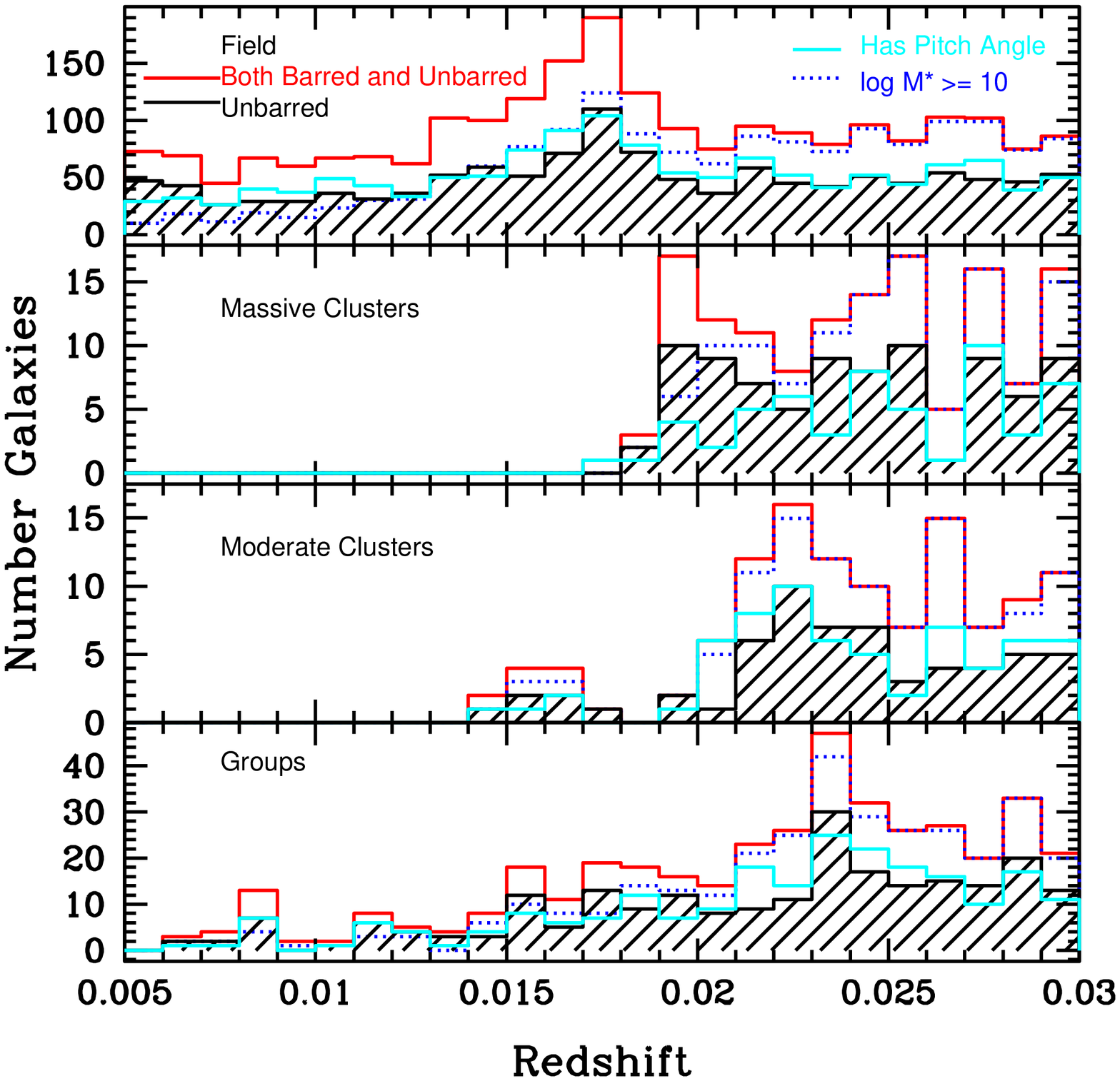}
\caption{
Histograms of redshifts for each set of galaxies.
Top panel: field galaxies.  Second panel: galaxies in massive clusters.
Third panel: galaxies in moderate clusters.  Fourth panel: Galaxies in groups.
Red histograms: all galaxies.
Hatched black histograms: unbarred galaxies.  
Dotted blue lines: galaxies with stellar mass $\ge$10$^{10}$ M$_{\sun}$.
Histograms in 
cyan:
galaxies with pitch angle measurements.
\label{fig:redshift}}
\end{figure}

Figure \ref{fig:redshift} shows histograms
of the redshifts of the galaxies in the field (top panel), in massive clusters (second
panels), in moderate clusters (third panel), and in groups (fourth panel).
No strong redshift bias
is apparent in the galaxies with pitch angle measurements.
Cluster galaxies in the sample
are at higher redshift than the field and group galaxies.

\section{Cluster Galaxies vs. Field Galaxies} \label{sec:clusters}

\subsection{Histograms of Parameters for Field vs.\ Cluster Galaxies}

\begin{figure}[ht!]
\gridline{\fig{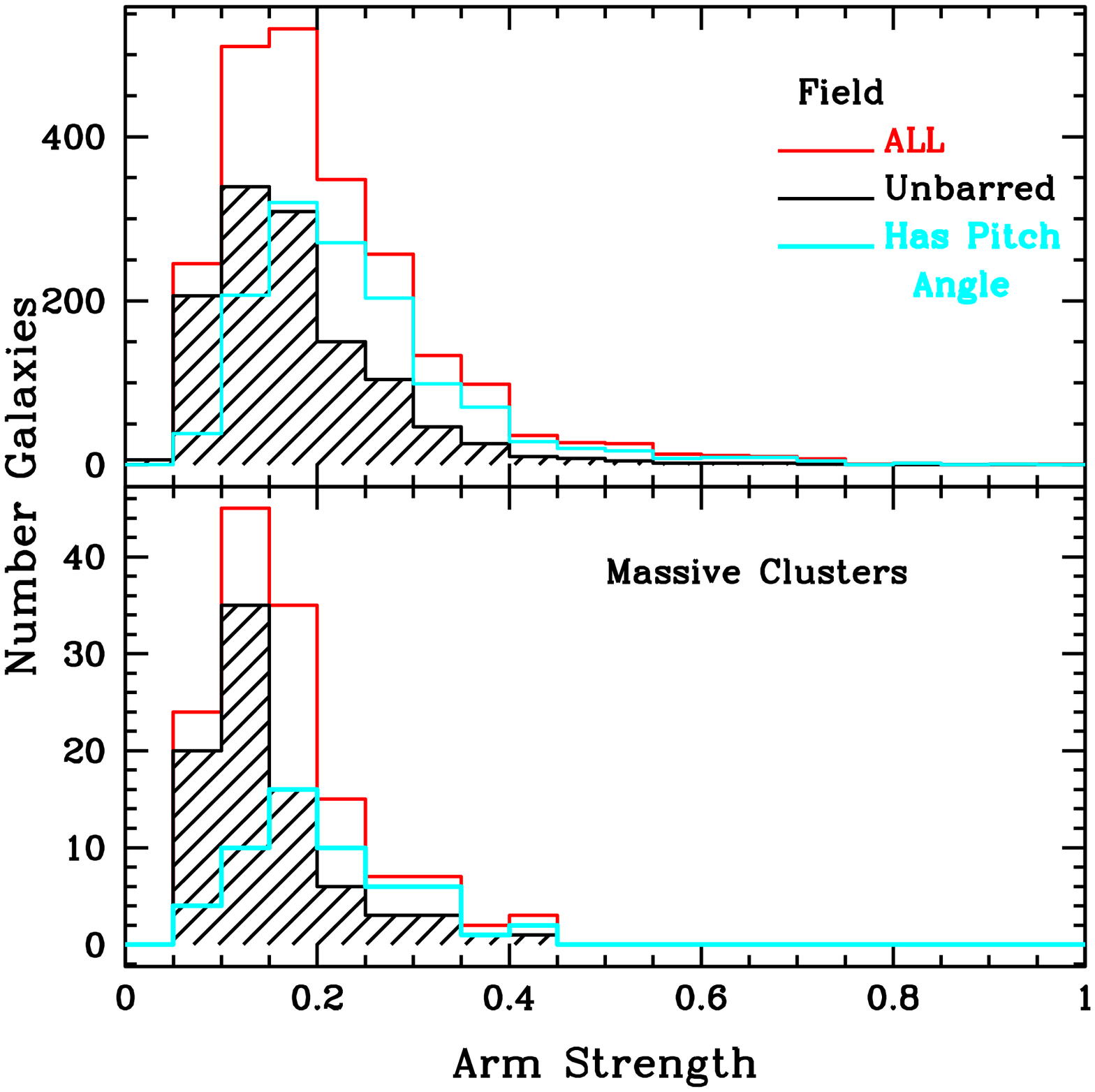}{0.33\textwidth}{(a)}
\fig{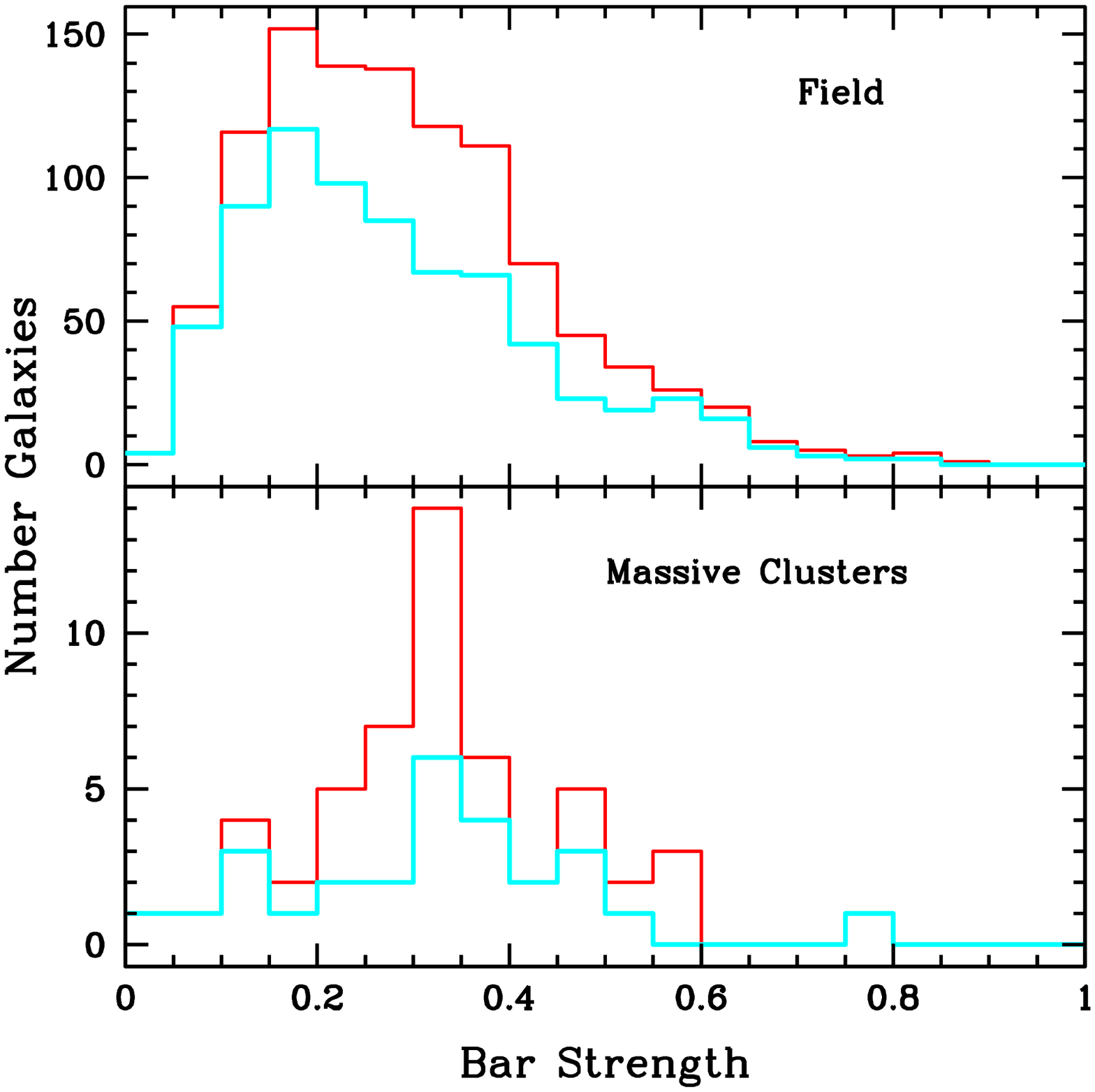}{0.33\textwidth}{(b)}
\fig{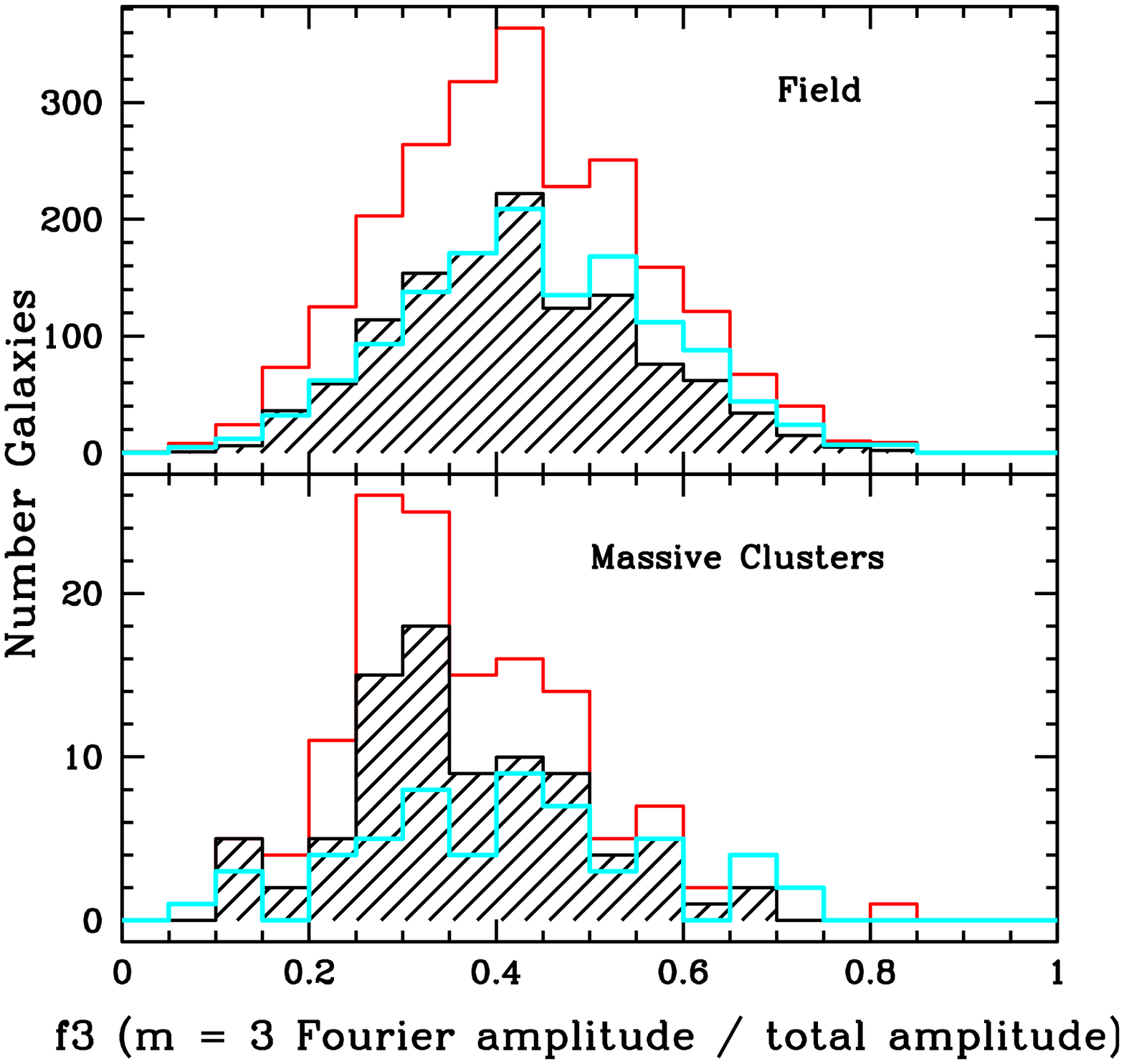}{0.33\textwidth}{(c)}
}
\gridline{\fig{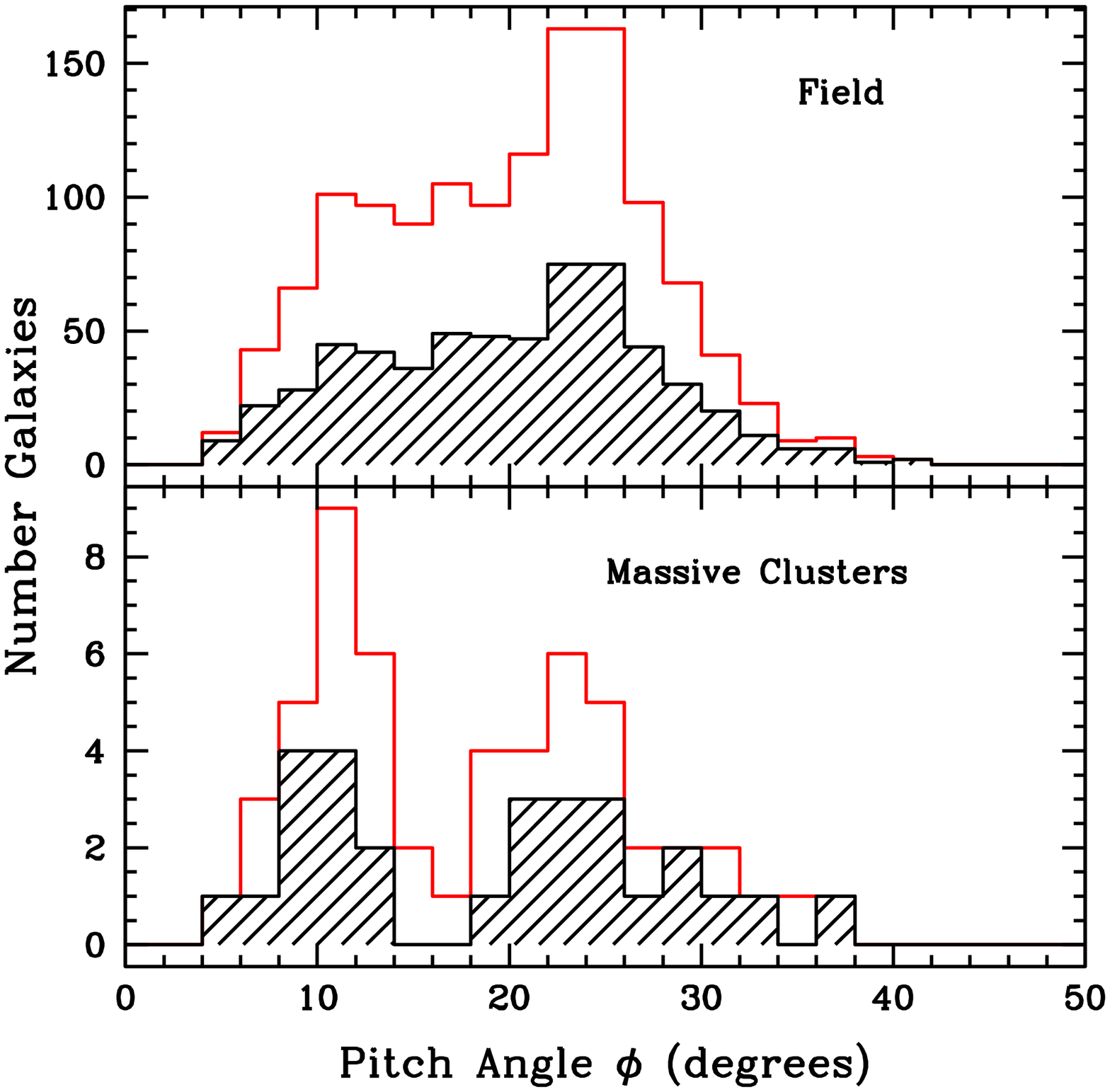}{0.33\textwidth}{(d)}
\fig{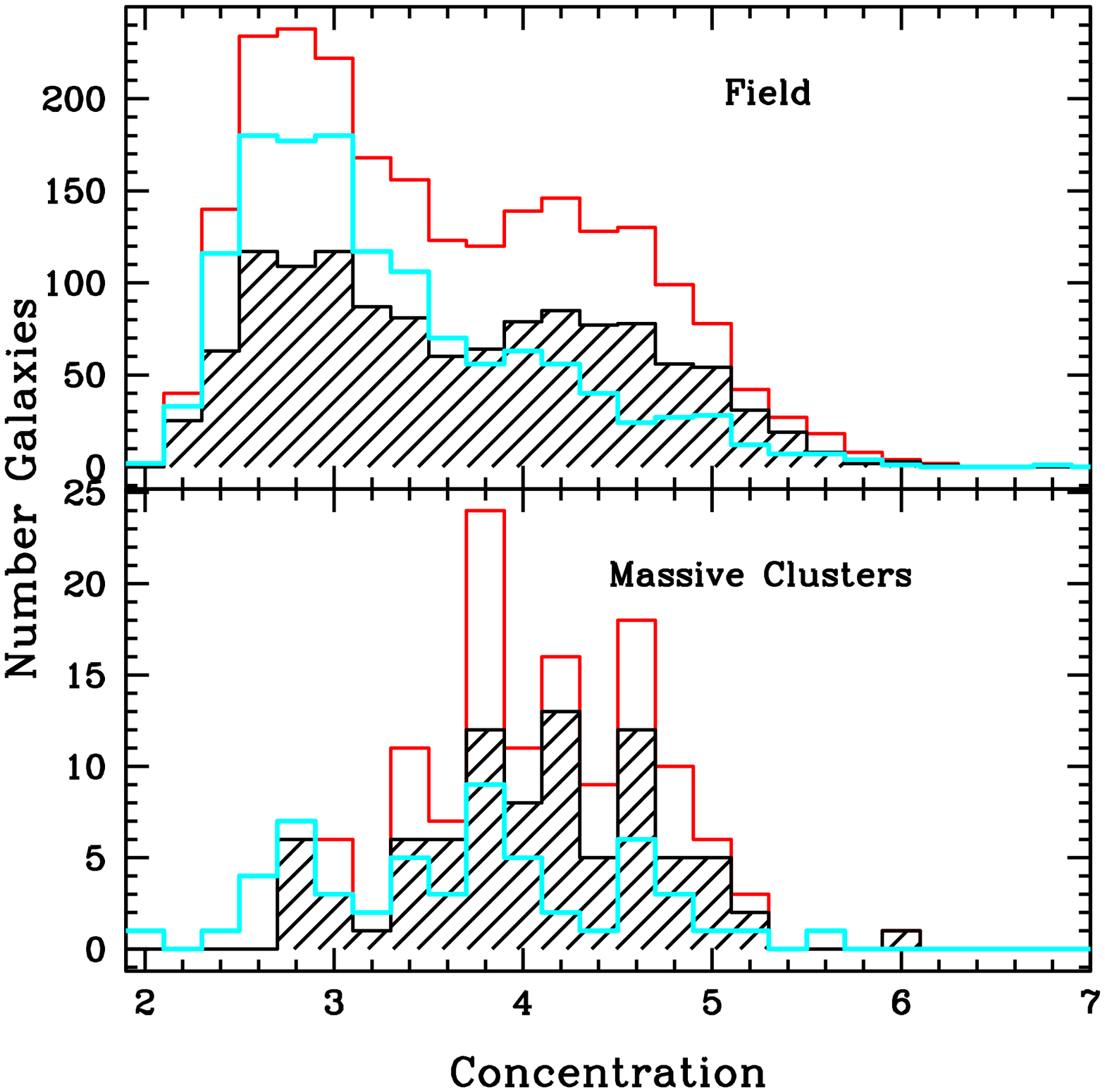}{0.33\textwidth}{(e)}
\fig{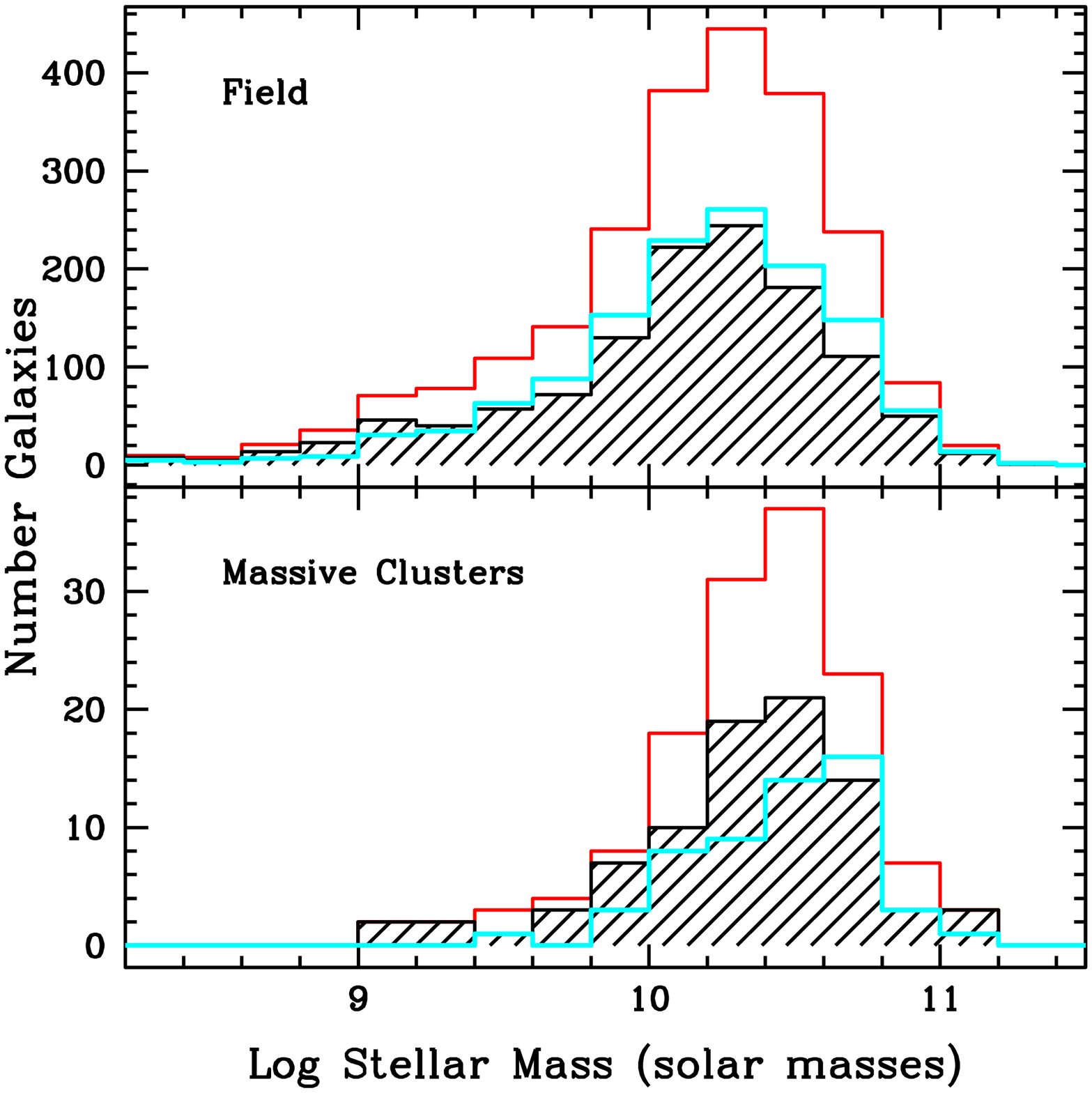}{0.33\textwidth}{(f)}
}
\caption{Histograms of spiral galaxy properties for field galaxies (top
panels) and galaxies in massive clusters (bottom panels).
a) arm strength,
b) bar strength,
c) f3,
d) pitch angle $\phi$,
e) concentration,
and 
f) stellar mass.
The red histograms give the full sample.
Hatched histograms indicate unbarred galaxies.
The cyan histograms mark galaxies with pitch angle measurements.
\label{fig:armstrength}}
\end{figure}

We start by comparing galaxies in massive clusters (halo
masses $\ge$ 10$^{14}$ M$_{\sun}$)
with those in the field.
Figure \ref{fig:armstrength} shows histograms of arm strength
(top left), bar strength (top middle), f3 (top right), $\phi$
(bottom left), 
concentration (bottom middle), and stellar
mass (bottom right) for the galaxies in the field 
(top panel in each plot) vs. galaxies
in massive clusters (bottom panel in each plot).
For the sample as a whole, there are significant differences in the 
spiral parameters of the galaxies in massive clusters
vs. the field galaxies.
Cluster galaxies tend to have lower arm strengths, 
smaller f3, and smaller pitch angles 
(Figure \ref{fig:armstrength}).
Kolmogorov-Smirnov (KS) tests for arm strength,
f3, and $\phi$ give probabilities of
2 $\times$ 10$^{-5}$, 2 $\times$ 10$^{-7}$, and 0.019, respectively,
that the cluster and field galaxies come from the same parent sample.
The differences are highly significant for 
arm strengths
and f3, and moderately significant for pitch angles. 

Among field galaxies with log M* $\ge$ 10,
738 out of 1566 (47 $\pm$ 2\%) are barred.  For cluster galaxies, 
49 out of 119 are barred (41 $\pm$ 7\%).
Among barred galaxies, cluster spirals have stronger bars
than field galaxies 
(Figure \ref{fig:armstrength}b), a difference of moderate significance
(KS 
probability of 0.020).
For both cluster and field galaxies,
the arm strengths of barred galaxies are generally stronger than
those of unbarred galaxies
(Figure \ref{fig:armstrength}a).  This difference 
was noted for the sample
as a whole by 
\citet{2020ApJ...900..150Y}.
We investigate differences in the arm strengths of barred
galaxies further 
in Section 
\ref{subsec:armstrengthclusters}.
Barred and unbarred galaxies have similar values of f3 and $\phi$
for both cluster and field galaxies
(Figure \ref{fig:armstrength}c and 
Figure \ref{fig:armstrength}d).
Galaxies with pitch angle determinations tend to have
higher arm strengths, but are not biased in f3.
Among barred field galaxies, galaxies with pitch angle measurements
tend to have slightly smaller bar strengths
(Figure \ref{fig:armstrength}b).
Among barred cluster galaxies,
there is little difference in the bar strengths of galaxies with and without
pitch angle determinations.

Cluster galaxies typically have larger concentrations
and larger stellar masses than the field galaxies 
(Figures 
\ref{fig:armstrength}e and
\ref{fig:armstrength}f). 
These differences are highly
significant (KS probabilities of 5 $\times$ 10$^{-13}$ and 
1 $\times$ 10$^{-6}$, respectively).
A large population of barred galaxies with low concentration indices
are seen in the field sample, a population not seen in the cluster sample.
Galaxies with pitch angles tend to be biased towards lower
concentrations but are not biased in stellar mass.

\subsection{Concentration vs.\ Mass for Field vs.\ Cluster Galaxies}

In the rest of Section \ref{sec:clusters}, we investigate correlations
between concentration, M*, bar strength, and the three arm parameters
for cluster vs.\ field galaxies,
searching for significant differences
with environment.
We focus on the stellar mass range 10 $\le$ log M* $<$ 11,
since there are few cluster galaxies with log M* $<$ 10 in our sample.
In Table 1, we provide the best-fit linear
relationships between various parameters for field vs.\ cluster galaxies
with 10 $\le$ log M* $<$ 11,
and compare 
unbarred galaxies with all galaxies.  
In some cases, we also investigate correlations for barred galaxies alone.
We consider a correlation/anti-correlation to be
present if either the Pearson or Spearman correlation coefficient 
is greater than 0.3 or less than $-$0.3.
Plots corresponding to these relations
are
discussed below.  
Although linear fits are an over-simplification in some cases,
they provides a first look at possible differences between the samples. 

In Figure \ref{fig:C_vs_mass_barred_unbarred},
we plot C vs.\ stellar mass for all field
galaxies (upper left panel), unbarred field galaxies (upper right),
all galaxies in massive clusters (lower left), and unbarred cluster galaxies
(lower right).
There is a `bend' in the distribution of
points for field galaxies 
(Figure \ref{fig:C_vs_mass_barred_unbarred}, top left), in that there 
is a deficiency of galaxies with log M* $<$ 10 and C $>$ 3.5
(see Section
\ref{sec:why_weaker}).
Concentration is weakly correlated with stellar mass 
for field galaxies,
and for the same stellar mass the concentration
is larger for cluster galaxies. 
The offset between the best-fit C-to-log-M*
lines for the cluster and field galaxies
is most pronounced when the full mass range is included, because
of the 
large population
of low C, low M* field galaxies.
However, this offset
is still present when only galaxies in the range 10 $\le$ log M* $<$ 11
are included.
For the same environment,
barred and unbarred galaxies follow 
approximately the same C vs.\ mass relations.

The observed differences in arm strength, f3, and pitch angle 
between cluster and field galaxies 
(Figure \ref{fig:armstrength}) 
may be caused by 
dependencies of these three parameters on 
concentration, in concert with the larger
concentrations of 
cluster spirals vs. those in the field.
In the sections below, 
we 
compare the 
bar strength, arm strength, f3, and pitch angle of cluster spirals
with those of galaxies in the field,
taking into account the differences in concentration.

\input table1_spiral_paper.tex

\subsection{Bar Strength in Field vs.\ Cluster Galaxies} \label{subsec:barstrengthclusters}

\begin{figure}[ht!]
\plotone{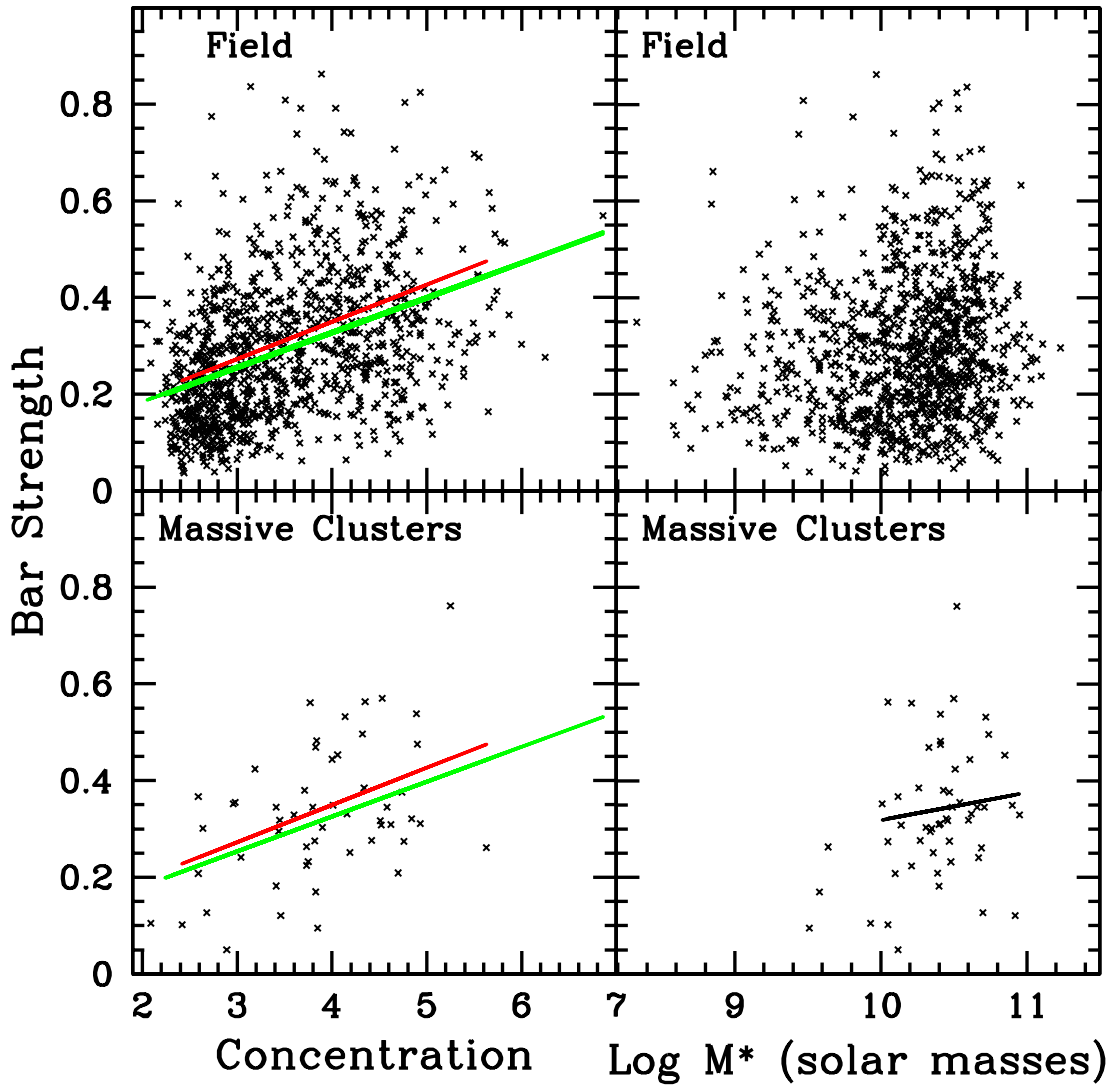}
\caption{Bar strength vs.\ concentration (left panels)
and bar strength vs.\ log M* (right panels) for barred
galaxies in the field
(top row) and in massive clusters (bottom row).
In the left panels, the green and red lines are the best-fit 
linear 
bar strength vs. C relations
for all field and cluster galaxies, respectively, with 
10 $\le$ log M* $<$ 11.
No correlation between bar strength and log M*
is found for unbarred field galaxies (upper right panel).
In the lower right panel,
the black line is the best-fit bar strength vs. log M* 
relation for unbarred cluster galaxies.
See Table 1 for more information about the best-fit relations.
\label{fig:bar_strength_vs_C}}
\end{figure}

In Figure 
\ref{fig:bar_strength_vs_C}, 
we 
plot bar strength vs.\ C (left panels) and
bar strength 
vs.\ log M* (right panels) for field galaxies (top panels) vs.\ cluster
galaxies (bottom panels).
Bars in galaxies with larger concentrations tend to be stronger 
than bars in galaxies with smaller concentrations.
The best-fit bar-strength-to-C relations for field galaxies 
and cluster galaxies 
agree within the uncertainties.
When galaxies without pitch angle measurements are
omitted, consistent results are obtained.

For field galaxies, there is no correlation between bar strength
and stellar mass 
(Figure \ref{fig:bar_strength_vs_C}, upper right panel).
For cluster galaxies, there is a very weak
correlation between bar strength and stellar mass. 

\subsection{Arm Strength in Field vs.\ Cluster Galaxies} \label{subsec:armstrengthclusters}

\begin{figure}[ht!]
\plotone{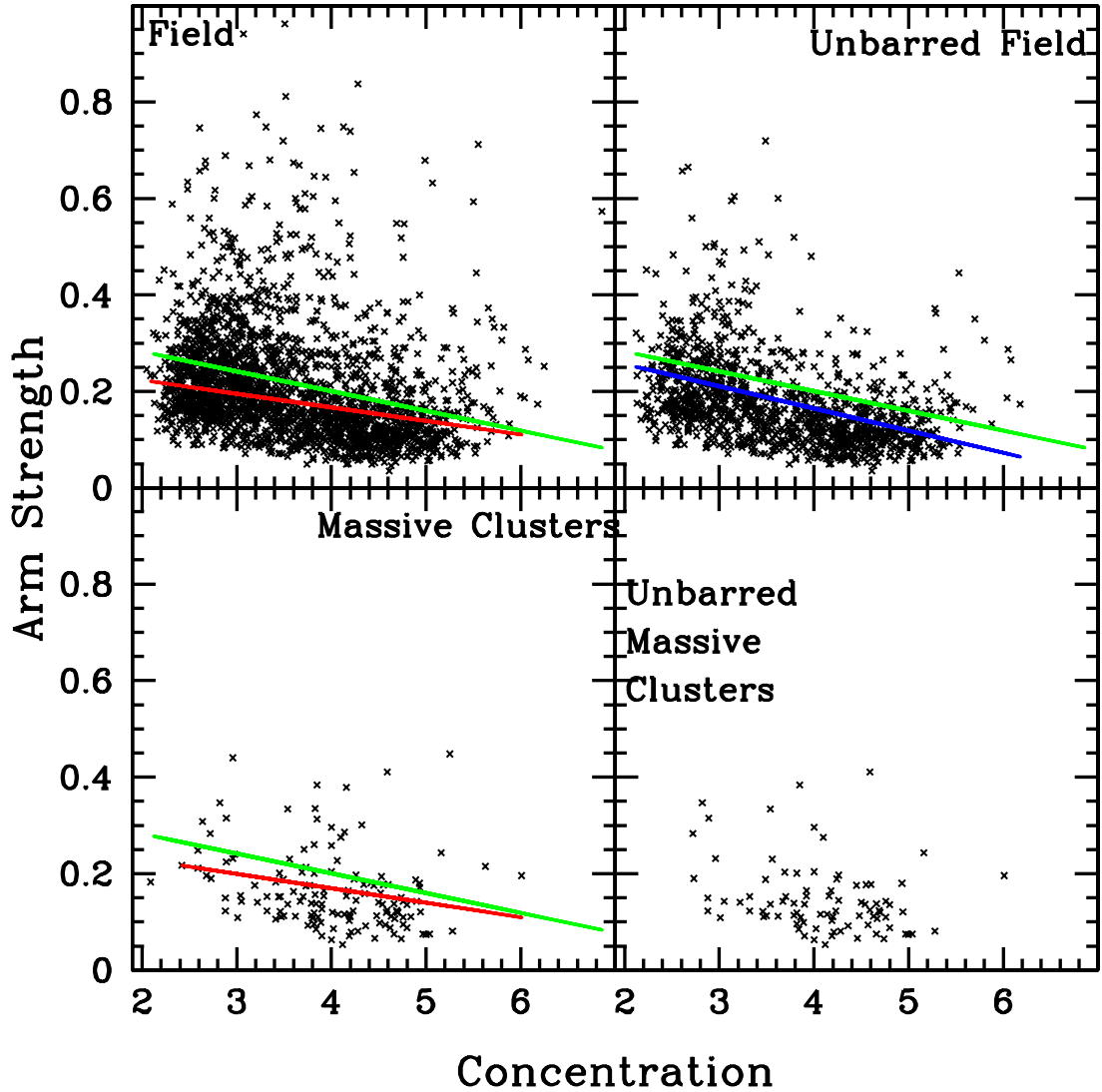}
\caption{Comparison of s vs. C for all field galaxies (top left)
all galaxies in massive clusters (bottom left),
unbarred field galaxies (top right), and unbarred galaxies 
in massive clusters (bottom right).
The green line in the two left panels and the upper right panel is
the best-fit line for all field galaxies with 10 $\le$ log M* $<$ 11.
The red line in the two left panels is the best-fit relation
for all galaxies with 10 $\le$ log M* $<$ 11 in massive clusters.
The blue line in the upper right panel is the best-fit
lines for unbarred field galaxies
for 10 $\le$ log M* $<$ 11.
The parameters of the best-fit lines are given
in Table 1.
\label{fig:s_vs_C_barred_unbarred}}
\end{figure}

For the sample as a whole, 
\citet{2020ApJ...900..150Y}
found that arm strength
is weakly anti-correlated with C.  
When we separate the galaxies by environment
(Figure \ref{fig:s_vs_C_barred_unbarred}),
similar trends are seen in each subset, with similar slopes
(see Table 1).
For a given concentration, there appears to be 
a small positive offset in s for field galaxies relative
to cluster galaxies
(Figure \ref{fig:s_vs_C_barred_unbarred}).
However, 
this offset in s (about 0.05) is small, smaller than
the median 
measurement uncertainty in s
of 0.07
from 
\citet{2020ApJ...900..150Y}.
We assess the significance of this offset below.
When galaxies without pitch angle measurements are omitted from
the samples, the anti-correlation weakens.   This may be a selection
effect, because high concentration galaxies with weak arm
strengths are more likely to lack pitch angle measurements.

\begin{figure}[ht!]
\plotone{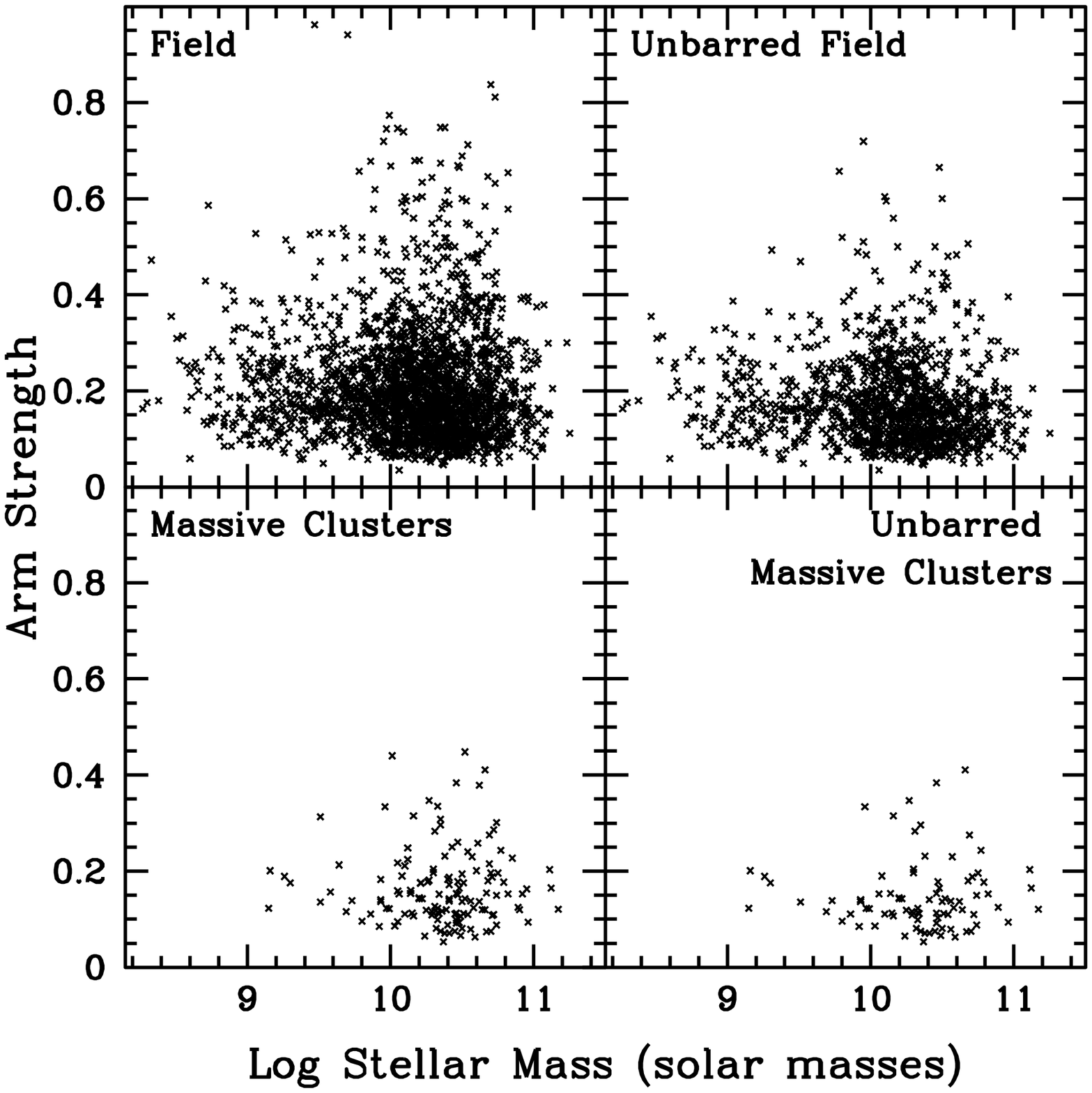}
\caption{
Comparison of s vs. stellar mass for the full set of field galaxies (top left)
and the full set of galaxies in
massive clusters (bottom left).
In the upper right, the data for the unbarred field spirals is shown,
and the lower right gives the unbarred galaxies in massive clusters.
No correlations are present (see Table 1).
\label{fig:s_vs_mass}}
\end{figure}

Among field galaxies, both the full set of galaxies 
and the unbarred galaxies show anti-correlations
between s and C,
however, 
for the same concentration,
there is an apparent upward offset in the full sample 
compared to the unbarred sample
(top two panels in Figure \ref{fig:s_vs_C_barred_unbarred}).
This implies
that barred galaxies tend to have larger arm strengths for the same
concentration.
This offset was also present in the global sample of 
\citet{2020ApJ...900..150Y}.  We investigate the significance
of this offset below.
The anti-correlation between arm strength and concentration is stronger
for unbarred galaxies alone than for the combined set, and it weakens further
when only barred galaxies are included (see Table 1).
This difference between barred and unbarred galaxies is discussed
further in Section 
\ref{sec:dis_bar}.
Note that arm strength s is not correlated with stellar mass 
in this sample 
(Figure \ref{fig:s_vs_mass} and Table 1).

\begin{figure}[ht!]
\plotone{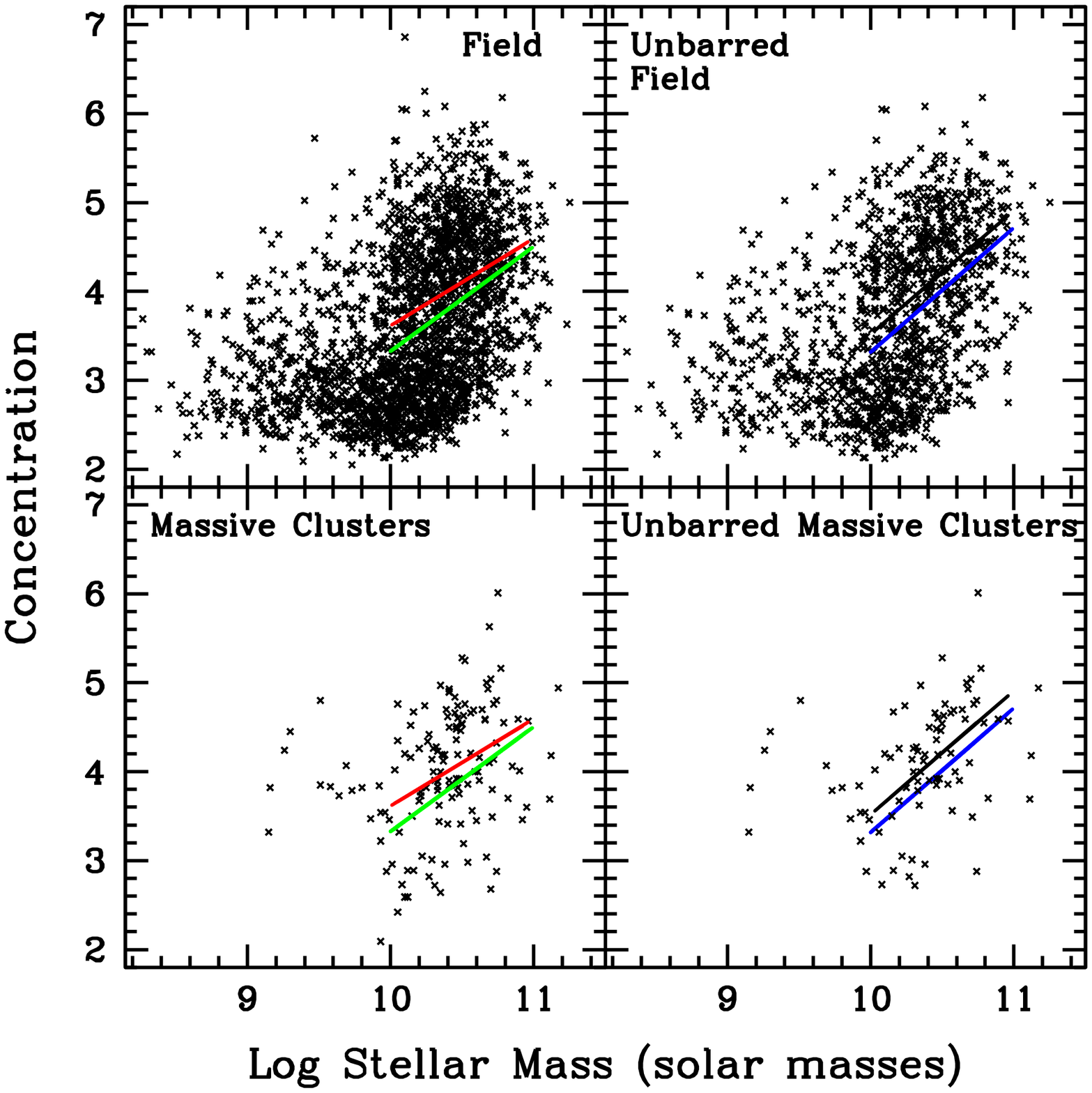}
\caption{a) Concentration vs.\ log(stellar mass)
for all field galaxies (top left),
all galaxies in massive clusters (bottom left), 
unbarred field galaxies (top right), and unbarred galaxies in massive 
clusters (bottom right).
In the left panels,
the green and red lines give the best-fit lines for 
10 $\le$ log M* $<$ 11 for 
all field
and cluster galaxies, respectively.
In the right panels,
the blue and black lines give the best-fit lines for 
10 $\le$ log M* $<$ 11 for 
unbarred field
and cluster galaxies, respectively.
The best-fit parameters are given in Table 1.
\label{fig:C_vs_mass_barred_unbarred}}
\end{figure}

We used 
both
KS
and Anderson-Darling (AD)
tests
to determine the significance of the cluster-vs.-field offset in the arm
strength-vs.-C relation noted above. 
For the field vs.\ cluster subsamples we only included
galaxies with 
10.0 $\le$ log M* $<$ 11.0
and 2.8 $\le$ C $<$ 4.8, since few cluster galaxies lie outside
of this range
 (see 
Figure \ref{fig:armstrength}).
We tried to minimize the effect of the dependence of s on C by
dividing the samples into small ranges of C, to compare
field and cluster galaxies with similar concentrations.
We settled on intervals of 0.4 in C, to ensure sufficient
galaxies were present in each bin.
We did no binning in log M* since there is no apparent mass dependence
in s. 
We then looked for evidence that the cluster and field galaxies were
drawn from the same parent sample.  

The results of these KS/AD tests are given in Table 2,
where we provide the KS and AD probabilities that
the two subsets were selected from the same parent sample.  
We also provide
the number of galaxies, the median
arm strength, and the first and third quartile
in each subset.
As the concentration increases, the median
arm strength in the bin tends to decrease.
We used the usual cutoff of $\le$5\% to distinguish samples coming from
different distributions (i.e., the probability that the two
samples came from the same parent sample is $\le$5\%).
Excluding bins with $\le$5 galaxies in one of the subsets,
in no case is there a significant difference between the two
samples.  
Because of the different concentration vs. mass relations for cluster
compared to field galaxies 
(Figure \ref{fig:C_vs_mass_barred_unbarred}), within a given mass range
field galaxies show a different distribution of concentrations
than cluster spirals.  These differences in concentration 
appear to be responsible for the global difference in arm strengths between
galaxies in the two environments (Figure 
\ref{fig:armstrength}e).

Our sample contains
only 95 galaxies in massive clusters with z $<$ 0.03, 
10 $\le$ log M* $<$ 11, and 2.8 $\le$ C $<$ 4.8.
Of these 95 galaxies, 29 have 2.8 $\le$ C $<$ 3.8 and 66 have 
3.8 $\le$ C $<$ 4.8.
Of these 95 galaxies, 60 are in the 
10.0 $\le$ log M* $<$ 10.5 range, and 
35 in the 10.5 $\le$ log M* $<$ 11 range.
These low numbers limit our statistics.

Since cluster galaxies in this sample have higher
redshifts on average than the field galaxies (Figure \ref{fig:redshift}), to 
minimize possible differences in arm strength measurements
due to resolution effects,
we repeated the calculations while
limiting both samples to
z $\ge$ 0.019, again with our upper limit of z $<$ 0.03.
Above z $\ge$ 0.019, the distribution in redshift
for the field galaxies resembles that of the cluster galaxies (Figure \ref{fig:redshift}).
We get consistent results for the 0.019 $\le$ z $<$ 0.03 subset as for
the 0 $\le$ z $<$ 0.03 set (Table 2).

As an alternative method of comparing galaxies in the two environments, 
for each subset of cluster galaxies we extracted a `concentration-matched'
sample of field galaxies.   For each cluster galaxy in the 10 $\le$ log M*
$<$ 11 range,
we randomly selected a field galaxy with similar concentration (within 0.2 in C)
for our comparison sample.
We then compared the distribution of arm strengths
for the cluster galaxies vs. those in the matched set of field galaxies using 
KS/AD tests.
We repeated this random selection of matched samples 1000 times per subset,
and calculated the median KS/AD probabilities for each case as well as 
the first and third
quartile probabilities.
We then divided the sample into smaller ranges of C and repeated.
In no case was the median probability for the 1000 runs less than 5\%,
so we cannot rule out that our cluster and field
galaxies come from the same parent sample.  

We also ran KS/AD tests comparing barred and unbarred galaxies
for the same concentration range, this time comparing field to field
galaxies.   
These results are presented in Table 3.
These reproduce the results that 
\citet{2020ApJ...900..150Y}
found for the sample
as a whole: barred galaxies in this sample have significantly stronger
arms than the unbarred galaxies, for the same concentration.
For all concentration bins larger than the 2.8 $\le$ C $<$ 3.2 bin,
there are significant differences in the arm strengths of the barred
vs.\ unbarred galaxies.

\begin{figure}[ht!]
\plotone{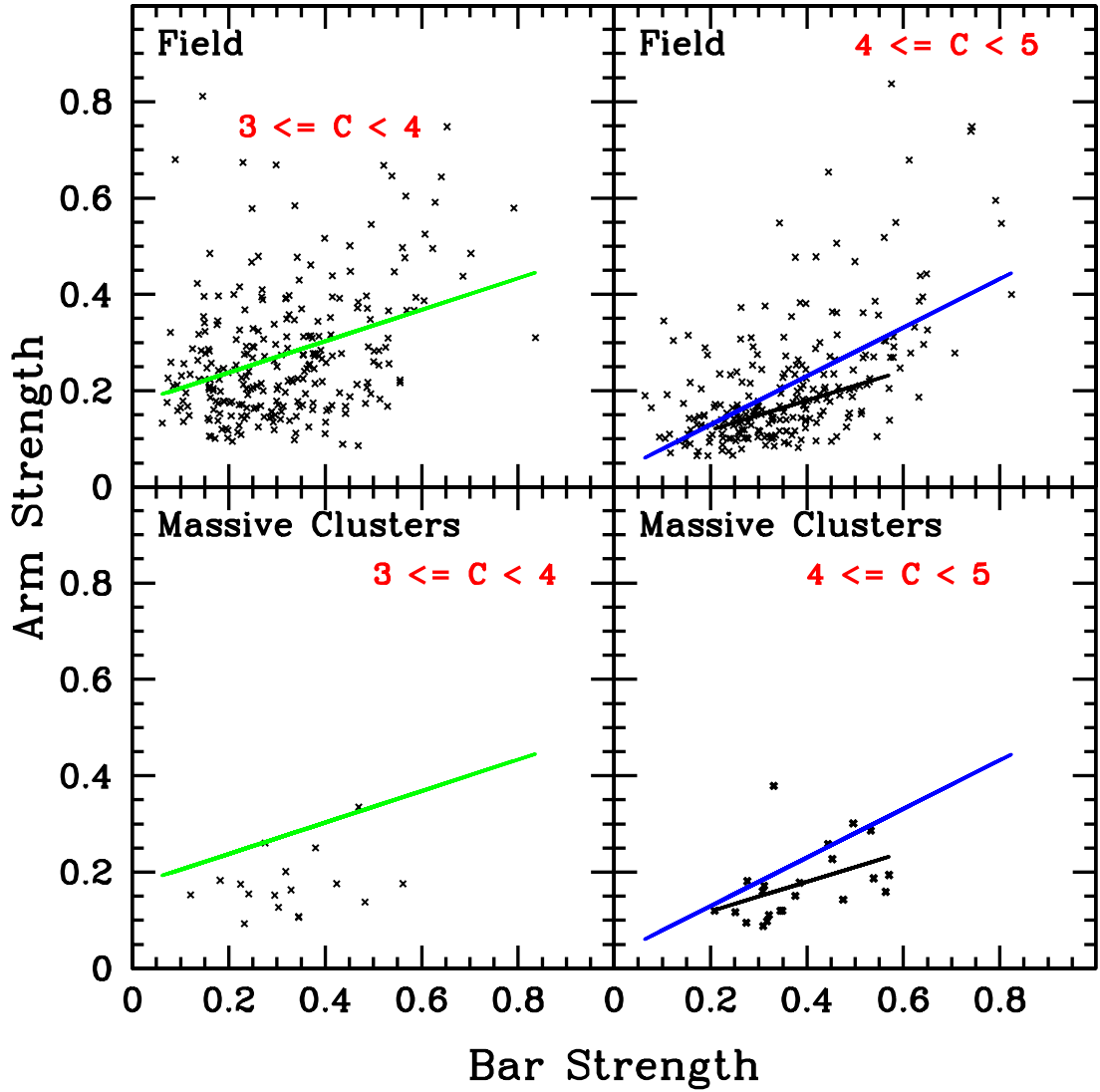}
\caption{
Comparison of arm strength s vs. bar strength for field galaxies
(top row) and galaxies in massive clusters (bottom row).
The left panels include galaxies with 3 $\le$ C $<$ 4,
while the right panel includes
galaxies with 4 $\le$ C $<$ 5.
In all panels, the sample is limited to galaxies with
10.0 $\le$ log M* $<$ 11.0.
The green line in the left panels is the best-fit line for
field galaxies with 
3 $\le$ C $<$ 4.
The blue and black lines in the right panels are the best-fit
lines for 
field and cluster galaxies, respectively, for
4 $\le$ C $<$ 5.
Best-fit parameters of these lines are provided in Table 1.
\label{fig:s_vs_bar}}
\end{figure}

As noted earlier, the bars of barred cluster galaxies are stronger
on average than the bars of barred galaxies in the field
(Figure \ref{fig:armstrength}).
In Figure \ref{fig:s_vs_bar}, 
we compare arm strength to bar strength
for the galaxies with
bars. 
We divide the
galaxies into field galaxies (top panels) 
and cluster galaxies (bottom panels), and plot galaxies
with smaller concentration (3 $\le$ C $<$ 4) (left panels) 
vs.\ galaxies with
larger concentrations (4 $\le$ C $<$ 5) (right panels),
limiting the stellar mass range
to 10.0 $\le$ log M* $<$ 11.0.
Figure \ref{fig:s_vs_bar} shows
a positive correlation for field galaxies between
arm strength and bar strength, with a stronger correlation for
galaxies with larger concentrations (see Table 1).
For cluster galaxies, no correlation is seen for the lower concentration
bin, but one is present for the 4 $\le$ C $<$ 5 bin (Table 1).
The best-fit lines for field galaxies lie at higher arm strengths
than most of the data points for cluster galaxies.
To test whether this difference is significant,
we ran KS/AD tests comparing arm strengths for subsets of field vs.\ cluster
galaxies
within relatively narrow ranges of C and bar strength.
In most cases, no significant differences are detected.
The one exception is 
for the range 3 $\le$ C $<$ 4 and 0.2 $\le$
bar strength $<$ 0.4, where 
KS/AD tests give probabilities of 0.016/0.0015
that the two subsets come from the same parent sample.
The number of galaxies in these subsets 
are small (125 field and 11 cluster
galaxies), however.
Therefore,
for the 
set of 
41 barred cluster galaxies with 2.8 $\le$ C
$<$ 5, 
we constructed a `matched' field sample 
by randomly selecting
a `matching' 
field galaxy for each cluster galaxy, i.e., 
finding a field galaxy
with a similar concentration (within 0.2) and a similar
bar strength (within 0.1) as the cluster galaxy.
We ran KS/AD tests comparing the arm strengths
of the `matched' sample with the cluster
sample.  We repeated this process 1000 times, and find
median KS/AD probabilities of 0.16/0.11, thus 
we cannot rule out that cluster and field galaxies have
similar arm-to-bar strengths for a given concentration.

\input table2_spiral_paper.tex
\input table3_spiral_paper.tex

\subsection{f3 in Cluster vs. Field Galaxies} \label{subsec:f3clusters}

\begin{figure}[ht!]
\plotone{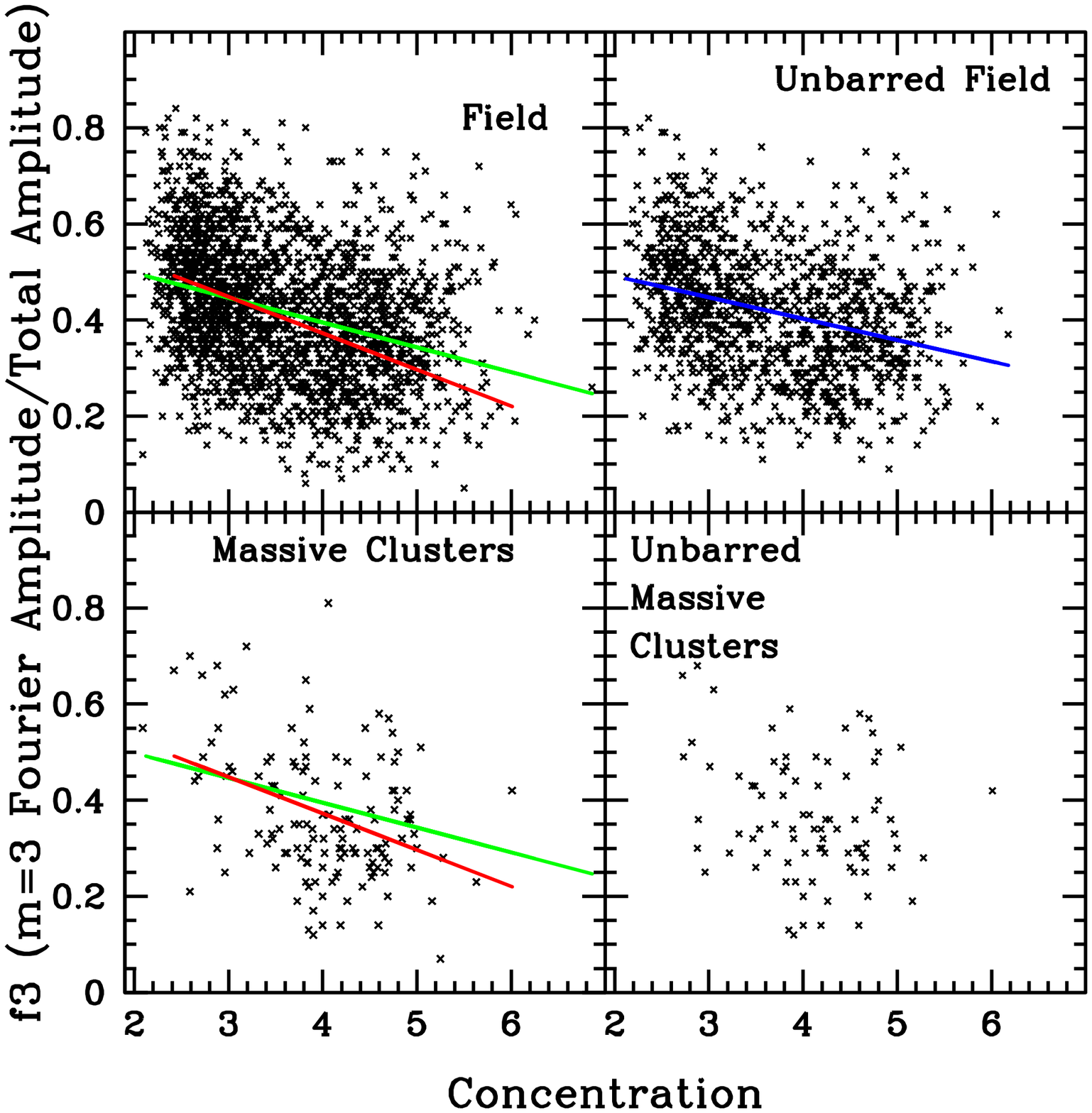}
\caption{Plots of f3 vs.\ concentration for all field galaxies (top left),
unbarred field galaxies (top right), 
all galaxies in massive clusters (bottom left),
and unbarred galaxies in massive clusters (bottom right).
The green and red lines in the two left panels
are the best-fit lines for the full set of field and cluster galaxies,
respectively, with 10 $\le$ log M* $<$ 11.
The blue line in the upper right panel is the best-fit line for
10 $\le$ log M* $<$ 11
unbarred field galaxies.
No significant correlation is seen for unbarred cluster galaxies.
Best-fit parameters are available in Table 1.
\label{fig:f3_vs_C}}
\end{figure}

\citet{2020ApJ...900..150Y}
found that 
the normalized m=3 Fourier amplitude f3 
is weakly inversely correlated with C.
They also found that,
for
high concentrations (C $>$ 4), unbarred galaxies
have slightly higher f3 than barred galaxies.
In Figure \ref{fig:f3_vs_C}, we compare f3 to C for cluster galaxies
separately from field galaxies. 
Weak anti-correlations are visible in both cases.
The best-fit relation for field galaxies is slightly
flatter than that for cluster galaxies; for high concentrations,
field galaxies have higher f3 values than cluster galaxies.
However, the slopes of the best-fit relations 
are consistent within the uncertainties (Table 1).
The f3 vs. C relation unbarred field galaxies
are also similar (top right panel).
For unbarred cluster galaxies the correlation weakens. 
Omitting galaxies without pitch angles does not change
the results significantly.
The parameter f3 is not correlated with stellar mass (see Table 1;
plots not shown).

Cluster galaxies have
higher concentrations for a given mass 
(Figure \ref{fig:C_vs_mass_barred_unbarred}),
and larger C means lower f3
(Figure \ref{fig:f3_vs_C}).
We search for significant differences between
cluster and field galaxies
accounting for this C dependency by again subdividing the sample
into bins of 0.4 in C.
Within these bins, we compared the f3 values using KS and AD 
tests, limiting
the sample to only galaxies with 0.019 $\le$ z $<$ 0.03 to
minimize resolution differences as before,
and separating barred from unbarred galaxies.
For most of the bins, the KS/AD probabilities exceed 0.05
(table of probabilities not shown).
KS/AD probabilities of 0.003/0.013 are obtained for 
barred galaxies with 4.0 $\le$ C $<$ 4.4, with 
field and cluster sample sizes of 62 and 9, respectively.
For 
unbarred galaxies with 
4.4 $\le$ C $<$ 4.8, the KS/AD probabilities are
0.006/0.02 (with 81 field galaxies and 18 cluster galaxies).
For these 
cases, 
the field
galaxies have slightly higher median f3 values than 
cluster galaxies,
with 
differences in f3 of 0.095 and 0.09, respectively.   
These differences are similar
to the median measurement uncertainty in f3 for 
the sample of 0.07.

As a further test of the significance of this result,
we then selected concentration-matched subsets of field galaxies.
We started by combining barred and unbarred galaxies together and
focused on galaxies with 0.019 $\le$ z $<$ 0.03
and 10 $\le$ log 
M* $<$ 11.
For each of the 
59 cluster galaxies with large
concentration (4 $\le$ C $<$ 5),
we randomly-selected field galaxies with concentrations that matched with a difference
of less than
0.2.   
We then compared the f3 distribution
of this concentration-matched sample with that of the cluster galaxies.
We ran this test 1000 times with different random numbers.  
For the KS test, the median probability was 0.061, while the AD test
gave a median probability of 0.047.   
When we treated the 21 barred and 38 unbarred galaxies separately,
we obtained median KS/AD probabilities of 0.23/0.16 and 0.26/0.17, respectively.
Although there 
may be a weak difference in f3 at
high concentrations, 
we conclude that the f3-to-concentration 
relations for 
cluster and field galaxies are likely consistent within the uncertainties, even
for high concentration galaxies.
A larger sample of galaxies would be useful
to investigate this issue further.

We also compared f3 to arm strength for galaxies in the field
and in clusters; no correlations are present 
(see Table 1).
\citet{2020ApJ...900..150Y}
showed that f3 is weakly
anti-correlated with bar strength (Pearson correlation
coefficient = $-$0.32).   When we restrict the sample
to 10 $\le$ log M* $<$ 11, separate into field vs.\ cluster galaxies,
and separate into 3 $\le$ C $<$ 4 and 4 $\le$ C $<$ 5, 
the correlation
weakens further (see Table 1).
This very weak anti-correlation may be
an indirect consequence of the f3 vs. concentration anti-correlation,
and the bar strength to concentration correlation.

\subsection{Pitch Angle $\phi$ in Cluster vs. Field Galaxies} \label{subsec:phiclusters}

\begin{figure}[ht!]
\plotone{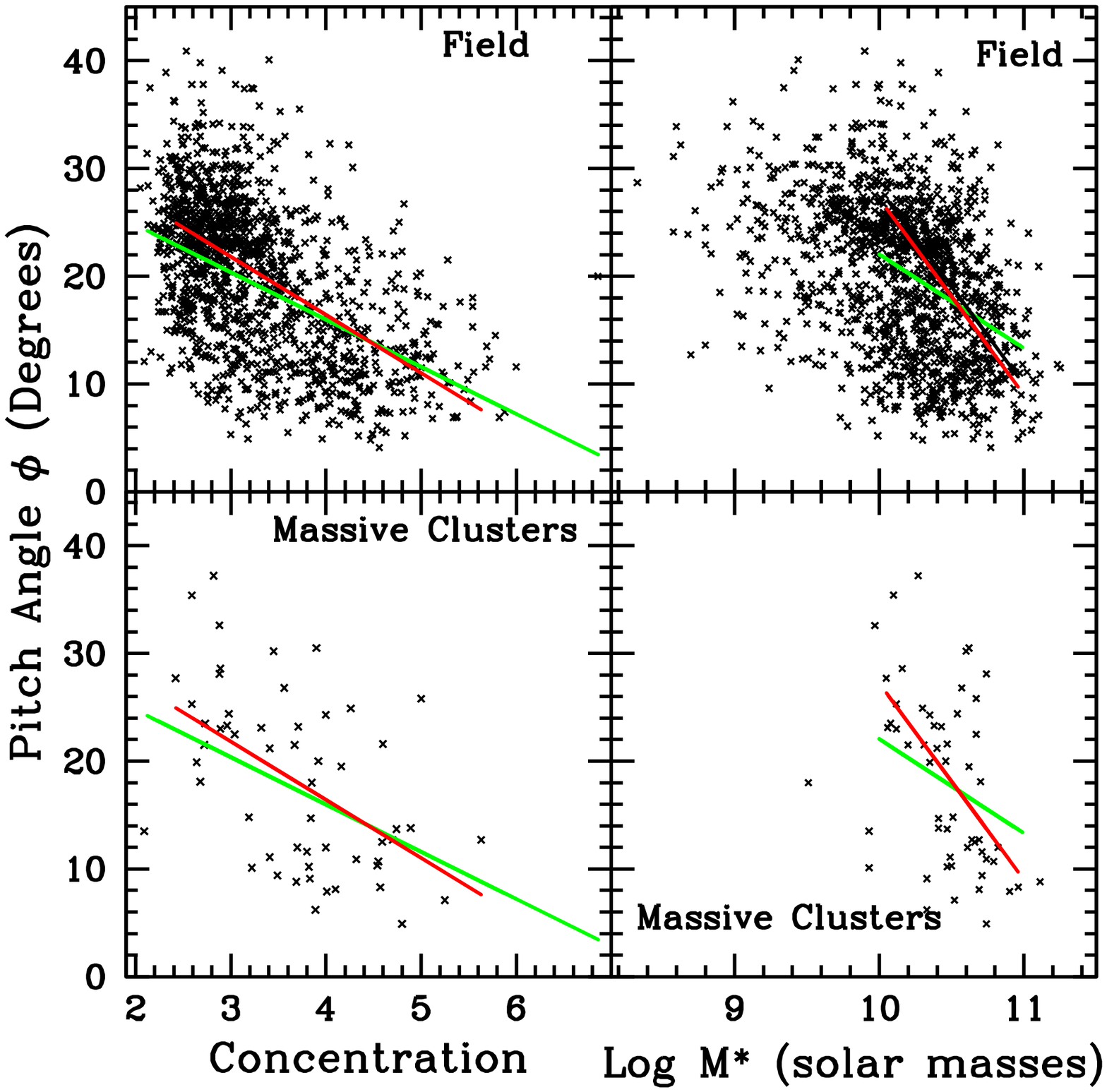}
\caption{
Pitch angle $\phi$ vs.\ concentration (left panels) and 
$\phi$ vs.\ log M* (right panels) for all
field galaxies (top row) and
all 
galaxies in massive clusters (bottom row).
The green and red lines in the left panels are the best-fit lines
for $\phi$ vs.\ C for all
field and cluster galaxies, respectively, 
with 10 $\le$ log M* $<$ 11.
The green and red lines in the right panels are 
the best-fits for $\phi$ vs.\ log M* for
all field and cluster galaxies, respectively,
with 10 $\le$ log M* $<$ 11.
See Table 1 for the parameters of these fits.
\label{fig:phi_vs_C_and_M}}
\end{figure}

As 
\citet{2020ApJ...900..150Y} note,
for the entire sample 
the pitch angle is inversely
correlated with concentration index.
In  
Figure \ref{fig:phi_vs_C_and_M}, we separate the field galaxies (top panels)
from the cluster galaxies (bottom panels), and 
plot $\phi$ vs.\ C (left panels) and $\phi$ vs.\ log M* (right panels).
The best-fits
$\phi$-to-C lines 
for field and cluster galaxies
agree within the uncertainties (Table 1).
Barred and unbarred field galaxies have similar $\phi$ vs.\ C relations,
as do barred and unbarred cluster galaxies
(Table 1; plots not shown).
The larger observed pitch angles of field galaxies 
(Figure
\ref{fig:armstrength}d)
can likely be accounted for
by their smaller concentrations
(Figure \ref{fig:armstrength}e).
We test this hypothesis further below.

Cluster galaxies on average are more massive 
(Figure \ref{fig:armstrength}), and
pitch angle is anti-correlated with mass 
(Figure \ref{fig:phi_vs_C_and_M}, right panels).  
The $\phi$ vs.\ mass relation for cluster galaxies
appears steeper 
than for field galaxies (Figure
\ref{fig:phi_vs_C_and_M}, right panels).
Barred and unbarred field galaxies follow the same 
$\phi$ vs.\ mass relation; barred and unbarred cluster galaxies
also appear to lie along the same path in a $\phi$ vs.\ mass diagram
(Table 1, plots not shown).

To search for statistically-significant differences between
the pitch angles of cluster vs.\ field galaxies, 
we again separated the sample into the same subsets in C, lumping
together
all galaxies with 10 $\le$ log M* $<$ 11.  We did not see any
statistically-significant differences for both the KS and AD tests 
for any bin with more than five galaxies.
As the concentration goes up, the median pitch angle
in the bins tends to decrease.
Since fewer galaxies
have $\phi$ measurements compared to f3 and arm strength, the statistics
are poorer for $\phi$.
We also tried
dividing the sample into two bins in C (3 $-$ 4 and 4 $-$ 5),
two bins in log mass (10.0 $-$ 10.5 and 10.5 $-$ 11.0), and
barred vs. unbarred.  
In no case did we
find a statistically-significant difference when we had more than
5 members in a sample.  
This may indicate that global differences in pitch angle between
field and cluster galaxies 
(Figure \ref{fig:armstrength})
are due to different concentrations
and different masses in these two populations of galaxies.

\citet{2020ApJ...900..150Y}
did not find a significant correlation between pitch
angle and arm strength (Pearson correlation coefficient of 0.23), but saw a very
weak correlation between pitch angle and f3 (Pearson correlation coefficient 0.30). 
When we consider only field galaxies with
10 $\le$ log M* $<$ 11, this weak correlation remains (Table 1).
These statistics might be biased because some galaxies
lack pitch angle measurements, which tend to have weaker arms
and lower f3 values.
Of the 10 $\le$ log M* $<$ 11 galaxies with
arm strengths $>$0.15, only 25\% are lacking $\phi$, but 
68\%
of the galaxies 
with arm strengths $\le$0.15 are missing $\phi$.
Of the galaxies in this mass and z range with f3 $>$ 0.4, 30\% are missing $\phi$,
while 50\% of the galaxies with f3 $\le$ 0.4 lack $\phi$.
\citet{2020ApJ...900..150Y} also saw a 
weak anti-correlation between 
pitch angle and bar strength (Pearson coefficient $-$0.36).  
This weak anti-correlation remains when separated into field
vs.\ cluster galaxies, but only for lower concentration (3 $\le$ C $<$ 4) 
galaxies (see Table 1).
Cluster and field galaxies give consistent best-fit relations, within the relatively 
large uncertainties.

\subsection{Galaxies in the Interiors of Clusters vs. Galaxies in Cluster Outskirts}

One possible reason we do not find strong differences between the spiral
parameters of
the cluster vs.\ field galaxies might be that we have been
liberal in our designation of cluster galaxies, including
galaxies out to 3R$_{\rm 200}$ in radius and 
velocity difference from the cluster velocity of $|$$\Delta$(V)$|$ = 
3$\sigma$$_{\rm V}$.  The inclusion of
galaxies in the outskirts may have diluted the sample sufficiently
so that a difference is not detectable.
To assess this possibility and to
search for trends with position within the clusters, we compared
the 27 galaxies in the inner regions of the massive
clusters (R/R$_{\rm 200}$ $<$ 1, $|$$\Delta$(V)$|$/$\sigma$$_{\rm V}$
$<$ 1) and the 31 galaxies in the cluster outskirts (2 $\le$ 
R/R$_{\rm 200}$ $<$ 3) to the field galaxies.
We do not see significant differences in the best-fit relations for
s vs.\ C, f3 vs.\ C,
and $\phi$ vs.\ C for these subsets compared 
to each other or to the field galaxies.

\subsection{Groups and Moderate Mass Clusters} \label{sec:groups}

We also compare the C vs.\ stellar mass relations
of galaxies in groups and galaxies in moderate mass clusters with the relations
for field galaxies and galaxies in massive clusters.
Within the uncertainties, galaxies in groups follow the same trend as field galaxies.
Galaxies in moderate clusters may have properties in between those
of field galaxies and massive clusters, but this is uncertain
because of the relatively small number of galaxies in these moderate clusters.

The s vs. C relation for group galaxies is similar to that for field galaxies
within the uncertainties.
Galaxies in moderate clusters have a best-fit relation that is slightly
steeper than that of field galaxies, but consistent within 1.5$\sigma$.
Both group galaxies and galaxies in moderate clusters follow 
the same f3 vs.\ C trend as field galaxies.
The best-fit f3-to-C relations 
are not changed significantly when galaxies without
pitch angles are omitted.
The $\phi$ vs.\ C relations for groups and moderate
clusters matches those of field and massive cluster galaxies.

\section{Specific SFR vs.\ Arm Parameters and AGN Activity} \label{sec:sSFR}

\begin{figure}[ht!]
\plotone{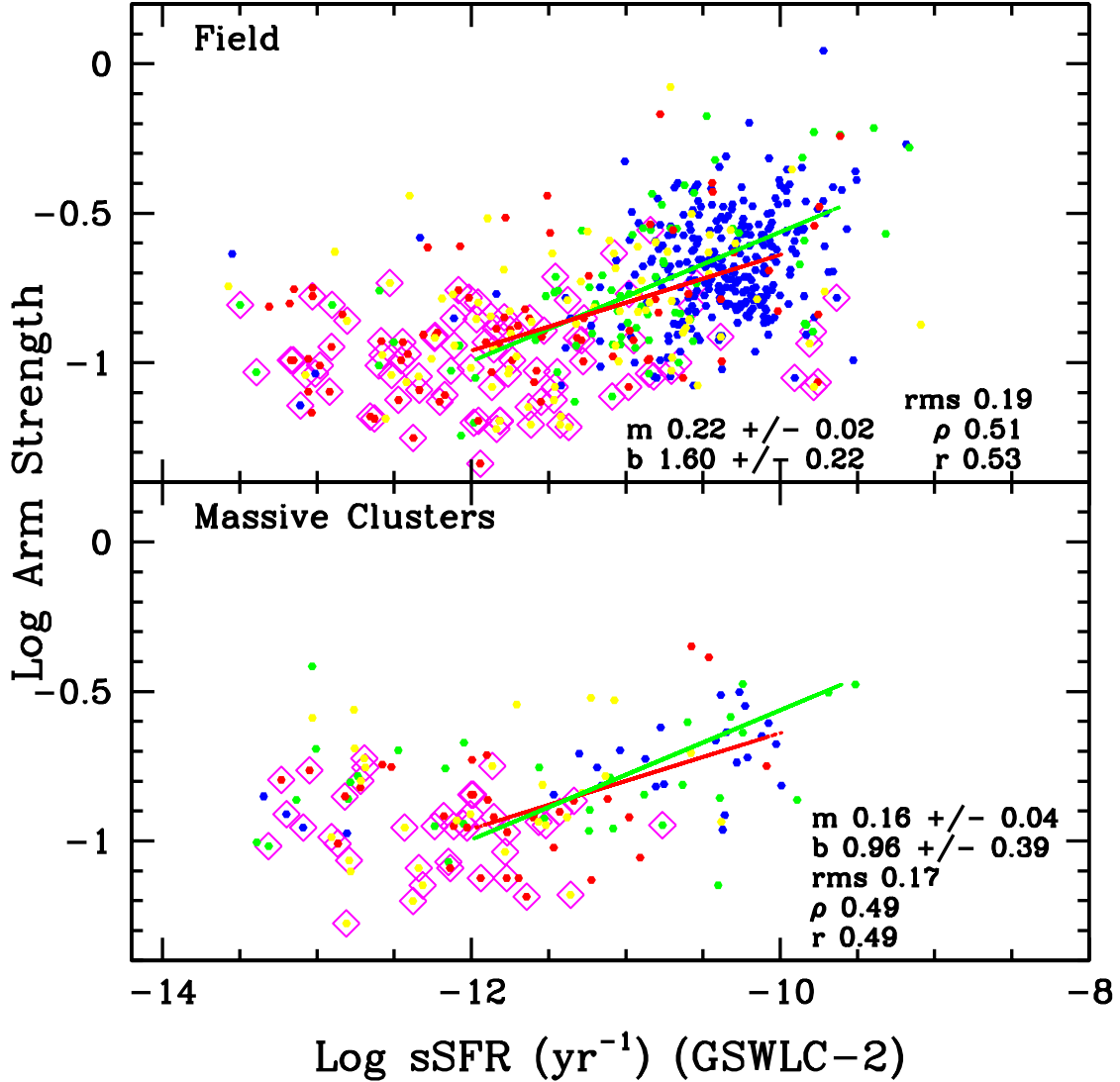}
\caption{The log of the spiral arm strength vs.\ log sSFR 
for 
field galaxies (top panel) and galaxies in massive
clusters (bottom panel).
The sSFRs in these plots come from the GSWLC-2.
The green and red lines in both plots are the best fits for
field and cluster galaxies, respectively, with log sSFR $\ge$ -12,
only including galaxies with 10 $\le$ log M* from the NSA $<$ 11.
The slope (m), y-intercept (b), and rms of the best-fit lines 
are printed on the corresponding plot,
along with the Spearman ($\rho$) and 
Pearson (r) correlation coefficients.
The data points are color-coded based
on concentration (red: C $\ge$ 4.5; yellow: 
4.0 $\le$ C $<$ 4.5; green: 3.5 $\le$ C $<$ 4.0;
blue: C $<$ 3.5).
The galaxies marked by open magenta diamonds were classified
by 
\citet{2020ApJ...900..150Y}
as either S0-, S0, or SB0.
\label{fig:sSFR}}
\end{figure}

\citet{2021ApJ...917...88Y}
showed that spiral arm strength 
tends to increase 
with increasing sSFR.
We 
cross-correlated the sample
with the 
GALEX-Sloan-WISE 
Legacy Catalog
version 2 
(GSWLC-2)\footnote{https://salims.pages.iu.edu/gswlc/}
\citep{2016ApJS..227....2S,
2018ApJ...859...11S},
using a 5$''$ search radius.
The GSWLC-2 provides SFRs and stellar masses
from UV/optical/IR population synthesis using the CIGALE population
synthesis code\footnote{https://cigale.lam.fr} 
\citep{2009A&A...507.1793N,
2019A&A...622A.103B}.
In Figure \ref{fig:sSFR}, we plot arm strength vs.\ sSFR
from the GSWLC-2
for field galaxies (top panel) and galaxies in clusters (bottom
panel).
A trend is apparent,
in that galaxies with higher sSFR have larger arm
strengths.
The cut-off between `quenched' and `star-forming' galaxies 
is frequently set to log sSFR = -11 
\citep{2011MNRAS.413..996M,
2012MNRAS.424..232W, 2013MNRAS.432..336W}.
Below
log sSFR = -12 estimates of the sSFR are very uncertain 
\citep{2007ApJS..173..315S, 2016ApJS..227....2S}, thus in
deriving best-fit relations between sSFR and other parameters 
we only consider galaxies with log sSFR $\ge$ -12.
In Figure 
\ref{fig:sSFR}, 
the best-fit line for field galaxies is consistent
within the uncertainties with that of cluster galaxies.
Including only galaxies with pitch angle measurements
does not change these results significantly.
The S0-, S0, and SB0 galaxies 
(open magenta diamonds)
tend to have weak arm strengths
and low sSFRs.
In Appendix \ref{sec:appendix_sSFR_A}, 
we derive a second estimate of sSFR, and 
use this new estimate in 
an alternative version of 
Figure \ref{fig:sSFR} in
Appendix \ref{sec:appendix_sSFR_B}.

Figure \ref{fig:sSFR}
shows that both the
field and cluster samples contain numerous galaxies that are quenched
yet retain their disk structure and a weak spiral pattern.   
As noted by 
\citet{2021ApJ...917...88Y},
these quenched galaxies tend to have red optical colors.
In Appendix 
\ref{sec:appendix_sSFR_A},
we plot the sample galaxies on the galaxy main sequence,
and confirm that many of these galaxies lie below the main sequence.
Based on Figure \ref{fig:sSFR}, 
we conclude that during the quenching process,
field and cluster galaxies follow approximately the same
arm strength vs.\ sSFR relation.  
We reach the same
conclusion when we use our alternative derivation of sSFR
(Appendix \ref{sec:appendix_sSFR_B}).

\begin{figure}[ht!]
\plotone{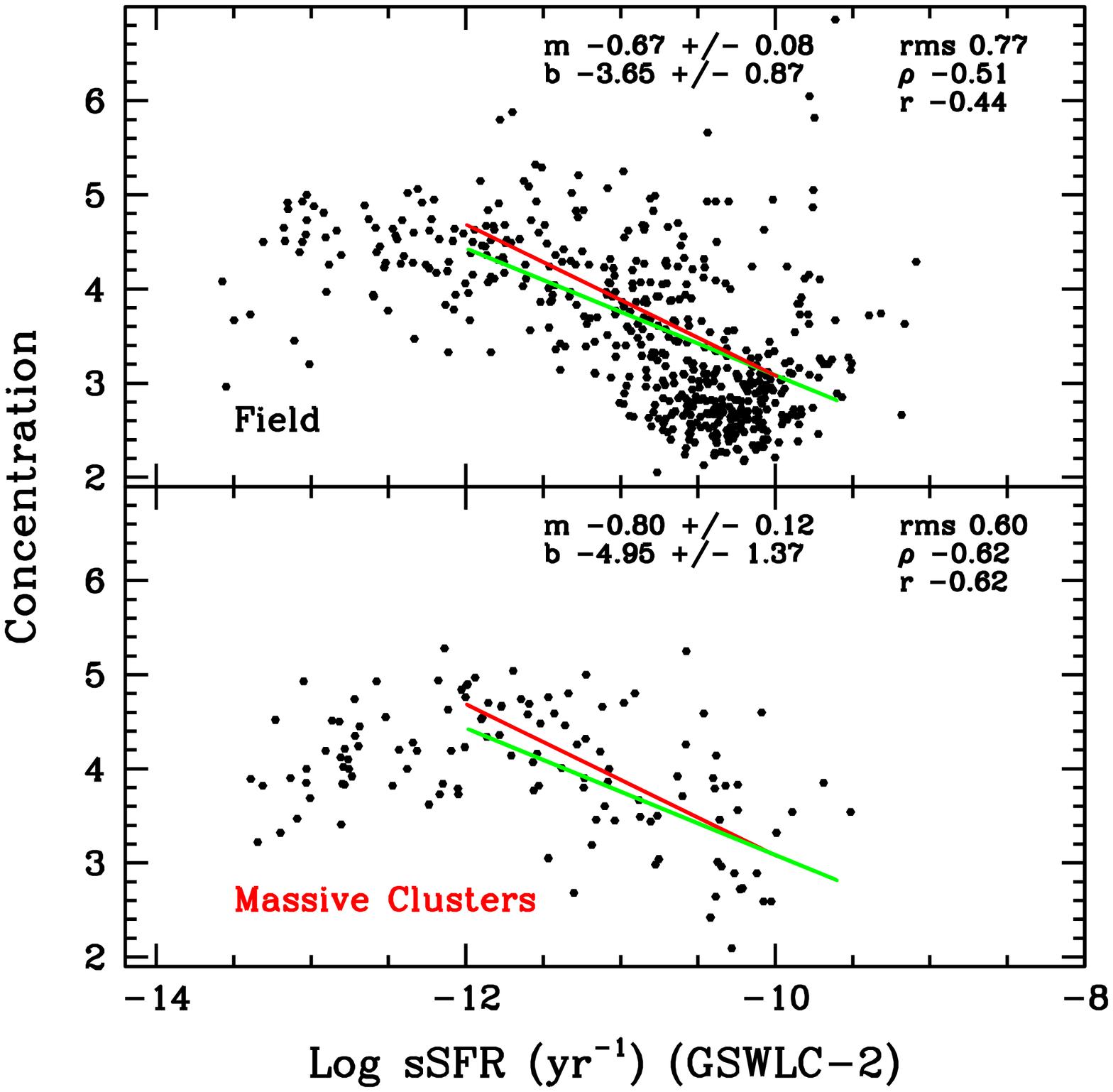}
\caption{Concentration vs.\ log sSFR 
for 
field galaxies (top panel) and galaxies in massive
clusters (bottom panel).
The sSFRs in these plots come from the GSWLC-2.
The green and red lines in these plots are
the best fits for field and cluster galaxies, respectively,
for
log sSFR $\ge$ -12 and 10 $\le$ log M* $<$ 11.
The slope (m), y-intercept (b), and rms of the best-fit lines 
are printed on the corresponding plot,
along with the Spearman ($\rho$) and 
Pearson (r) correlation coefficients.
\label{fig:sSFR_C}}
\end{figure}

Arm strength is anti-correlated with concentration 
(Figure \ref{fig:s_vs_C_barred_unbarred})
and correlated with sSFR 
(Figure \ref{fig:sSFR}).
As shown in Figure 
\ref{fig:sSFR_C}, 
concentration is in turn anti-correlated with sSFR, and
this anti-correlation holds for both field and cluster galaxies. 
The best-fit lines agree within the uncertainties for galaxies with log sSFR
$\ge$ -12.

\begin{figure}[ht!]
\plotone{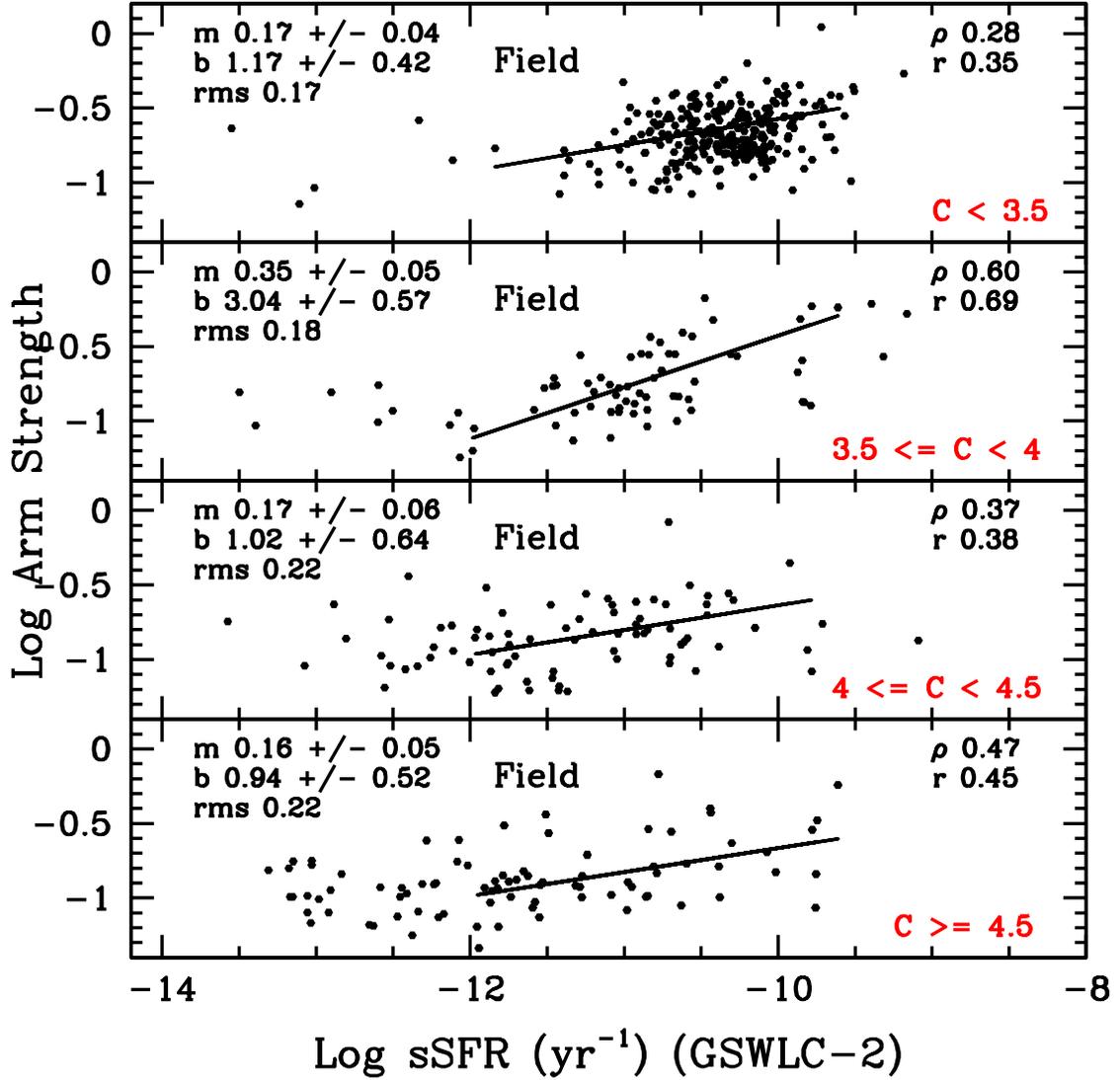}
\caption{The log of the spiral arm strength vs.\ log sSFR 
for 
field galaxies in four concentration bins:
C $<$ 3.5 (top panel), 3.5 $\le$ C $<$ 4 (second panel), 
4 $\le$ C $<$ 4.5 (third panel),
and C $\ge$ 4.5 (bottom panel).
The sSFRs in these plots come from the GSWLC-2.
The black line in all plots is the best fit
for that concentration range for 
log sSFR $\ge$ -12 and 10 $\le$ log M* $<$ 11.
The slope (m), y-intercept (b), and rms of the best-fit lines 
are printed on the corresponding
plot,
along with the Spearman ($\rho$) and 
Pearson (r) correlation coefficients.
\label{fig:sSFR_via_C}}
\end{figure}

\citet{2021ApJ...917...88Y} showed that when the C dependence is
removed, the arm strength is still correlated with sSFR. 
This result is illustrated 
in Figure \ref{fig:sSFR_via_C}, where 
we separate the field galaxies into four
ranges of concentration: C $<$ 3.5, 3.5 $\le$ C $<$ 4, 4 $\le$ C $<$ 4.5,
and C $\ge$ 4.5.   In each concentration bin, we see a correlation 
between arm strength and sSFR.
Galaxies in all four concentration bins span the full range
of sSFR and arm strength, though the median sSFR and arm strength
in each bin decreases with increasing concentration.

\begin{figure}[ht!]
\plotone{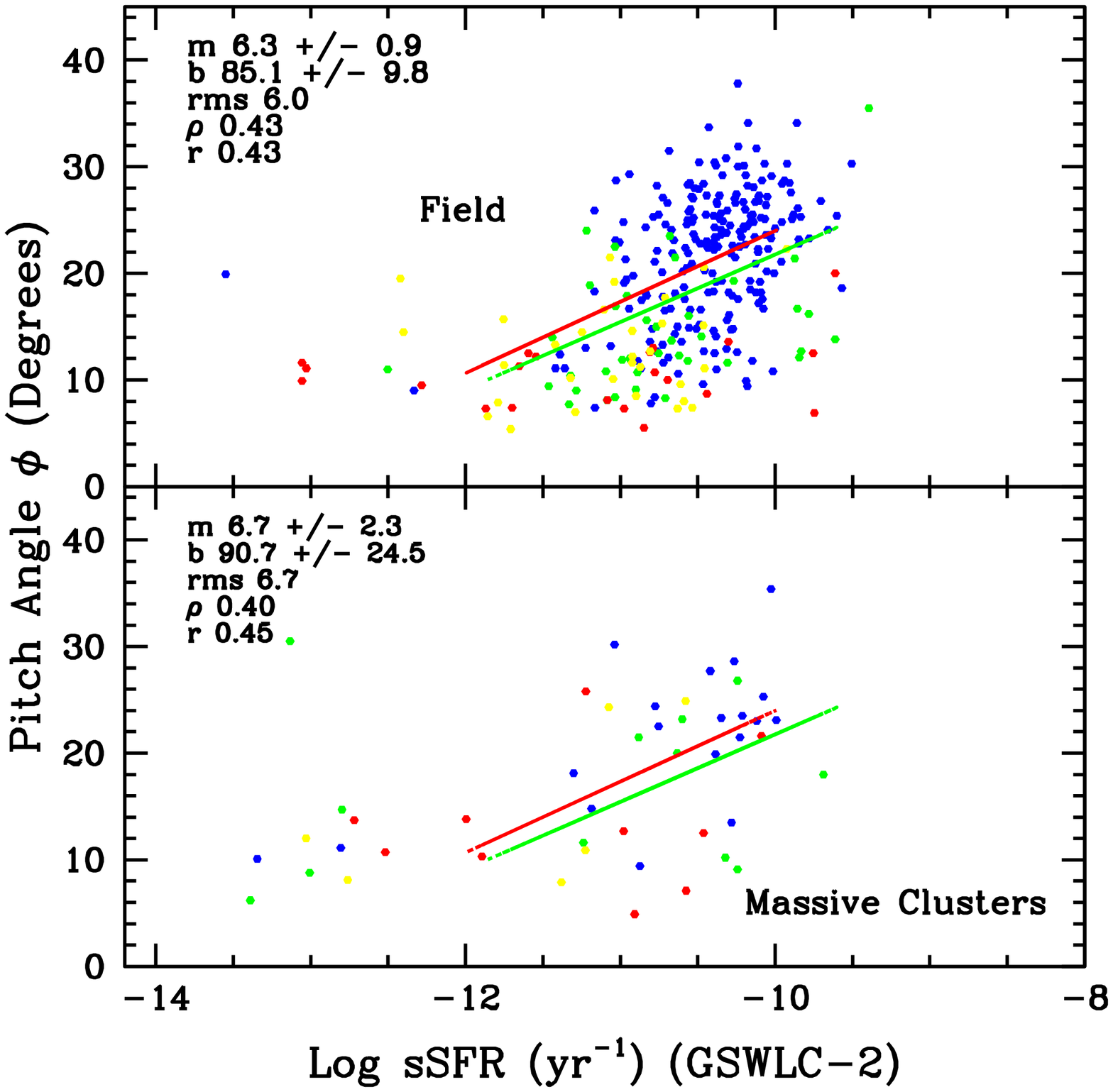}
\caption{Pitch angle $\phi$ vs.\ log sSFR 
for 
field galaxies (top panel) and galaxies in massive
clusters (bottom panel).
The sSFRs in these plots come from the GSWLC-2.
The green and red lines
in both plots are the best fits for
field and cluster galaxies, respectively, with log sSFR $\ge$ -12
and 10 $\le$ log M* $<$ 11.
The slope (m), y-intercept (b), and rms of the best-fit lines 
are printed on the corresponding
plot, 
along with the Spearman ($\rho$) and 
Pearson (r) correlation coefficients.
The data points are color-coded based
on concentration (red: C $\ge$ 4.5; yellow: 
4.0 $\le$ C $<$ 4.5; green: 3.5 $\le$ C $<$ 4.0;
blue: C $<$ 3.5).
\label{fig:sSFR_phi}}
\end{figure}

\begin{figure}[ht!]
\plotone{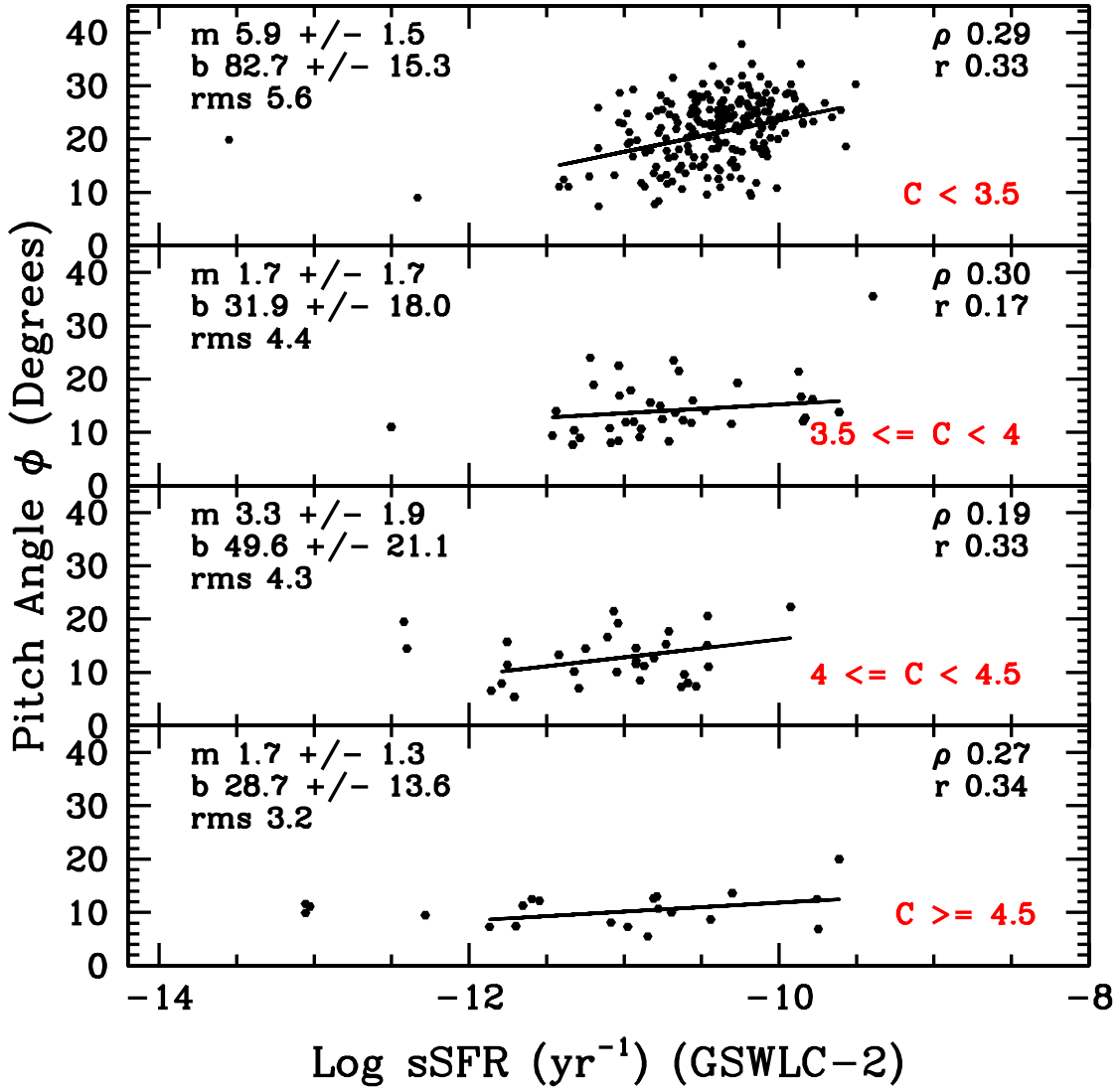}
\caption{The pitch angle vs.\ log sSFR 
for 
field galaxies in four concentration bins:
C $<$ 3.5 (top panel), 3.5 $\le$ C $<$ 4 (second panel), 
4 $\le$ C $<$ 4.5 (third panel),
and C $\ge$ 4.5 (bottom panel).
The sSFRs in these plots come from the GSWLC-2.
The black line in each plot is the best fit for 
log sSFR $\ge$ -12 and 10 $\le$ log M* $<$ 11.
The slope (m), y-intercept (b), and rms of the best-fit lines 
are printed on the corresponding
plot,
along with the Spearman ($\rho$) and 
Pearson (r) correlation coefficients.
\label{fig:sSFR_phi_via_C}}
\end{figure}

Pitch angle
is also correlated with 
sSFR for both field and cluster galaxies
(Figure \ref{fig:sSFR_phi});
galaxies with more open spiral patterns
tend to have higher sSFR. 
The best-fit relationships for the field
and cluster galaxies agree within the uncertainties.
Few of the galaxies with log sSFR $\le$ $-$12
have measurements of pitch angle available 
(Figure \ref{fig:sSFR_phi}), although arm strength
is measured
(Figure \ref{fig:sSFR}).
Based on 
Figure \ref{fig:sSFR_phi},
we conclude that as galaxies quench, field and cluster
galaxies follow approximately the same $\phi$-to-sSFR
relation.   When we separate the field galaxies into subgroups
binned by concentration, pitch angle still is weakly
correlated with
sSFR (Figure \ref{fig:sSFR_phi_via_C}).

\begin{figure}[ht!]
\plotone{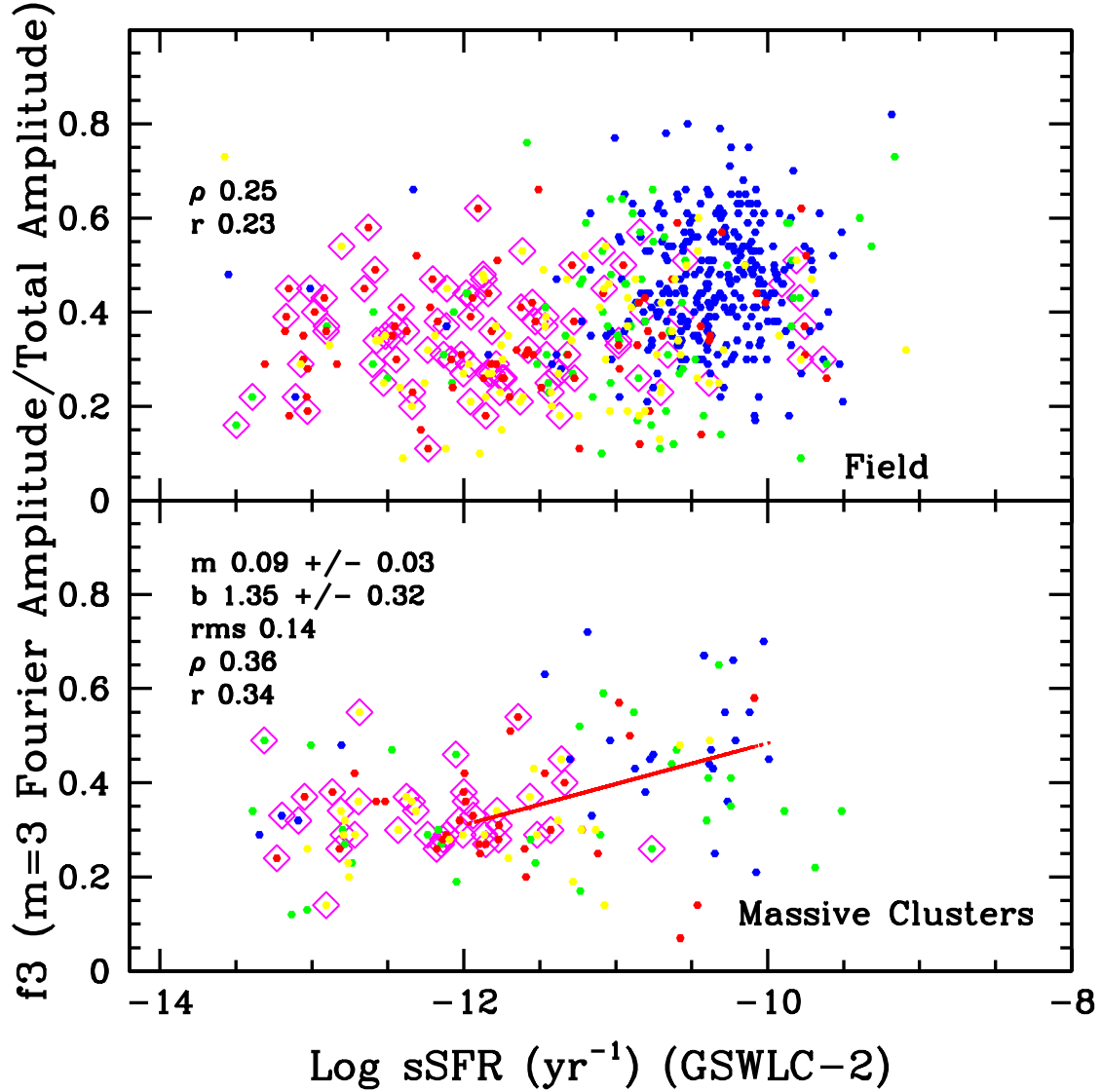}
\caption{Plots of f3 vs.\ the log of the sSFR 
for 
field galaxies (top panel) and galaxies in massive
clusters (bottom panel).
The sSFRs in these plots come from the GSWLC-2.
The red line
in the bottom plot is the best fit for
cluster galaxies with 
log sSFR $\ge$ -12 and 10 $\le$ log M* $<$ 11.
The slope (m), y-intercept (b), and rms of the best-fit lines 
are printed on the corresponding
plot, 
along with the Spearman ($\rho$) and 
Pearson (r) correlation coefficients.
The data points are color-coded based
on concentration (red: C $\ge$ 4.5; yellow: 
4.0 $\le$ C $<$ 4.5; green: 3.5 $\le$ C $<$ 4.0;
blue: C $<$ 3.5).
The galaxies marked by open magenta diamonds were classified
by 
\citet{2020ApJ...900..150Y}
as either S0-, S0, or SB0.
\label{fig:sSFR_f3}}
\end{figure}

\begin{figure}[ht!]
\plotone{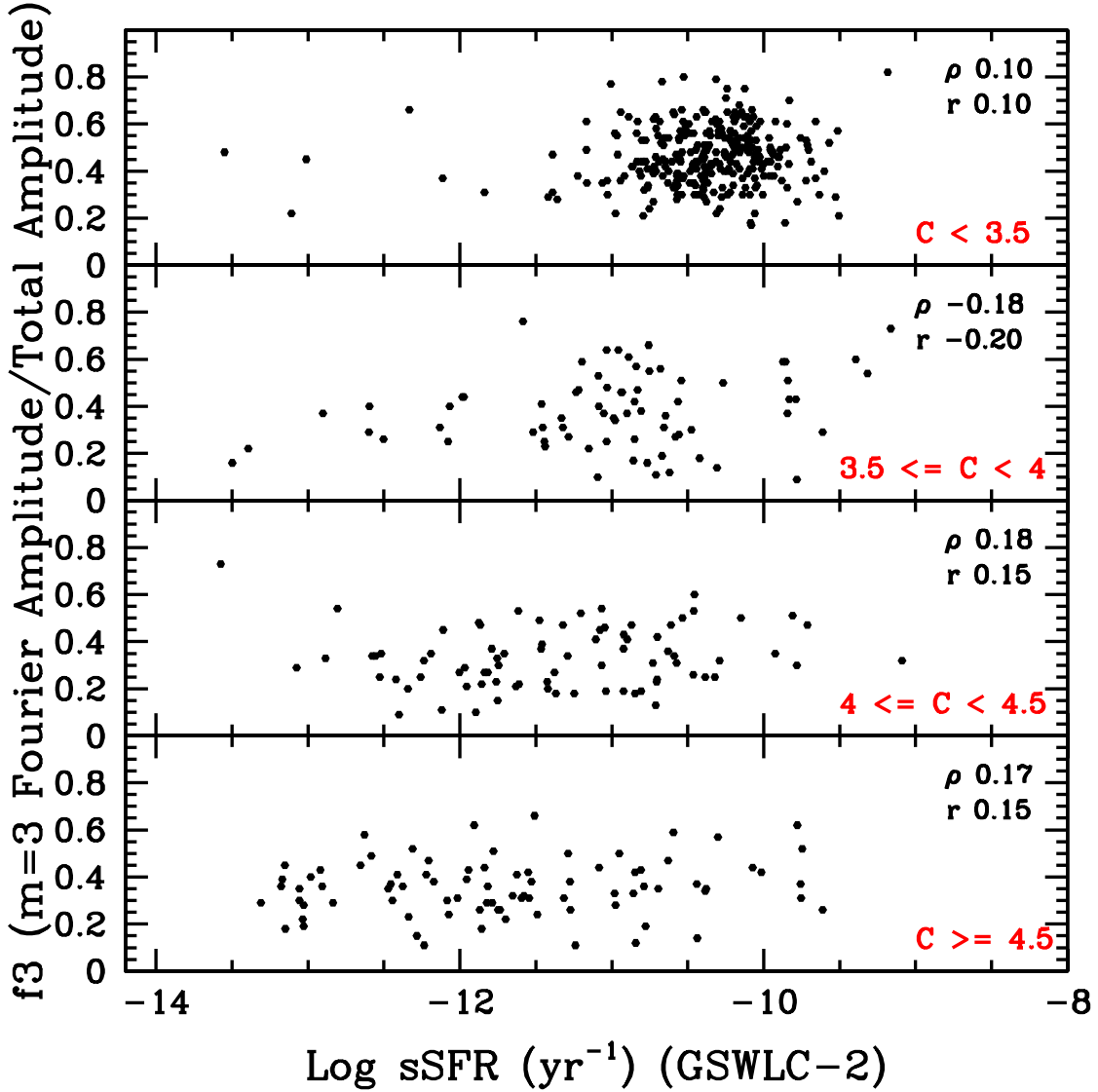}
\caption{The f3 parameter vs.\ log sSFR 
for 
field galaxies in four concentration bins:
C $<$ 3.5 (top panel), 3.5 $\le$ C $<$ 4 (second panel), 
4 $\le$ C $<$ 4.5 (third panel),
and C $\ge$ 4.5 (bottom panel).
The sSFRs in these plots come from the GSWLC-2.
The Spearman ($\rho$) and 
Pearson (r) correlation coefficients 
for log sSFR $\ge$ -12 and 10 $\le$ log M* $<$ 11 are
provided on the corresponding plot.
\label{fig:sSFR_f3_via_C}}
\end{figure}

For the full mass and sSFR range of the sample, 
f3 is correlated with sSFR, but when we limit the
sample to 10 $\le$ log M* $<$ 11 and log sSFR $\ge$ -12,
the correlation weakens
and the 
correlation coefficients drop below our nominal cutoff of 0.3 
for a correlation
(Figure \ref{fig:sSFR_f3}).
When we divide the sample further into narrow ranges of concentration,
the relation weakens even further
(Figure \ref{fig:sSFR_f3_via_C}).
This suggests that the very weak overall correlation between f3 and sSFR
(Figure \ref{fig:sSFR_f3})
is driven by both being dependent upon concentration,
rather than a direct relationship between f3 and sSFR.
In contrast, concentration and sSFR both independently affect
arm strength and $\phi$.

\begin{figure}[ht!]
\plotone{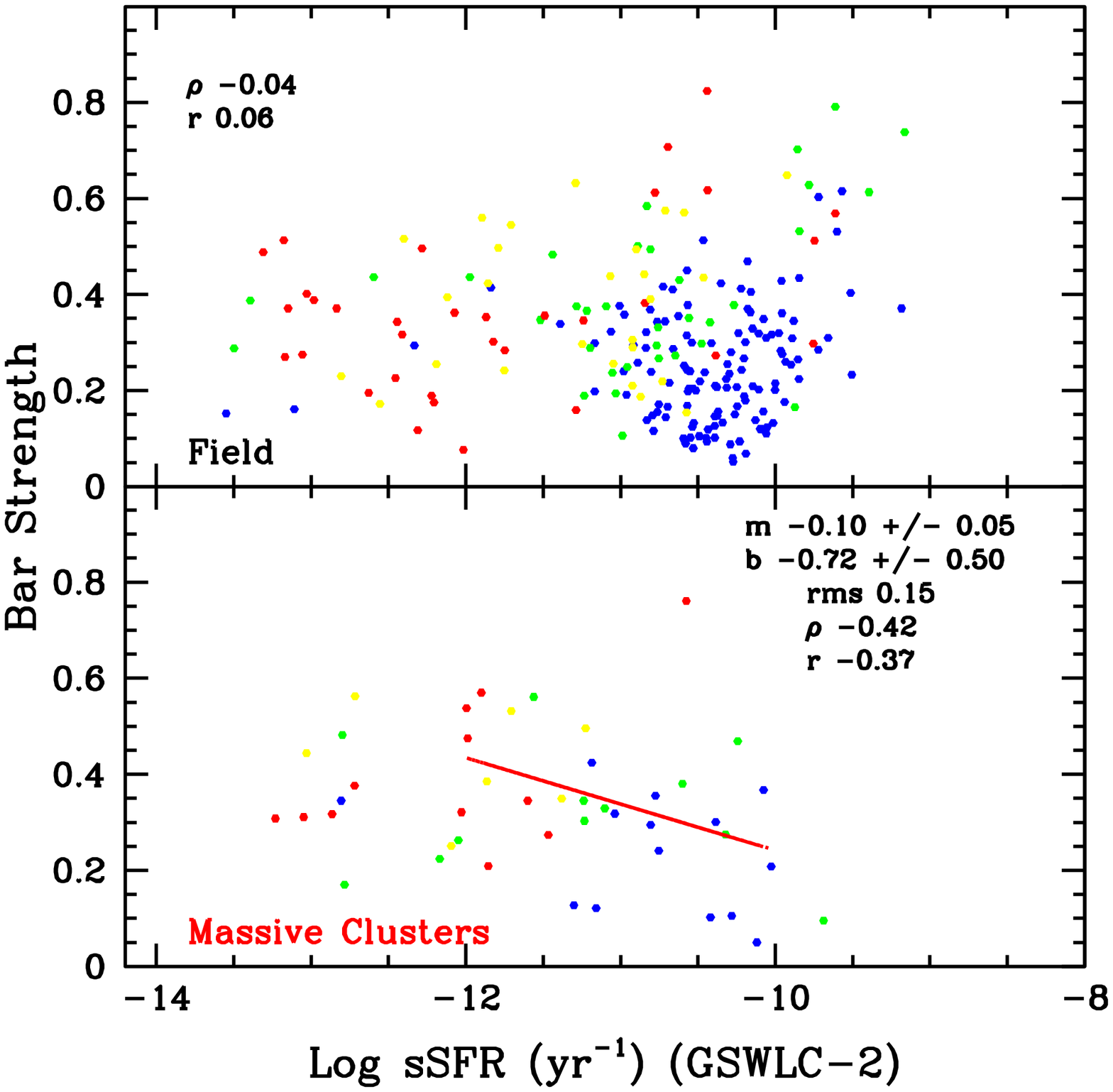}
\caption{The bar strength plotted against the GSWLC-2 sSFR
for 
field galaxies (top panel) and galaxies in massive
clusters (bottom panel).
In the top panel, the 
Spearman ($\rho$) and 
Pearson (r) correlation coefficients 
for the field galaxies with 
log sSFR $\ge$ -12 and 10 $\le$ log M* $<$ 11
are printed.
The red line in the bottom panel 
is the best fit
for the 
log sSFR $\ge$ -12 and 10 $\le$ log M* $<$ 11
galaxies 
in massive clusters.
The slope (m), y-intercept (b), rms,
Spearman ($\rho$) and 
Pearson (r) correlation coefficients for that line
are also provided.
The data points are color-coded based
on concentration (red: C $\ge$ 4.5; yellow: 
4.0 $\le$ C $<$ 4.5; green: 3.5 $\le$ C $<$ 4.0;
blue: C $<$ 3.5).
\label{fig:sSFR_bar}}
\end{figure}

We also compared bar strength with sSFR and found no correlation
for field galaxies, but an anti-correlation for
cluster galaxies
(Figure \ref{fig:sSFR_bar}).  
However, there are only 30 barred cluster galaxies in 
the GSWLC-2
with 
log sSFR $\ge$ $-$12 and 10 $\le$ log M* $<$ 11, 
and when we
use the alternative derivation of sSFR, this anti-correlation
disappears (see Appendix). 
Thus the relation between sSFR and bar strength remains uncertain.

\begin{figure}[ht!]
\plotone{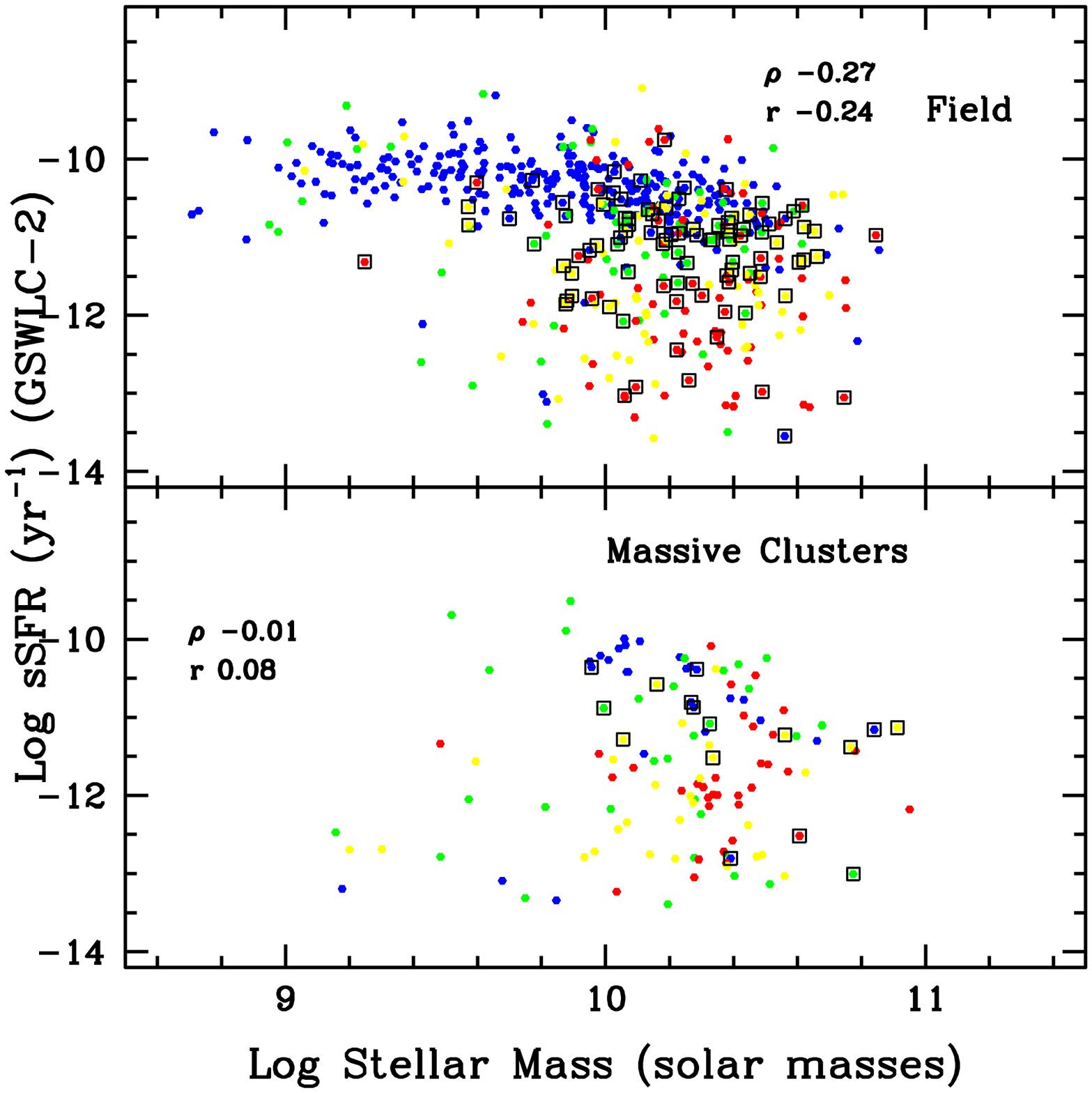}
\caption{
Log sSFR vs.\ log stellar mass
for 
field galaxies (top panel) and galaxies in massive
clusters (bottom panel).
The sSFRs in these plots come from the GSWLC-2, while
the stellar masses on the y-axis come from the NSA.
Correlation coefficients for galaxies with 10 $\le$ log M* $<$ 11
and log sSFR $\ge$ -12 are displayed on the corresponding plot.
The data points are color-coded based
on concentration (red: C $\ge$ 4.5; yellow: 
4.0 $\le$ C $<$ 4.5; green: 3.5 $\le$ C $<$ 4.0;
blue: C $<$ 3.5).
The galaxies marked by open black squares are classified
as AGN (see text for details).
\label{fig:sSFR_AGN}}
\end{figure}

In Figure \ref{fig:sSFR_AGN} we 
plot sSFR vs.\ stellar mass. 
For the full mass range, 
the sSFR is weakly inversely correlated with stellar mass. 
However, when we limit the sample to 10 $\le$ log M* $<$ 11
and log sSFR $\ge$ -12, the correlation weakens further, with 
correlation coefficients outside our defined range for a reliable 
correlation.  
In Figure \ref{fig:sSFR_AGN}, we
identify the optically-selected
Seyfert galaxies by open black squares.  
The Seyferts tend to
have moderately low sSFR, near the nominal quenching threshold
of log sSFR = $-$11 or below.  
The Seyferts have a large range in stellar mass,
from 9.8 $\le$ log M* $<$ 11, with a few lower.
The fraction of Seyferts 
in cluster galaxies
is relatively similar 
to that of field galaxies (12 $\pm$ 3\% vs.\ 
8 $\pm$ 1\%, respectively, for galaxies with
0.019 $\le$ z $<$ 0.03 and 10 $\le$ log M* $<$ 11).

\section{Comparison with Galaxy Zoo} \label{sec:galzoo}

\subsection{The Galaxy Zoo Parameters} \label{sec:galzoo_params}

The 
\citet{2020ApJ...900..150Y}
dataset represents a very large sample of
carefully classified galaxies.  It is only with such large samples
that one can begin to explore the effects of environment
on the evolution of galaxies with any statistical precision.
Another large sample of classified galaxies has been assembled
as part of the Galaxy Zoo project.   As we summarize below,
there is substantial overlap in galaxies between the two samples.
Some of the 
classifications in Galaxy Zoo are expected to be strongly related to parameters
in the 
\citet{2020ApJ...900..150Y}
study (f3 vs.\ number of arms, pitch
angle vs.\ winding parameter, and concentration vs.\ bulge class).

We explored the extent to which these parameters are related by
cross-correlating our list with the 
Galaxy Zoo sample of 
\citet{2013MNRAS.435.2835W}.
Galaxy Zoo posed a nested set of questions for each galaxy.
The first question
asked participants to decide whether the 
galaxy is `smooth', has `features/disk', or has an `artifact'. 
If `features/disk' was chosen, 
they were asked if it is or is not an edge-on disk.
If not edge-on was selected, then the participant was asked
whether it has spiral arms.  If it had spiral arms, 
they were asked whether the arms are tightly wound,
loosely wound, or medium wound.  
Finally, they were 
asked how many spiral arms are present.
If a viewer did not select `features/disk'
initially, then they were not asked the number of spiral arms.
If they selected the `features/disk' option for the first question,
they were asked about the bulge prominence, and were
given four choices: no bulge, just noticeable,
obvious, or dominant.

\citet{2016MNRAS.461.3663H}
define a `securely identified' spiral
in Galaxy Zoo 
as one for which 
p$_{\rm feature}$$\times$p$_{\rm not~edge-on}$$\times$p$_{\rm spiral}$ $>$ 0.5
where p$_{\rm feature}$ is the fraction of votes for features/disk, 
p$_{\rm not~edge-on}$ is the fraction of votes for not being edge-on,
and p$_{\rm spiral}$ the fraction of votes for having spiral arms.
Of the 4062 Yu/Ho galaxies with z $<$ 0.03, 2401 are in Galaxy Zoo  
with at least 20 classifications.
Of those 2401 galaxies, only 1437 (60\%) are 
`securely identified' as spirals
in Galaxy Zoo.
Of the 964 `not securely identified spirals', 
the median p$_{\rm feature}$ is 0.352,
the median p$_{\rm not~edge-on}$ = 1.0, and the median p$_{\rm spiral}$ = 0.125.
Many Galaxy Zoo participants did not notice disks or spiral
patterns in the SDSS images of some of the 
\citet{2020ApJ...900..150Y}
galaxies.
This indicates that the Fourier analysis software was able to 
extract information about a spiral pattern for galaxies
for which it is difficult to discern the spiral by eye.

As discussed in detail by 
\citet{2016MNRAS.461.3663H},
morphological classification of higher redshift galaxies is less 
reliable
due to resolution
effects. 
They conclude that 
at higher
redshifts, viewers are more likely to classify a multi-armed
galaxy as a two-armed galaxy, thus there is an artificial
increase in the fraction of two-armed spirals at higher redshifts
compared to low redshifts. 
Also, as the redshift increases the fraction of galaxies 
classified as tightly wound decreases.  
Spirals
in some tightly wound galaxies are missed by viewers, 
who classify them as 
`smooth' or as having `features/disks' but not spirals.
\citet{2016MNRAS.461.3663H}
correct for this bias with redshift,
and provide bias-corrected vote fractions for
each galaxy in the sample.  If the corrected vote fraction in
a particular category 
exceeds 80\%, they flag that galaxy for that property.

\subsection{The f3 Parameter vs. the Number of Spiral Arms}

In Figure \ref{fig:f3_vs_number_arms},
as black histograms we display the f3 values for the
`secure' spirals with 1 arm (top panel), 2 arms (2nd panel),
3 arms (3rd panel), 4 arms (4th panel), 
and more than 4 arms (5th panels).  
The sixth panel gives galaxies that were ranked as spirals,
but Galaxy Zoo participants could not count the number of arms.
The black histograms
only include galaxies for which more than half of the reviewers who
counted arms selected the corresponding number of arms, using
the
original weighted vote counts from 
\citet{2013MNRAS.435.2835W}
without a correction for classification bias.
The blue dotted histograms 
mark the galaxies that were flagged by 
\citet{2016MNRAS.461.3663H}
as having bias-corrected vote fractions greater than 80\%.

Figure \ref{fig:f3_vs_number_arms} 
shows that the f3 values for reliably-identified two-arm
spirals are generally lower than those for galaxies with other numbers
of arms, although the spread 
is large. 
There is not a clean correlation between f3 and the number 
of arms, but a low f3 means a higher probability that the galaxy 
has two arms.
The galaxies that the Galaxy Zoo 
contributors could not reliably identify as spirals 
have an f3 distribution
similar to that of two-armed spirals.

\subsection{Galaxy Zoo Winding Parameter vs.\ Pitch Angle}

\begin{figure}[ht!]
\plotone{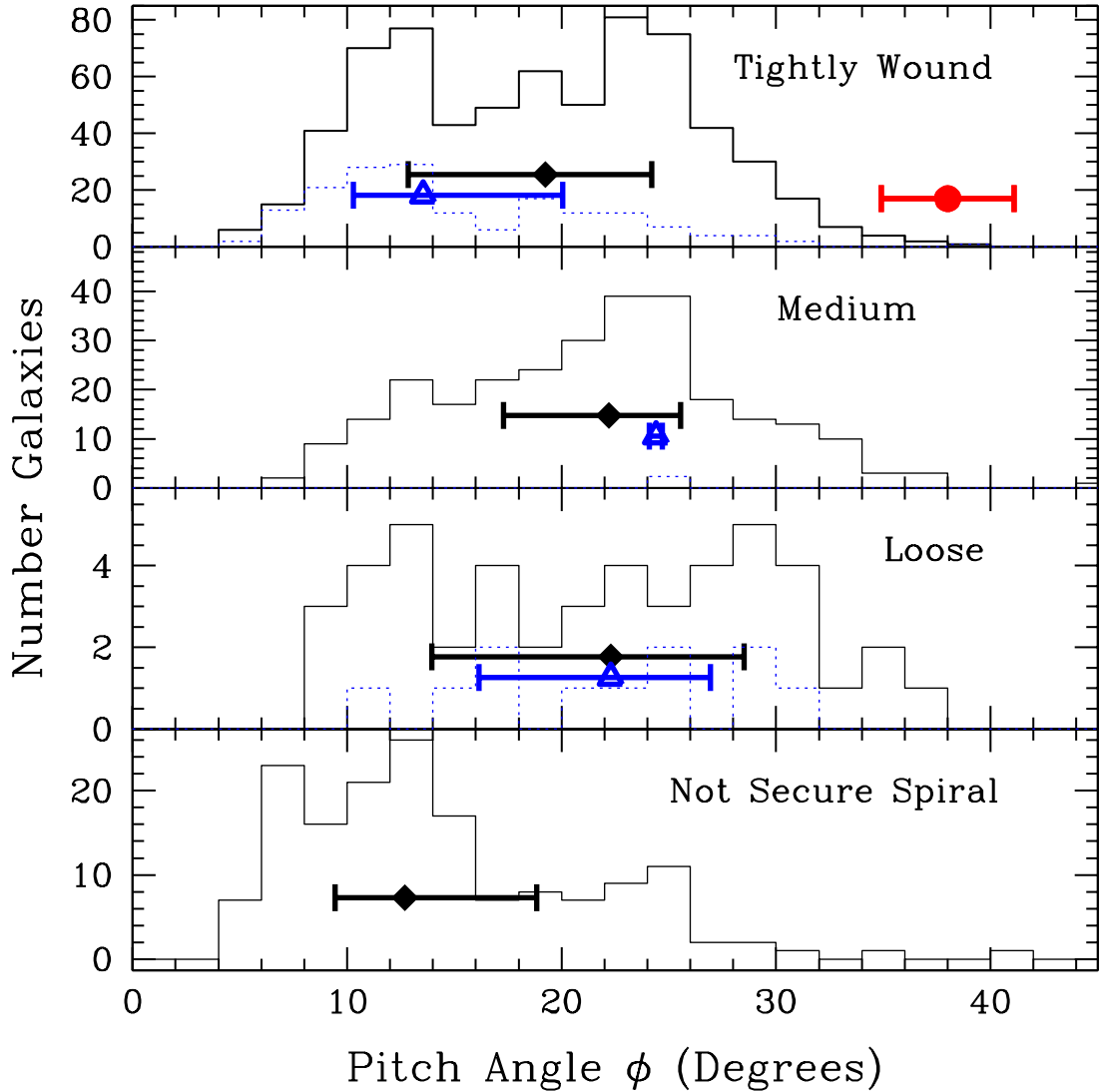}
\caption{Histograms of the number of 
\citet{2020ApJ...900..150Y}
z $<$ 0.03
galaxies as a function of pitch angle,
for galaxies identified by Galaxy Zoo as having arms
that are `tightly wound'
(top panel), `medium wound' (2nd panel), and `loosely
wound' (3rd panel).
The black histograms in the top three panels are based on vote fractions
uncorrected for classification bias, and include
only `securely-identified' spirals 
with at least 20 Galaxy Zoo classifications, and 
only include galaxies for which at least half of the participants agreed
on the winding class.
The blue dotted histograms 
include only galaxies 
flagged 
by 
\citet{2016MNRAS.461.3663H} as 
having
a bias-corrected vote fraction greater than 80\%. 
The bottom panel gives the distribution of $\phi$ for galaxies not considered
`secure spirals' in Galaxy Zoo.
The black filled diamonds and blue open triangles give the median $\phi$
value for 
the black and blue histograms, respectively,
with the errorbars 
marking the range from first quartile to third
quartile. 
The red circle with
errorbars illustrates the
median measurement uncertainty on $\phi$.
\label{fig:phi_vs_winding_class}}
\end{figure}

In Figure \ref{fig:phi_vs_winding_class}, 
we provide histograms of the 
\citet{2020ApJ...900..150Y}
pitch angles
for each of the three winding classes
as determined by the Galaxy Zoo participants.
The black histograms in 
the top three panels only include 'secure spirals' with 
p$_{\rm tightly~wound}$
$>$ 0.5 (top panel), p$_{\rm medium~winding}$ $>$ 0.5 (second panel),
and p$_{\rm loosely~wound}$ $>$ 0.5 (third panel), where
the vote fractions have not been corrected for
classification bias.
The blue dotted histograms give the galaxies that have been
flagged by 
\citet{2016MNRAS.461.3663H}
as having a vote fraction greater than 80\%
after correction for bias.
The bottom panel 
of Figure \ref{fig:phi_vs_winding_class}
gives the histogram
for the galaxies that are not `secure spirals'.

This plot shows extensive scatter in the 
\citet{2020ApJ...900..150Y}
pitch
angles among each of the three Galaxy Zoo winding classes.   
As expected, the median pitch angle for the
tightly wound group is less than for the other two groups,
especially after correction for classification bias.  
However, there is still considerable
overlap in values with the other groups.
The medium and loosely wound groups have very similar
medians and spreads, and cannot be distinguished
in terms of pitch angle.
The galaxies that are not classified as `secure spirals'
typically have small pitch angles.   This is consistent
with the idea that Galaxy Zoo viewers are more likely to miss
spiral patterns when the arms are tightly wound.
Another possible contributor to the scatter 
is that the 
\citet{2020ApJ...900..150Y}
parameters are measured on deprojected images, 
while the Galaxy Zoo classifications were not.   
Even with the 
\citet{2016MNRAS.461.3663H}
correction
for classification bias, there is not a good correspondence between
the Galaxy Zoo winding class and the 
\citet{2020ApJ...900..150Y}
pitch angle.

\begin{figure}[ht!]
\plotone{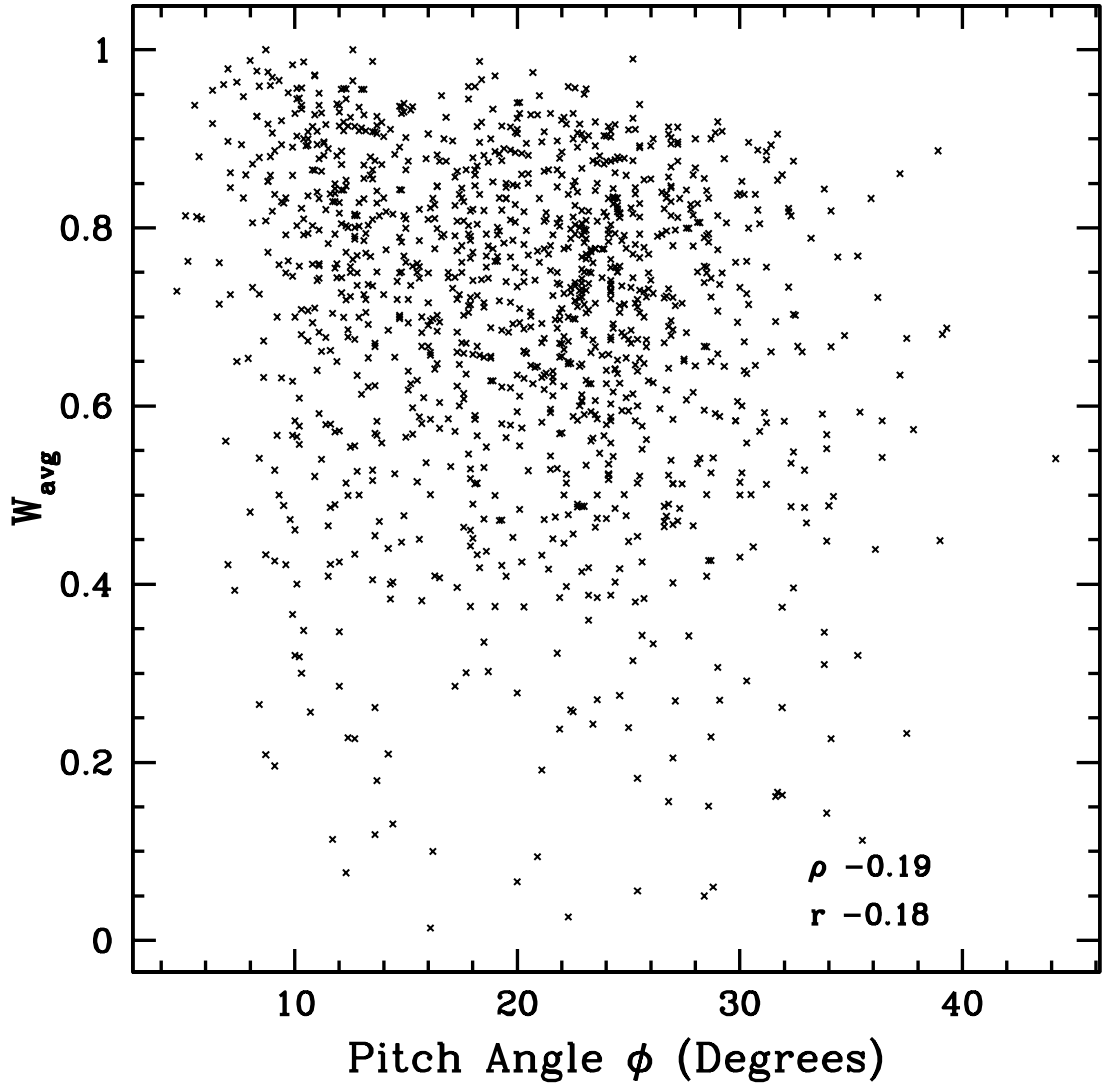}
\caption{The Galaxy Zoo `winding score' w$_{avg}$ 
= 0.5 $\times$ p$_{\rm medium}$ + 1.0 $\times$ p$_{\rm tight}$,
plotted against the pitch angle from 
\citet{2020ApJ...900..150Y}.  
Here, 
p$_{\rm medium}$ is the vote fraction for `medium wound',
and p$_{\rm tight}$ the vote fraction for `tightly wound';
a larger w$_{avg}$ means more tightly wound arms.
The Spearman ($\rho$) and Pearson (r) correlation coefficients
are provided.
Only galaxies with at least 20 Galaxy Zoo classifications
are included in this plot.
\label{fig:phi_vs_Wavg}}
\end{figure}

\citet{2019MNRAS.487.1808M}
define a parameter called the
`arm winding score' w$_{avg}$ = 0.5 $\times$ p$_{\rm medium}$ +
1.0 $\times$ p$_{\rm tight}$.  This provides a single number
for each galaxy in the sample, 
with larger 
w$_{avg}$ meaning tighter arms.
In Figure \ref{fig:phi_vs_Wavg},
for all of the `secure spirals' in both samples,
we plot w$_{avg}$ vs.\ the 
\citet{2020ApJ...900..150Y}
pitch angle.
No correlation is apparent.

\subsection{Galaxy Zoo Bulge Prominence vs. Concentration}

\begin{figure}[ht!]
\plotone{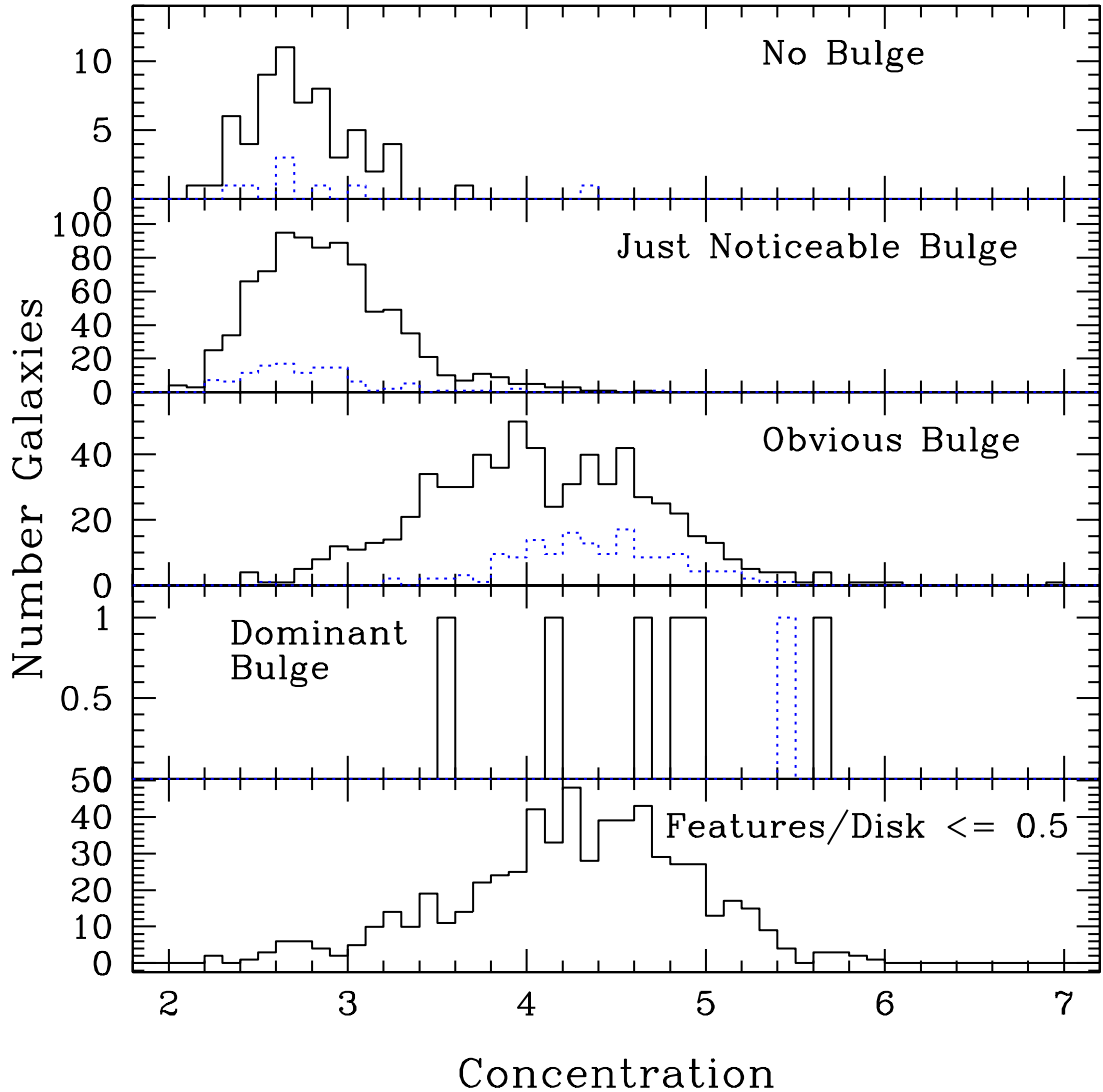}
\caption{Histograms of the number of 
\citet{2020ApJ...900..150Y}
z $<$ 0.03
galaxies as a function of 
concentration.
In the black histograms in the top four panels, 
only galaxies with p$_{\rm disk}$ $>$ 0.5 and 
at least 20 Galaxy Zoo classifications are included, 
and only galaxies for which at least half of the participants agreed
on the bulge class.  The blue dotted histograms include galaxies
with bias-corrected vote fractions greater than 80\% as tabulated
by 
\citet{2016MNRAS.461.3663H}.
The bottom histogram gives the distribution of concentrations 
for galaxies with vote fractions of $\le$0.5 for `features/disk'.
Only galaxies with at least 20 classifications in Galaxy
Zoo are included in these plots.
\label{fig:C_vs_bulge}}
\end{figure}

We also compared 
the bulge prominence as indicated by the Galaxy Zoo 
participants with the 
\citet{2020ApJ...900..150Y}
central concentration.
In Figure \ref{fig:C_vs_bulge},
for the galaxies with at least 20 classifications in Galaxy
Zoo 
we plot histograms of the concentration for galaxies
within the four Galaxy Zoo prominence classes: no bulge (top panel),
just noticeable bulge (second panel), obvious bulge (third panel), 
and 
dominant bulge (fourth panel). 
The bottom
panel shows a histogram of the concentrations of the galaxies
with `disk/feature' vote fraction less than or equal to 0.5.
Figure \ref{fig:C_vs_bulge} 
shows a clear distinction in concentration 
between galaxies with a `just noticeable
bulge' and those with an `obvious bulge', 
with the vast majority of the former having
2 $<$ C $<$ 3.3, and the 
latter mainly having 3.3 $<$ C $<$ 5.2.   
In contrast, the distribution of concentrations for the `no bulge'
and the `just noticeable bulge' classes are very similar.
Only a handful of galaxies were classed as `dominant bulge',
and these have a range in C that overlaps with the `obvious
bulge' set.
The galaxies with low features/disk vote fractions
(Figure \ref{fig:C_vs_bulge}, bottom panel) 
tend to have large concentrations.
While Galaxy Zoo is successful
at separating low concentration disk galaxies from
those with high concentrations, finer separation into
additional bulge subclasses is uncertain.

\begin{figure}[ht!]
\plotone{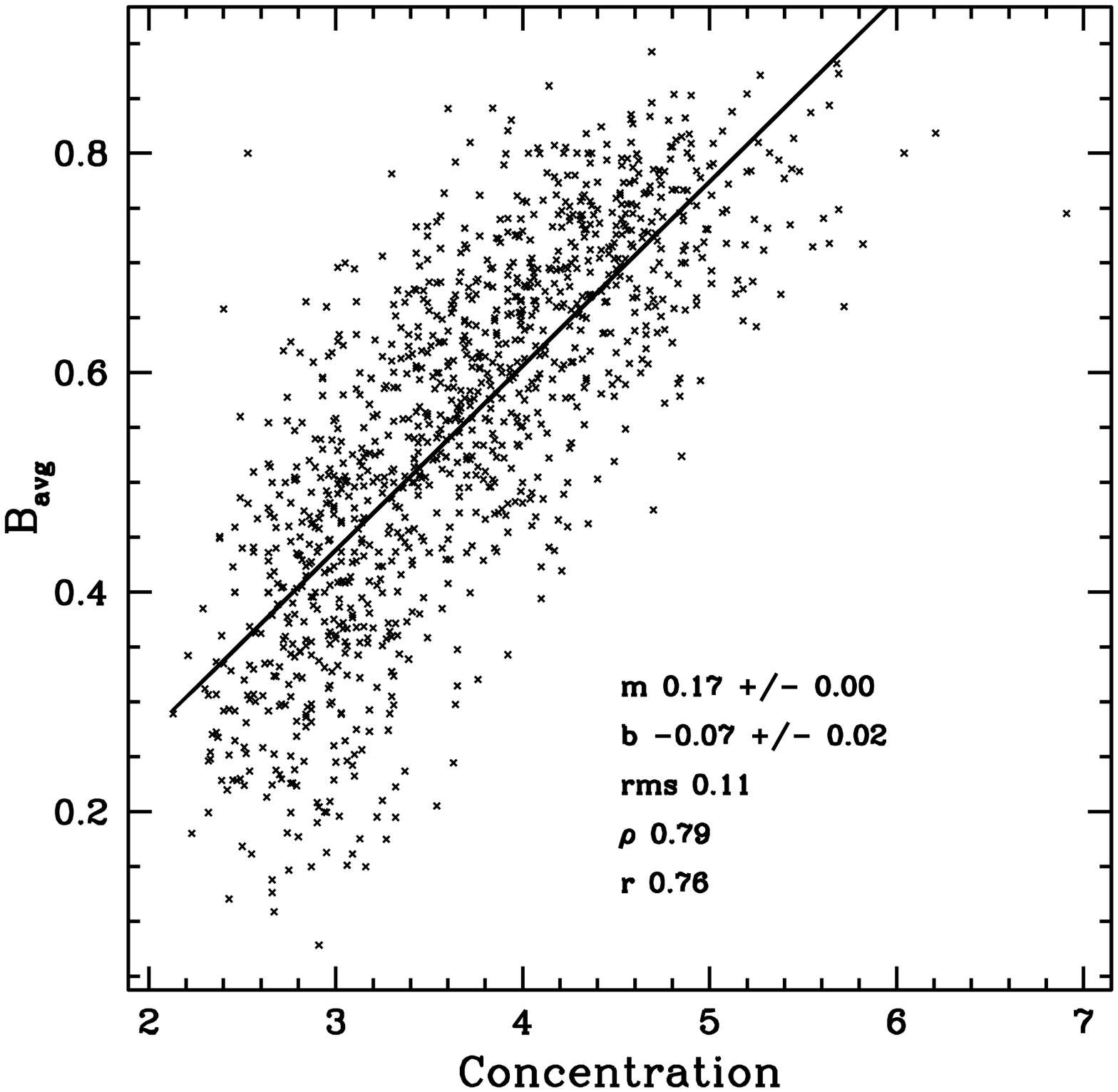}
\caption{A comparison of 
the Galaxy Zoo bulge prominence parameter
B$_{avg}$ 
\citep{2019MNRAS.487.1808M, 2021MNRAS.504.3364L}
and concentration, where
$B_{avg}$ = 0.2 $\times$ p$_{\rm just~noticeable}$ + 0.8 $\times$
p$_{\rm obvious~bulge}$ + 1.0 $\times$ p$_{\rm dominant~bulge}$.  
Here, 
p$_{\rm just~noticeable}$,
p$_{\rm obvious~bulge}$,
and p$_{\rm dominant~bulge}$ are the Galaxy Zoo vote fractions 
for a `just noticeable' bulge, an `obvious bulge', and a `dominant
bulge', respectively.
Only galaxies with at least 20 classifications in Galaxy Zoo
are included.
The slope (m), y-intercept (b), and rms are provided, along with the 
Spearman ($\rho$) and 
Pearson (r) correlation coefficients.
\label{fig:C_vs_bulge2}}
\end{figure}

\citet{2019MNRAS.487.1808M}
and 
\citet{2021MNRAS.504.3364L}
define
a new variable called the `bulge prominence' B$_{avg}$,
which combines the various vote fractions for
a galaxy into a single variable:
$B_{avg}$ = 0.2 $\times$ p$_{\rm just~noticeable}$ + 0.8 $\times$
p$_{\rm obvious~bulge}$ + 1.0 $\times$ p$_{\rm dominant~bulge}$.  
For all the galaxies with a `features/disk' vote fraction
greater than 0.5, 
in Figure \ref{fig:C_vs_bulge2}
we plot B$_{avg}$ against the 
\citet{2020ApJ...900..150Y}
concentration
index.   A strong correlation is seen, although with
significant scatter.

Overall, these comparisons
show coarse agreement between two independent
measures of spiral and disk structure.  A more detailed
comparison is hampered partly by the challenges of characterizing
observer's bias in classification, and also by the fact that the
two studies did not measure the same parameters in the same way.

\begin{figure}[ht!]
\plotone{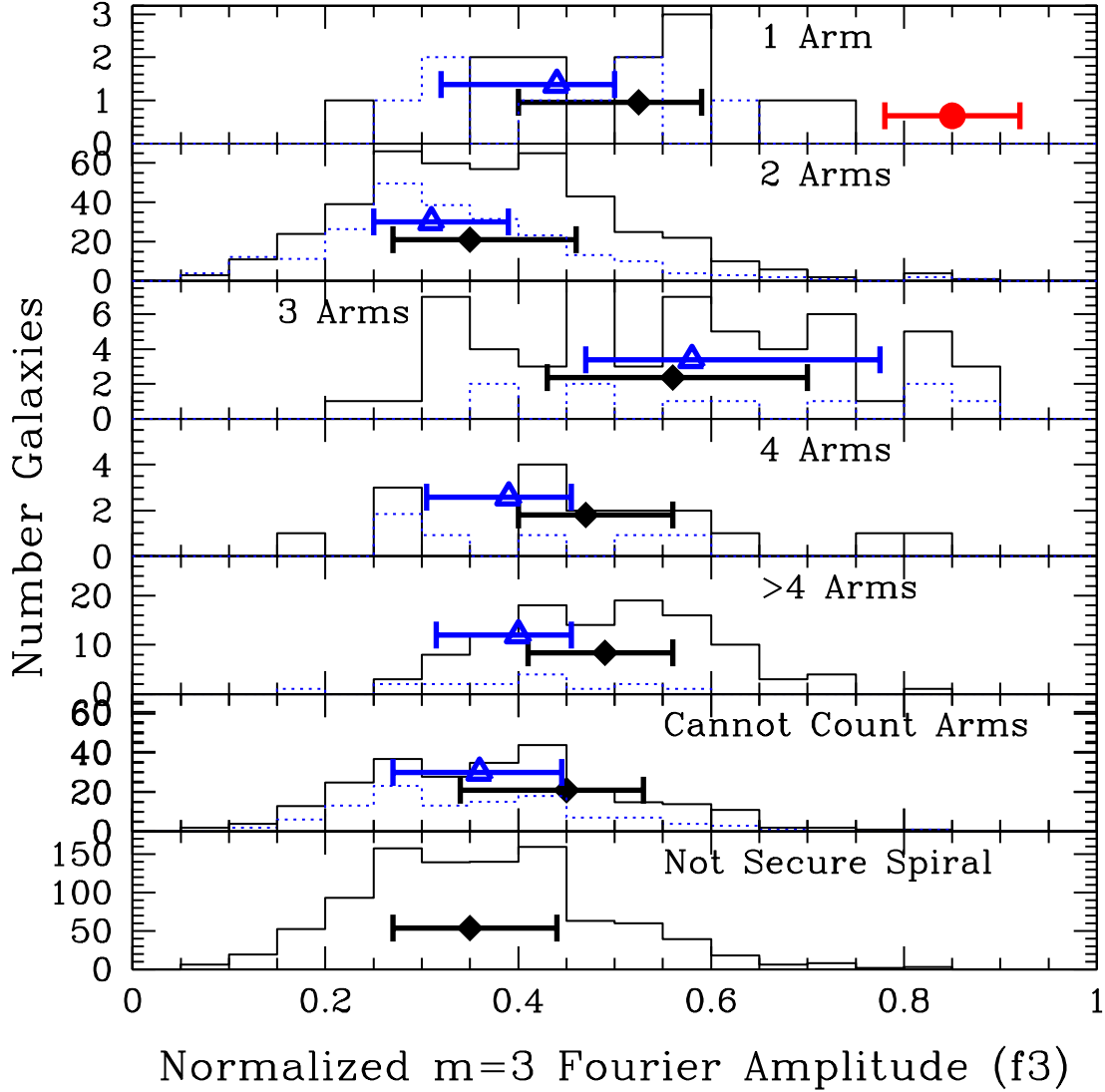}
\caption{
Histograms of the number of galaxies as a function of
f3 
for 
\citet{2020ApJ...900..150Y}
galaxies with z $<$ 0.03 identified by Galaxy Zoo
as having one arm (top panel), two arms (2nd panel), three arms
(3rd panel), four arms (4th panel), more than 4 arms (5th panel),
or `cannot count arms' (6th panel).  
The histograms in black in the top six panels include
only `securely-identified' spirals which have at least 20 classifications
in Galaxy Zoo, and for 
which at least half of the participants agreed
on the number of arms based on vote fractions uncorrected for
classification bias.  
The blue dotted histograms mark the
galaxies which were `flagged' 
by \citet{2016MNRAS.461.3663H}
as having reliable arm counts after
correction for classification bias.
The bottom panel gives the distribution of f3 for galaxies not considered
`secure spirals' in Galaxy Zoo. 
The black filled diamonds and blue open triangles give the median f3 values
for the black and blue histograms, respectively, 
with the errorbars marking the range 
from first quartile to third
quartile. 
The red circle with
errorbars illustrates the
median measurement uncertainty on f3.
\label{fig:f3_vs_number_arms}}
\end{figure}

\section{Discussion} \label{sec:discussion}

\subsection{Overview} \label{sec:discussion_overview}

We compared the 
\citet{2020ApJ...900..150Y}
central concentrations, 
spiral arm strengths, pitch angles, and normalized m=3 Fourier amplitudes
of galaxies in clusters with those in the field
using the 
\citet{2017MNRAS.470.2982L}
catalog of galaxy clusters and groups.
Cluster galaxies in this sample have larger concentrations
for the same stellar mass, compared to the field galaxies.
Spirals in clusters tend to have weaker arms, smaller
pitch angles, and lower f3 than field galaxies. 
However,
when we account for the larger central concentrations of 
galaxies in clusters
and compare galaxies with similar central concentrations,
we do not find significant differences between the spiral arm 
parameters 
of cluster and field galaxies.
Within the uncertainties, 
spirals in clusters follow the same
arm-strength-to-concentration, f3-to-concentration, and 
pitch-angle-to-concentration 
anti-correlations as spirals in the field.
The correlations between the arm parameters
and the sSFRs in cluster galaxies are also 
similar to those of field galaxies.
We find that 
barred galaxies in clusters have a similar bar-strength-to-concentration
correlation
within the uncertainties
as field galaxies.
We reproduce the earlier results of 
\citet{2020ApJ...900..150Y}
that
the arm strengths of barred galaxies are higher than those of
unbarred galaxies.
In the following sections, we discuss these 
results in light of theories of
galaxy evolution and spiral arm production.

\begin{figure}[ht!]
\plotone{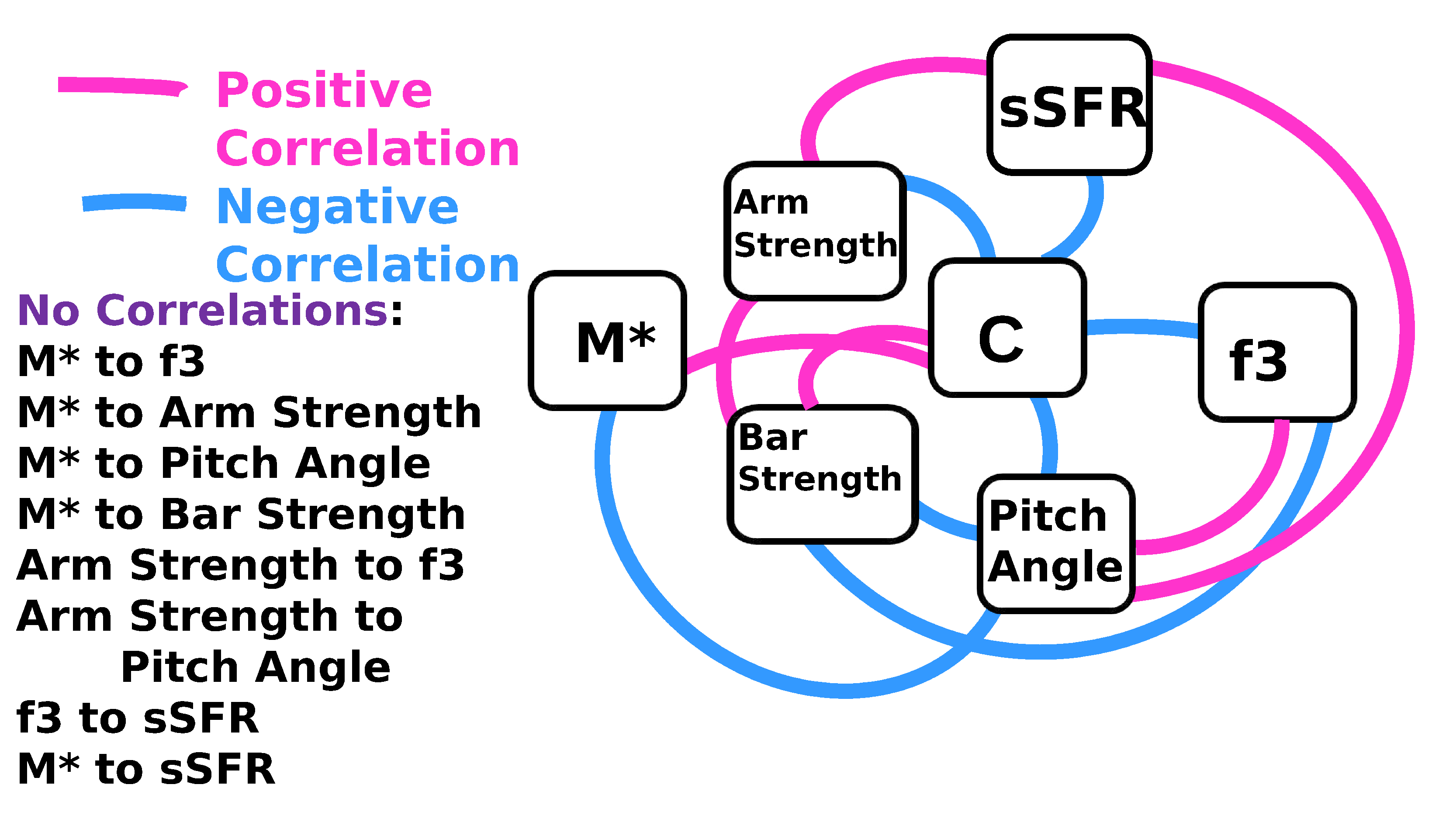}
\caption{Diagram illustrating the relationships between
the various parameters discussed in this paper.
We define correlations as relations with either the Pearson
or Spearman correlation coefficient larger than 0.3, and an anti-correlation
as one with either coefficient less than -0.3, 
limiting the sample to field galaxies with 10 $\le$ log M* $<$ 11. 
\label{fig:diagram}}
\end{figure}

In Figure \ref{fig:diagram} we diagram the relations between the parameters
concentration, stellar mass, arm strength,
bar strength, pitch angle, f3, and sSFR.
In this diagram, a pink curve connecting two variables designates
a positive correlation, while a blue line indicates a negative correlation.
Uncorrelated parameters 
are not connected in this plot.  The non-correlations
are listed at the bottom of the diagram.
The relations marked in Figure 
\ref{fig:diagram} 
include both the results discussed in this paper, 
and the
results of 
\citet{2020ApJ...900..150Y}
and 
\citet{2021ApJ...917...88Y}.
We set a correlation coefficient of $\ge$0.3 or $\le$$-$0.3 for field
galaxies as
the limit for a weak correlation or anti-correlation,
using the larger of the absolute value of the Spearman or
Pearson correlation coefficient, and limit the sample
to galaxies with 10 $\le$ log M* $<$ 11.
There are disagreements in the literature about
some of these trends; these cases are discussed below.
Stellar mass is somewhat isolated on this diagram,
only connecting to concentration and pitch angle.
In the mass range we are focusing on in this study
(10 $\le$ log M* $<$ 11) stellar mass is neither
correlated or anti-correlated with the rest of the parameters.
This diagram does not include some other parameters 
of interest, including gas content, environment, and AGN activity.

In this diagram, there are numerous `cycles' (closed paths) of three variables
connected by three lines (for example, the C-sSFR-arm-strength cycle).
In Figure  
\ref{fig:diagram},
most cycles of three paths connecting three variables
have two negative correlations (two blue connecting lines)
and one positive correlation (a pink connecting line).
In these cases, the two anti-correlations `cancel out', 
and the cycle is internally consistent.
One cycle of three, however, is internally inconsistent: the
concentration-bar-strength-arm-strength cycle, which has two positive correlations
and one negative correlation.   
This apparent contradiction is discussed further in Section
\ref{sec:dis_bar}.

\subsection{Concentration, Quenching, Galaxy Evolution, and Environment} \label{sec:dis_concentration}

For this sample of disk galaxies, 
the cluster galaxies have larger
central concentrations for a given stellar mass than field galaxies
(Figure \ref{fig:C_vs_mass_barred_unbarred}).
This is in agreement with the results of 
\citet{2009MNRAS.394.1213W},
who found that late-type galaxies in clusters
have smaller radii and larger concentrations for the same stellar mass
than a control sample of galaxies in lower density environments.
This result is also consistent with the results of 
\citet{2003MNRAS.346..601G},
who
found that a larger fraction of spirals
in clusters have early-type morphologies compared to field spirals.
Earlier studies have found that 
different types of spirals have different spatial
distributions within clusters.
Blue (star-forming) spirals tend to reside 
further out in clusters, while the fraction of red (quenched) 
spirals peaks at intermediate ranges, inside the virial radius 
but not in the core 
\citep{2009MNRAS.399..966S,
2009MNRAS.393.1324B,
2009MNRAS.393.1302W,
2010MNRAS.405..783M}.
Late-type spirals tend to be found further out in clusters, 
compared to early-type spirals 
\citep{2003MNRAS.346..601G,
2004AJ....127.1344A}.
Why the galaxy population varies with environment in this manner is
not settled.

In theory, it may be possible to distinguish between 
evolutionary theories by determining the timescale of quenching
vs. that of morphological change.  However,
there are conflicting results in the literature on whether
galaxy
quenching in clusters occurs simultaneously
with 
morphological transformation,
or if they happen on different timescales.  
\citet{2022MNRAS.509..567S}
conclude that when a galaxy falls into a cluster, 
morphological transformation 
happens first, followed by a gradual
decrease in sSFR.   In contrast, 
\citet{2009MNRAS.393.1324B}
conclude that blue-to-red evolution 
occurs faster than evolution from late-type morphology to early-type. 
\citet{2017MNRAS.471.2687B}
find that morphology correlates most strongly
with sSFR, independent of environment, and conclude that the same physical
process is responsible
for both the morphology change and the decrease in sSFR.
\citet{2014MNRAS.441..599B,
2022arXiv220107814B}
conclude that quenching is most
closely correlated with the growth of the bulge 
and/or the central stellar velocity dispersion.
\citet{2014MNRAS.440..889S}
suggest that there are two different evolutionary paths
for star-forming disk galaxies: some become spheroidal 
quickly via mergers and quench rapidly, while others retain their disk-like
structure as their sSFRs slowly decrease, becoming red spirals.

In addition to quenching and central concentration,
additional information that may help distinguish between
evolutionary scenarios during cluster infall
are the pitch angles, number of arms, and strengths of the spiral
arms, and how they vary with environment.  
In the following sections, we discuss our results on the arm
parameters of cluster vs.\ field spirals in terms of these ideas.

\subsection{Arm Strengths and Quenching in Cluster Spirals vs.\ Field Spirals} \label{sec:why_weaker}

\citet{2020ApJ...900..150Y}
found an anti-correlation between arm
strength and concentration.  They explain this relation by 
an increasing Toomre Q factor in galaxies with larger bulges.
We find that cluster and field galaxies follow approximately the same
trend.
For the sample as a whole, cluster spirals have weaker spiral arms than
field galaxies.   This effect can be accounted for by 
the fact that cluster spirals have larger concentrations;
the 
arm-strength-to-concentration relation for cluster galaxies is consistent
with that of field galaxies within the uncertainties.  

\citet{2021ApJ...917...88Y}
also
found that the strength of spiral
arms is correlated with sSFR; galaxies above
the main sequence have stronger arms and those below have weaker arms.
We reproduce this result 
in Figure \ref{fig:sSFR_via_C}, where 
we plot arm strength
vs.\ sSFR for narrow ranges of C, and see correlations in the 
subsets.
Galaxies with stronger arms also tend to be bluer 
\citep{2020ApJ...900..150Y}.
In addition,
\citet{2021ApJ...917...88Y}
conclude that arm strength positively correlates
with gas fraction.  
We find that cluster and field galaxies
follow the same arm-strength-to-sSFR relation 
(Figure \ref{fig:sSFR}).  This supports
the idea that the arm-weakening mechanisms are the same in field and cluster
galaxies.
Beyond the relations with concentration and sSFR,
we find no evidence for an additional environmental 
influence on arm strength.
\citet{2021ApJ...917...88Y}
suggest that spiral arms fade away in
the absence of cold gas.  Without gas, spiral patterns 
tend to heat the disk, increasing random stellar motions
\citep{2008gady.book.....B,
2011ApJ...730..109F,
2021arXiv211005615S}.
We do not find strong evidence for a time lag between the build-up
of the bulge and the weakening of the arms, or vice versa.

Observationally, galaxies with less cold gas form proportionally
less stars 
\citep{1959ApJ...129..243S,
1989ApJ...344..685K,
1998ApJ...498..541K},
and have less prominent spiral arms 
\citep{2021ApJ...917...88Y}.
Therefore, the weakening of the arms in spirals may be 
due at least in part to removal, depletion, or heating of the gas in
the disk, lowering the SFR.  
Our results suggest that the
quenching of the disk and the weakening
of the spiral arms happen approximately simultaneously with the
increase in concentration.  
Such evolution happens to both field and cluster spirals, 
since quenched spirals with weak arms are seen in the field
as well as in clusters (Figure \ref{fig:sSFR}).  
The hypothesis
of disk gas removal/depletion
and disk fading is consistent with this scenario; the reduction
of the SFR will lead to less prominent arms,
and an overall diminishing of the starlight of the disk.  

In field spirals,
ram pressure stripping of cold interstellar
disk
gas is unlikely to be the 
dominant cause of gas removal.
Close to the core of a galaxy cluster, 
this may 
be an important process,
but not outside the
virial radius or in the field 
\citep{1999MNRAS.308..947A,
2021arXiv210913614B}.
The excess numbers of red spirals measured out to 3 $-$ 5 times
the virial radius in clusters 
\citep{2009MNRAS.393.1324B}
cannot be attributed
to ram pressure stripping of cold interstellar gas in the cluster 
potential. 
Although
ram pressure stripping of cold gas may operate in 
groups 
\citep{2012MNRAS.427.2841F,
2013MNRAS.436...34C,
2017MNRAS.466.1275B},
it is not likely to be the main gas removal process in more isolated galaxies.
Red spirals are found in both clusters and the field 
\citep{2009MNRAS.393.1324B, 2010MNRAS.405..783M}, therefore
other processes beside
ram pressure stripping of cold gas must produce these objects.
Either different gas removal processes dominate
in different environments but have similar consequences in terms
of quenching 
and morphological change, 
or another process besides ram pressure stripping of cold gas
dominates gas
removal in both clusters and the field.

Strangulation (the stripping of hot circumgalactic gas followed
by a slow quenching) is more likely 
in our sample galaxies than ram pressure stripping of cold gas, 
as strangulation is thought
to be active in less dense environments.
Strangulation 
may operate out to 
three virial radii in clusters or beyond
\citep{2013MNRAS.430.3017B,
2018MNRAS.475.3654Z,
2021MNRAS.502.1051A}
and 
may also 
occur in small groups 
\citep{2009MNRAS.399.2221B,
2008ApJ...672L.103K} including groups falling into clusters 
\citep{2004PASJ...56...29F,
2013MNRAS.435.2713V}.
Based on the sizes, concentrations, and colors of cluster vs. field galaxies,
\citet{2009MNRAS.394.1213W}
argue that strangulation and disk fading are
the main cause of quenching and morphological change for spirals in clusters.
Measurements of metallicity and stellar ages also favor strangulation
as a primary cause for quenching 
in cluster galaxies
\citep{2015Natur.521..192P, 2019A&A...621A.131M}.
One possible explanation for the lack of environmental differences in
the arm-strength-to-sSFR correlation and the arm-strength-to-concentration
anti-correlation is that strangulation in a range of environments
removes hot halo gas, slowing the
rate of gas inflow onto the disk, weakening the spiral arms
and decreasing the disk SFR, causing the disk
to fade, and therefore increasing the central concentration. 

The spiral arms themselves might help enhance the concentration
of the galaxy by driving gas inwards \citep{2014MNRAS.440..208K} and triggering
central star formation.   
Galaxies with strong arms are more likely to have enhanced sSFR
in the central region \citep{2022arXiv220206932Y}, supporting this idea.
Stars can also migrate radially within a galaxy disk
due to resonances with spiral arms or bars, or interactions with massive
clumps, but 
this affects the outer
disk more than the central regions 
\citep{2002MNRAS.336..785S,
2012A&A...548A.127M, 2020MNRAS.499.2672W, 2022MNRAS.511.5639L}.
At high redshifts, the migration of massive clumps into the center
of a galaxy may also help to build a bulge 
\citep{2007ApJ...670..237B, 2008ApJ...688...67E, 2010MNRAS.404.2151C,
2012MNRAS.422.1902I},
however, clump migration is likely not the primary cause of bulge
growth at
z $<$ 1 \citep{2017ApJ...840...79S}.

Minor mergers may
also increase the central concentration of
galaxies, by building up the bulge of a galaxy 
\citep{1998ApJ...502L.133B,
2001A&A...367..428A,
2005A&A...437...69B}.
In addition, galaxy interactions may drive gas into the centers of galaxies, 
depleting the gas in the outer disk while
triggering a central starburst and bulge growth
\citep{2000MNRAS.317..667M, 2008MNRAS.385.1903L}.
These processes may be enhanced in galaxy groups falling into clusters
\citep{2003ApJ...582..141G,
2006asup.book..115S,
2008MNRAS.385L..38M}.
Mergers and interactions in infalling groups, possibly combined 
with stripping of hot gas in the group environment,
could potentially contribute to
morphological change and quenching in cluster galaxies
\citep{2013MNRAS.435.2713V,
2020MNRAS.498.3852B}.
The `bend' in the concentration-to-stellar-mass relation
(Figure \ref{fig:C_vs_mass_barred_unbarred})
was 
noted before by \citet{2020MNRAS.493.1686L}, who 
attribute the increase in the number of high concentration galaxies
above M* = 10$^{10}$ M$_{\sun}$ to a population of galaxies
with `classical' bulges (i.e., those with high Sersic indices).
Classical bulges may be caused by mergers 
\citep{2016ASSL..418..431K,
2016ASSL..418...41F}.
Figure \ref{fig:C_vs_mass_barred_unbarred} shows that 
cluster galaxies on average have higher 
concentrations than field galaxies
in the 10 $\le$ log M* $<$ 11 range, suggesting that 
at least some these may
be the product of mergers.
We therefore cannot rule out that mergers and interactions play a role
in the build-up of the central concentration of spiral disks in both
cluster and field galaxies.

The build-up of a central spheroidal component, 
regardless of its cause,
could potentially stabilize a gaseous 
disk and therefore decrease the
sSFR 
\citep{2009ApJ...707..250M,
2010MNRAS.404.2151C,
2020MNRAS.495..199G}, possibly weakening the spiral
arms \citep{2020ApJ...900..150Y}.
Simulations suggest that a central spheroid will increase the gas velocity
dispersion, thus suppressing star formation 
\citep{2020MNRAS.495..199G}.
In this scenario, the galaxy may still have considerable cold gas, but
a low SFR.  
An argument against disk stabilization being the primary quenching mechanism
is the observation
that quiescent spirals tend to be deficient in cold
gas \citep{2016MNRAS.462.1749S,
2021MNRAS.502L...6E}.
The weakening of spiral arms with decreasing sSFR and decreasing gas
content 
\citep{2021ApJ...917...88Y}
also argues against disk stabilization
by a central bulge as the sole process that quenches star formation
in spirals; gas depletion seems to play an important role.
However, both processes might operate simultaneously.  As spiral galaxies evolve,
the central concentration tends to grow, while the gas content of the disk
tends to decrease.  Both of these processes would tend to
cause decreased star formation
in the disk, and weaker arms.

For our sample of galaxies,
tidal stripping in the gravitational field of a cluster is probably
not the main process
that has increased the central concentration. 
Tidal stripping 
can remove stars from the
outer stellar disk of galaxies, increasing
the observed concentration
index as seen in optical light 
\citep{1998ApJ...495..139M, 1999MNRAS.304..465M,
2004AJ....127.1344A}.
However, this process is only expected to be important
for relatively low mass or low surface density galaxies.
For disk galaxies with stellar masses of a few times
10$^{10}$ M$_{\sun}$, in the range targeted in this study, 
simulations show that 
the size and mass of the stellar disk are little affected by tidal forces
from a host cluster 
\citep{2003ApJ...582..141G,
2020A&A...638A.133L}.
Tidal disruption
may be important 
for lower mass galaxies 
\citep{2005MNRAS.364..607M}, but 
it may not have played a big role in shaping
the observed morphologies
of the galaxies in this sample.

Tidal stripping is expected to have a bigger
effect on the dark matter halo of galaxies falling into a cluster
than on the disk stars 
\citep{2016ApJ...833..109S,
2020A&A...638A.133L}.
Stripping of dark matter will change
the gravitational potential of the galaxy,
relative to the concentration observed in
optical light.  
However, we do not see a significance difference in spiral
arm strength or pitch
angle versus the observed
concentration for cluster galaxies 
relative to galaxies in the field.
If there are larger visible-light-to-dark-matter ratios 
in cluster galaxies than in field galaxies, it does not make an observable
difference in the 
arm-strength-to-concentration and 
pitch-angle-to-concentration relations.

\subsection{What About Quenching Due to AGN Feedback?} \label{sec:dis_agn}

\citet{2022arXiv220107814B}
argue that radio-mode AGN feedback is the dominant
quenching mechanism in galaxies.
In radio-mode AGN feedback (also known as kinetic AGN feedback), jets from
a central AGN heat the halo gas, prevent it from cooling,
and therefore inhibit star formation 
\citep{2012ARA&A..50..455F,
2016Natur.533..504C}.
Such jet activity has a short duration
(less than a few $\times$ 10$^8$ years)
and may be episodic with
a short duty cycle
\citep{2004MNRAS.351..169M,
2016Natur.533..504C},
thus it is
difficult to directly connect this activity with quenching in
individual galaxies, or statistically in large samples.
\citet{2022arXiv220107814B}
base their AGN quenching
argument on the observation that 
quenching correlates well with both 
central velocity dispersion and with bulge mass, and the fact that 
central velocity dispersion and bulge mass both
correlate with the mass of the central black hole.
In their argument, they assume that the time-averaged feedback 
from an AGN scales with black hole mass.  
Supporting the idea of AGN quenching or morphological quenching,
spatially-resolved stellar population synthesis shows that the disks
of spirals tend to have younger stellar ages at larger radii, implying
inside out quenching and/or inside out disk growth 
\citep{2013ApJ...764L...1P, 2016A&A...590A..44G, 2017A&A...607A.128G,
2018MNRAS.474.2039E, 2021MNRAS.504.3058M, 2022A&A...659A.160B}. 

In our galaxy sample, optically-selected Seyfert activity 
is most commonly seen in galaxies
that are quenched or quenching, rather than star-forming
(Figure \ref{fig:sSFR_AGN}).   
This is consistent with the results of 
\citet{2007ApJS..173..342M},
who
found that the fraction of AGNs peaks in the transition zone
between star-forming and quenched galaxies.
This supports the idea that there is 
a connection between quenching and nuclear activity.
An alternative explanation is that dust extinction and/or
emission lines associated with H~II regions
mask nuclear activity in some cases \citep{2017MNRAS.466.2879B}, 
so AGNs are missed among
the higher sSFR galaxies.
One issue with the AGN quenching hypothesis is timescale;
in the study by 
\citet{2009ApJ...692L..19S},
their AGN fraction peaks in transition galaxies which quenched about
10$^8$ years ago.  If this quenching was caused by 
the AGNs, then this implies
that there is a timelag between when quenching
occurs and when signatures of the AGN become observable,
or that AGN turn off and back on again on this timescale.

Another uncertainty is the relationship between Seyfert activity
and radio-mode feedback; 
optical signatures of AGN are not direct evidence for 
present or past 
jet activity.   As noted earlier, very few galaxies
in our sample 
(0.2\%)
are classified as radio-bright AGN 
in the 
\citet{2012MNRAS.421.1569B}
catalog.  
This means that
there is little direct evidence for radio-mode quenching in the
current
sample of galaxies.
Unless the timescale for this jet activity is very short, such that
almost all of the sample galaxies are now in the quiescent phase,
and the efficiency of quenching is high when the jets are turned on, 
radio-mode AGN feedback is likely not responsible for the quenching
in our sample galaxies.

Some studies have found a statistical connection between
signatures of 
ram pressure stripping and AGN activity, in that AGNs may 
be more common in cluster galaxies with 
prominent 
ram-pressure-stripped tails 
\citep{2017Natur.548..304P,
2021arXiv211102538P}.   
These authors suggest that ram pressure
stripping may feed AGNs.  Alternatively,
AGN feedback could potentially lead to heating and expansion of
halo gas, which may 
allow ram pressure stripping 
to more effectively remove gas.   
We see similar fractions of Seyferts in clusters
vs.\ the field in our sample, thus there is no strong evidence 
that ram pressure is enhancing Seyfert activity,
however, the number of Seyferts in our
sample is small.

\subsection{What About Bars?} \label{sec:dis_bar}

\citet{2007MNRAS.381..401L}
found that the bars in galaxies with larger
bulges tend to be stronger than those in galaxies with smaller bulges.
In Figure
\ref{fig:bar_strength_vs_C}, 
we showed that the same trend
also holds for 
the 
\citet{2020ApJ...900..150Y}
measurements, 
and that this tendency is present for field and cluster galaxies separately.
This result is consistent with the 
\citet{2011arXiv1111.1532G}
study that found a higher bar fraction
in early-type spirals.
However, other studies see the reverse situation: 
an increase in the fraction of
barred galaxies in less concentrated spirals and/or later-type spirals 
\citep{2008ApJ...675.1194B,
2009A&A...495..491A}
or no difference in
the bar fraction as a function of Hubble type 
\citep{2007ApJ...659.1176M}.
This discrepancy may be due to different bar selection criteria
in the different studies.
\citet{2019ApJ...872...97L}
searched for bars in a sample of z $<$ 0.01 SDSS
spiral galaxies using several methods, and concluded that 
automatic bar-finding techniques may miss some weaker bars,
particular in galaxies with prominent bulges.  
They also conclude that 
weak bars preferentially reside in later-type spirals,
while strong bars may be more likely in galaxies with earlier Hubble types
\citep{2019ApJ...872...97L}.
The relatively low spatial resolution of SDSS may mean bars
are missed in some galaxies \citep{2018MNRAS.474.5372E}.

It has been suggested that bar strengths are enhanced in clusters by
tidal forces from the cluster as a whole 
\citep{1990ApJ...350...89B,
2016ApJ...826..227L,
2020A&A...638A.133L}.
However, cluster galaxies follow the same bar-strength-to-concentration
relation as field galaxies (Figure 
\ref{fig:bar_strength_vs_C}), which argues against the idea
of an excess population of tidally-induced
bars in clusters.
Furthermore, we find similar percentages of barred galaxies in clusters
as in the field (41 $\pm$ 7\% vs.\ 47 $\pm$ 2\%, respectively)
for log M* $\ge$ 10.
This also argues that
the production of bars by tidal forces in clusters 
is not a major factor
for galaxies in this mass range.
Our lack of a difference in bar fraction is 
consistent with most earlier studies 
\citep{1986MNRAS.223..139K,
2009A&A...497..713B,
2009A&A...495..491A,
2010ApJ...711L..61M,
2011arXiv1111.1532G,
2021arXiv211111252S},
but not all
\citep{1981ApJ...244L..43T,
2012MNRAS.423.1485S}.

Earlier studies showed that 
bar strength and arm strength tend to be correlated 
\citep{2004AJ....128..183B,
2005AJ....130..506B,
2020ApJ...900..150Y}.
This correlation suggests that 
either 
bars drive spirals 
\citep{1976ApJ...209...53S,
1979ApJ...233..539K}, or alternatively, 
conditions in disks that favor bar production
also favor strong arms 
\citep{2010ApJ...715L..56S,
2019A&A...631A..94D}.
The observation that barred galaxies have stronger arm strengths
than unbarred galaxies 
\citep{2020ApJ...900..150Y}
may be a consequence of this correlation
between bar strength and arm strength; weaker bars may not be
detectable in the SDSS images, causing such galaxies to be classed
as unbarred.  
With our sample, 
we cannot rule out that cluster and field galaxies
obey the same arm-strength-to-bar-strength relation.

As noted earlier, there is 
an apparent contradiction between
three trends seen in the 
\citet{2020ApJ...900..150Y}
data:
as concentration increases, arm strength decreases and bar strength increases,
yet arm strength and bar strength are directly correlated.  
Bar strength is positively correlated with concentration (Figure 
\ref{fig:bar_strength_vs_C}),
but arm strength is inversely correlated with concentration
(Figure \ref{fig:s_vs_C_barred_unbarred}).
The resolution of this apparent inconsistency is that, when we look
at the subset of barred galaxies alone, the anti-correlation between
arm strength and concentration weakens enough to become doubtful (Table 1).
Thus there is no inconsistency.
Why there is a difference
between the s vs.\ C relation for barred and unbarred galaxies 
is uncertain.  As discussed above, there may be selection effects in identifying
bars in spiral galaxies, and these selection effects may be a function of
concentration.   This may lead to biases in the subset of galaxies classified
as barred.     
The small slope of the
bar strength to concentration relation and the relatively small number
of barred galaxies adds additional uncertainty in interpreting these relations.

Another complication is that 
arm strength is affected by both central concentration and sSFR (and
therefore gas content) (see Section \ref{sec:why_weaker}), 
while bar strength appears to be independent
of sSFR  
(Figure \ref{fig:sSFR_bar}).  Decreasing gas content weakens arms,
without having the same effect on bar strength \citep{2021ApJ...917...88Y}.
Variations in gas content affect arm strength and bar strength 
differently, while at the same time both arm strength and bar
strength are influenced by concentration and concentration is anti-correlated
with gas content.  This produces a complex relationship between 
arm strength, bar strength, and concentration.

\subsection{Pitch Angle is a Function of Concentration, Not of Environment} 

The 
\citet{2020ApJ...900..150Y}
trend of increasing pitch angle with decreasing
concentration is consistent with earlier studies that show trends
with concentration 
\citep{2013MNRAS.436.1074S}
and Hubble type
\citep{1981AJ.....86.1847K,
1999A&A...350...31M,
2002A&A...388..389M,
2018ApJ...862...13Y,
2020MNRAS.493..390S}.
The lack of an observable correlation
between 
bulge prominence and arm winding found with Galaxy Zoo data
\citep{2017MNRAS.472.2263H,
2019MNRAS.487.1808M,
2021MNRAS.504.3364L,
2022MNRAS.509.3966W}
may be due to uncertainties
in the Galaxy Zoo parameters.   When we compared the 
\citet{2020ApJ...900..150Y}
pitch angles with Galaxy Zoo arm winding classes, we found 
only a very weak trend with a large amount of scatter 
(Figure \ref{fig:phi_vs_winding_class}).  
The Galaxy Zoo bulge prominence parameter B$_{avg}$
correlates better with concentration, though there is some
scatter.
These uncertainties may contribute to the large scatter seen in the
Galaxy Zoo relations, and therefore the lack of an observed correlation
between Galaxy Zoo bulge prominence and arm winding.

Pitch angle also anti-correlates with stellar mass in the 
\citet{2020ApJ...900..150Y} dataset, though this is a weaker relation
than the trend with concentration.  The anti-correlation with stellar
mass may be an indirect consequence of the anti-correlation with 
rotation curve maximum velocity V$_{max}$ 
seen in earlier studies
\citep{1981AJ.....86.1847K,
1982ApJ...253..101K,
2017MNRAS.471.2187D,
2019ApJ...877...64D}, 
which implies a relationship between pitch angle and dynamical
mass. 

When concentration is held constant, pitch angle positively correlates
with sSFR 
(Figure \ref{fig:sSFR_phi_via_C}); galaxies with tighter arms tend
to have lower sSFR for a fixed concentration.    
This is consistent with the work of 
\citet{2015ApJ...802L..13D}, who found that pitch angle increases
with the surface mass density of atomic hydrogen gas in the disk,
since SFR increases with increasing gas surface density
\citep{1959ApJ...129..243S,
1989ApJ...344..685K,
1998ApJ...498..541K}.
\citet{2015ApJ...802L..13D}
suggested a `fundamental plane'
of spiral structure in disk galaxies, in which pitch angle depends
both on the central stellar bulge mass and the surface mass density of atomic
hydrogen gas in the disk.   They showed that 
as the gas content increases, the pitch
angle increases, while the pitch angle decreases with increasing bulge
mass.   This is consistent with our results, if 
sSFR is related to surface gas density, 
and concentration increases with increasing bulge mass.

Using the 
\citet{2020ApJ...900..150Y}
arm strengths and concentrations,
we found that cluster and field galaxies
follow the same basic pitch-angle-to-concentration relation
(Figure \ref{fig:phi_vs_C_and_M}).  
This suggests
that mass concentration and sSFR are the 
dominant factors that determine the tightness of spiral arms,
and environmental influences like tidal interactions are not major
factors.
Spirals in clusters tend to have larger concentrations, and therefore
have tighter arms, but otherwise, we do not detect strong environmental
differences between the two samples.
There is considerable scatter in the $\phi$-to-concentration relation, 
but the spread is similar
in both field and cluster galaxies. 

The anti-correlation between pitch angle and concentration is consistent
with classical 
density wave theory 
\citep{1964ApJ...140..646L,
1975ApJ...196..381R}.
The three-fold relationship between pitch angle, gas surface density,
and bulge mass noted by 
\citet{2015ApJ...802L..13D} is explained by those authors
in terms of the spiral density wave theory as outlined by
\citet{1964ApJ...140..646L}: pitch angle is determined
by the ratio of the density of matter in the disk to the mass of
the central bulge.  
The observed anti-correlation between
pitch angle and concentration can also be approximately reproduced with 
models of spiral arms produced by gravitational instabilities,
which produce 
a correlation between pitch angle and shear rate,
and therefore with mass distribution
\citep{2013A&A...553A..77G,
2014ApJ...787..174M,
2016ApJ...821...35M,
2014PASA...31...35D}.
However, 
the observed pitch-angle-to-concentration anti-correlation
for spirals cannot easily be explained 
using models when only interactions
are responsible for all spiral patterns.

Simulations show that 
flyby
gravitational interactions between
galaxies can produce very extended open spiral arms 
in some galaxies, with the
right 
orbital parameters 
\citep{2008ApJ...683...94O,
2011MNRAS.414.2498S}.
However, we see no statistical differences in the pitch angles of 
cluster vs.\ field spiral galaxies, as one might expect if fast
interactions in a cluster are perturbing cluster galaxies more frequently
than field galaxies.
Perhaps the combined effect of multiple encounters at random orientations
and random times 
cancels out the effect, producing no net difference in the
mean pitch angle.  
If spiral patterns wind up with time
after an initial perturbation as indicated
by both interaction models 
\citep{2008ApJ...683...94O,
2010MNRAS.403..625D,
2011MNRAS.414.2498S}
and models
of spiral arm generation by 
gravitational instabilities 
\citep{2019MNRAS.490.1470P},
a collection of galaxies undergoing
flyby interactions at random times may be in a range of stages of wind-up.
Thus a relatively random distribution of
pitch angles is expected for a large sample of galaxies.
In addition, 
ram pressure stripping in rotating cluster spirals may stretch and shear
outer spiral arms,
and therefore `unwind' the 
arms, increasing their pitch angle 
\citep{2001MNRAS.328..185S,
2021MNRAS.500.1285B}.
This may also increase the scatter in the observed pitch angles
of spirals.
Spiral arm wind-up may also contribute to the 
observed correlation between pitch angle
and sSFR;
as galaxies
quench with time and the arms weaken, the arms may wind up,
decreasing the pitch angle.

At present there is some uncertainty about the relationship between
pitch angle and 
bar strength.
\citet{2019A&A...631A..94D}
and \citet{2021MNRAS.504.3364L} found that 
pitch angle is not correlated
with bar strength, while 
\citet{2020ApJ...900..150Y}
found a weak anti-correlation.
\citet{2020ApJ...900..150Y}
found similar 
pitch-angle-to-C relations for barred
and unbarred galaxies, an argument against  
the idea that bars play a major role
in driving spirals.
Similarly, 
\citet{2019A&A...631A..94D}
found that for a given Hubble type the pitch angles of barred
and unbarred galaxies are similar.
However, using Galaxy Zoo data 
\citet{2019MNRAS.487.1808M}
found
that for a given bulge prominence barred galaxies have more open spiral
arms on average.   
Thus, although it appears as if bars may not play a big role in 
determining pitch angle, there is still some uncertainty.

\subsection{f3, Number of Arms, Concentration, and Environment} 

We compared the 
\citet{2020ApJ...900..150Y}
f3 parameter with the number
of spiral arms as measured by Galaxy Zoo.  Galaxies with two arms
tend to have lower f3 values than other spiral galaxies 
(Figure \ref{fig:f3_vs_number_arms}). 
Although f3
does not provide a clean separation between two-armed galaxies and other
spirals, there is a rough trend in that galaxies with lower f3 values
are more likely to have two arms, and higher f3 galaxies are proportionally
more likely to have a different number of arms.
This result is consistent with earlier work.
With a different sample of galaxies and different set of images,
\citet{2018ApJ...862...13Y}
found
that the mean relative m = 2 Fourier
amplitude for grand design galaxies tends to be larger than for other
spirals.  
Similarly, 
\citet{2011ApJ...737...32E}
found that grand design
galaxies have lower f3/f2 ratios compared to flocculent and 
multi-armed galaxies, where f2 is the normalized m = 2 Fourier amplitude.
These results support the idea that the arm parameter f3 is a rough
indicator of the number of arms.

In the 
\citet{2020ApJ...900..150Y}
dataset, we do not see a correlation between f3 and 
arm strength
(Table 1).
This suggests that statistically
two-armed spirals do not have measurably stronger arms than galaxies
with more arms, if f3 is indeed a reliable measure of the number of arms.
Classically, grand design galaxies are defined as 
having `two long symmetric arms dominating the optical disk' 
\citep{1987ApJ...314....3E}.
For a given
Hubble type grand design galaxies tend to have stronger
arms 
\citep{2011ApJ...737...32E}.
In the 
\citet{2020ApJ...900..150Y}
sample, there may be
some 2-armed galaxies with weaker and/or discontinuous arms that
would not meet the classical definition of grand design.
More investigations into the length, strength, and continuity of spiral arms
as a function of arm number and other disk parameters are needed to 
address this point.

As noted by 
\citet{2020ApJ...900..150Y},
the f3 parameter is 
weakly inversely correlated with concentration.
This is consistent with previous observations that galaxies with strong two-armed
patterns have larger B/D ratios than other spirals 
\citep{2017MNRAS.471.1070B},
and that the f3 parameter increases somewhat with increasing
Hubble type 
\citep{2020ApJ...900..150Y}.
It is also consistent with earlier studies 
based on
the Elmegreen arm class system.
Arm class is slightly correlated
with Hubble type, with grand design galaxies more likely to be early-type
spirals and flocculent galaxies late-type, though with a lot of scatter 
\citep{1982MNRAS.201.1021E,
2019A&A...631A..94D}.
This trend of arm class with morphological type is 
consistent with the Galaxy Zoo result that two-armed spirals
are redder than galaxies with more arms 
\citep{2016MNRAS.461.3663H},
since earlier
type spirals tend to be redder than later types (Roberts and Haynes 1994;
Buta 2011).

The f3 parameter is weakly positively correlated with sSFR
when the full mass and sSFR range is included, however, when
we limit the sample to 10 $\le$ log M* $<$ 11 and log sSFR $\ge$
-12 the correlation weakens further
(Figure \ref{fig:sSFR_f3}), and when the sample is subdivided
into subsets with narrow ranges of concentration, the correlation
disappears
(Figure \ref{fig:sSFR_f3_via_C}).  This suggests that the 
weak f3-to-sSFR
correlation is an indirect consequence of both quantities being indirectly
correlated with concentration (see diagram in Figure
\ref{fig:diagram}); for a fixed concentration f3 does not depend
upon sSFR.

When we compare cluster vs. field spirals
for the same concentration, we do not find a significant 
difference in the f3 parameters of cluster vs. field galaxies
(Figure \ref{fig:f3_vs_C}).
In other words,
we do not detect an environmental effect on f3, 
above and beyond what is expected based on the
larger concentrations of cluster spirals.  It appears that the number
of arms 
is determined more by the mass distribution in the galaxy
than by environment or by gas content.

In a past study,
\citet{1982MNRAS.201.1035E},
found similar fractions of grand design patterns in cluster and non-cluster
spirals.  In contrast, 
\citet{2011JKAS...44..161C}
found proportionally
more grand design spirals
in clusters.  
Using Galaxy Zoo data,
\citet{2016MNRAS.461.3663H}
found 
a larger fraction of
two-armed spirals in higher density regions.
\citet{2020MNRAS.493..390S}
concluded that non-isolated galaxies
are more likely to have two arms.
\citet{2011JKAS...44..161C},
\citet{2014JKAS...47....1A},
and 
\citet{1987ApJ...314....3E}
concluded that the fraction
of grand design spirals increases with galaxy density.

These earlier studies do not conflict with our results, because 
we correct for the larger central concentrations of cluster galaxies
while these earlier studies did not. 
For the sample as a whole we found
a larger fraction of cluster spirals have low f3 
compared to field spirals (Figure \ref{fig:armstrength}).
However, when we compare galaxies with similar concentrations we
find similar f3 values.   
We suggest that the main reason
that
more two-armed spirals are found in clusters
is because galaxies in clusters have larger central concentrations,
and galaxies with larger concentrations are more likely to have
two arms.

A quantitative comparison between our results and earlier
studies is difficult, 
since different parameters are being measured.
The 
\citet{2020ApJ...900..150Y}
f3 parameter is
not a perfect measure of the number of spiral arms, but instead has a lot
of scatter. Furthermore, since the f3 parameter is normalized by 
the sum of the m = 2, m = 3, and m = 4 
Fourier amplitudes, higher order components
are not taken into account, which might add uncertainty for flocculent galaxies.
The arm class system is also not a perfect measure of the number 
of spiral arms; in addition to having multiple arms,
the arms in flocculent galaxies are 
expected to be fragmented and irregular, while the arms in grand design
systems are expected to be long and continuous.   
This introduces additional uncertainty into the classification.

The interpretation of f3 and/or the number of spiral arms in terms of
theoretical models of spiral pattern production in galaxies is also 
challenging.   
\citet{2017MNRAS.471.1070B}
and 
\citet{2020ApJ...900..150Y}
conclude that 
classical spiral density wave theory
operates 
in galaxies with large bulges, producing long-lived two-armed
spirals, while 
gravitational instabilities 
produce spirals 
in galaxies with small bulges. 
They base this argument on the high Toomre Q parameter expected in the centers
of galaxies with large bulges; according to theory 
\citep{1985IAUS..106..513L,
1989ApJ...338..104B,
2016ApJ...826L..21S},
a high Q will reflect 
incoming waves, creating a stable spiral pattern.
In contrast,
a galaxy with a weak bulge and therefore a low expected Toomre 
Q parameter in its core will not 
produce a stable density wave pattern.
For galaxies with small bulges, 
\citet{2020ApJ...900..150Y}
suggest 
that the spiral patterns are 
instead
produced by random gravitational instabilities in the disk. 
Models of
this process
\citep{2011MNRAS.410.1637S,
2012MNRAS.421.1529G,
2013ApJ...766...34D,
2018MNRAS.478.3793D,
2021arXiv211005615S}
produce multiple short arm fragments similar to those seen
in flocculent galaxies. Flocculent galaxies
tend to have smaller bulges 
\citep{1982MNRAS.201.1021E,
2017MNRAS.471.1070B,
2019A&A...631A..94D},
consistent with this idea.

In models of spiral production via self-gravity, 
the number of arms depends upon the mass distribution of the 
galaxy 
\citep{2001A&A...368..107F,
2013ApJ...766...34D,
2015ApJ...808L...8D,
2016ApJ...821...35M}.
Since these kinds of models
typically
do not produce two-armed patterns 
\citep{2015ApJ...808L...8D},
two-armed galaxies need another explanation.
Since two-armed spirals can be produced in interactions
according to simulations
\citep{2008ApJ...683...94O,
2010MNRAS.403..625D,
2011MNRAS.414.2498S},
interactions are often invoked as the cause
of two-armed spiral patterns in galaxies
\citep{2014PASA...31...35D,
2017MNRAS.472.2263H}.
Another suggestion is 
that
ram pressure stripping of the interstellar gas
in rotating disk galaxies falling into clusters
produces flocculent and multi-armed spiral patterns,
depending on the orientation of the disk
\citep{2001MNRAS.328..185S}.

We do not see a trend in f3 with sSFR when the dependence on
concentration is removed.  Since sSFR is closely related to gas content,
this suggests that, for a fixed concentration, 
whether a galaxy has two arms or multiple arms
is not dependent on gas content.
This is consistent with the idea that
two-armed spirals are produced by a different mechanism
than flocculent and/or multi-armed galaxies.

Observationally, galaxies with larger concentrations are more likely
to have lower f3 and more likely to have two arms.
At the same time, galaxies with larger concentrations tend to be in
denser regions. 
The evolutionary connection
between concentration, environment,
and the number of arms in spiral galaxies is still unclear.
Simulations show that a gravitational 
interaction with the 
potential of a cluster can induce a two-armed spiral pattern 
\citep{1990ApJ...350...89B,
1993ApJ...408...57V,
2017ApJ...834....7S}.
Such interactions could produce two-armed patterns in galaxies
independent of whether the galaxy has a large concentration,
yet we see similar 
concentration to f3 relations in field and cluster galaxies.
Accounting for concentration,
we see neither an excess nor a deficiency
of two-armed galaxies in clusters relative to the field.

The similar concentration-to-f3 relations for cluster and
field galaxies introduces a number of intriguing questions.
Is concentration rather than environment the main factor
that drives the number of arms in spirals?
Galaxies with two arms are more frequently early-type spirals; is this
because early-type galaxies tend to be in denser environments and therefore
suffer more tidal interactions, or is it because
a larger central bulge favors a two-armed spiral pattern, independent
of environment?
Do tidally-induced grand design patterns persist 
longer if the galaxy has a large bulge?   
These are some questions about spiral galaxies that are still unanswered.

\section{Summary} \label{sec:summary}

\citet{2020ApJ...900..150Y}
conducted a careful Fourier analysis of the spiral patterns
in the SDSS images of more than 4000 spiral and S0 galaxies. 
We divided this sample up into subsets based on environment
using the 
\citet{2017MNRAS.470.2982L}
catalog of galaxy groups and clusters,
and searched for environmental differences in the parameters.
We investigated correlations between central concentration, 
spiral arm strength, bar strength, pitch angle, and the normalized
 m=3 Fourier amplitude (f3) in galaxies in clusters vs.\ field galaxies.
Cluster galaxies in this sample have larger concentrations
for the same stellar mass, compared to the field galaxies.
For the sample as a whole,
\citet{2020ApJ...900..150Y}
found anti-correlations between concentration
and arm strength, concentration and pitch angle, and concentration
and f3.  When we take into account the larger central concentrations of 
galaxies in clusters and compare galaxies with similar central concentrations,
we do not find significant differences between the spiral arm 
parameters 
of cluster and field galaxies.
We conclude that, within the uncertainties, 
spirals in clusters follow the same
arm-strength-to-concentration, f3-to-concentration, and 
pitch-angle-to-concentration 
anti-correlations as spirals in the field.
The correlations between the arm parameters
and the sSFRs in cluster galaxies are also similar to those as field galaxies.
Overall, cluster galaxies have weaker arm strengths, lower
f3 values (i.e., more two-armed spirals), 
and lower pitch angles than field galaxies. 
These differences can be accounted for by the larger concentrations
in cluster galaxies. 
We also find that 
barred galaxies in clusters have a similar bar-strength-to-concentration
correlation
within the uncertainties
as field galaxies.

We compared the 
\citet{2020ApJ...900..150Y}
galaxy parameters with related parameters
from Galaxy Zoo.  We found a weak trend between f3 and the number
of spiral arms, in that f3 
tends to be lower for galaxies identified as two-armed in Galaxy Zoo.
However, there is a large amount of scatter.
The Galaxy Zoo bulge prominence parameter is strongly
correlated with the 
\citet{2020ApJ...900..150Y}
concentration, but with significant scatter.  
Galaxy Zoo participants
clearly distinguished galaxies with large concentrations
from those with small concentrations, but finer gradations into multiple
bulge prominence classes are uncertain.
The Galaxy Zoo winding class (`tightly wound', `medium wound', and 
`loosely wound')
is only very weakly related to the pitch angle as
measured by 
\citet{2020ApJ...900..150Y}.
Galaxies flagged as `tightly wound' by Galaxy Zoo tend to have lower pitch
angles, but the other winding classes are not well-separated in terms 
of pitch angle.

We also compared the spiral parameters with sSFR, and investigated
correlations for narrow ranges of concentration. When concentration
is held fixed, arm strength and pitch angle are correlated with sSFR,
but f3 is not correlated with sSFR.   
Since sSFR depends in part on 
gas surface density, this implies that for a given concentration more gas 
leads to stronger and more open arms, but for a given
concentration, whether a galaxy has two arms
or multiple arms is independent of gas content. 
The relations between pitch angle, concentration, and sSFR 
support the suggestion by \citet{2015ApJ...802L..13D}
of a `fundamental plane' of spiral structure involving pitch
angle, bulge stellar mass, and gas surface density.

In sum, the evidence to date suggests that spiral waves are an internal phenomenon of disk galaxies, not strongly affected by environment in any direct or continuing way. An exception is strong M51-type interactions. The environment has an indirect influence on waves via the gas content and galaxy structure, specifically its concentration, which depend on bulge size and quenching history. These parameters depend, in turn, on environment, as detailed above. 
Many aspects of this `spirals are internal waves' scenario remain to be tested in detail. However, the prospects for such tests are good in this coming time of large surveys.

\vskip 0.1in

{\bf Acknowledgements:} 
We thank the anonymous referee for valuable suggestions
that greatly improved this paper.
We thank Si-Yue Yu for providing us with tables of data, and for 
helpful comments on this manuscript.  
This research has made use of the VizieR catalogue access tool
and the cross-match service 
at the Centre de Donn\'ees Astronomiques de Stasbourg,
CDS, Strasbourg, France.
This research has also made use of the NASA/IPAC Extragalactic Database (NED),
which is operated by the Jet Propulsion Laboratory, California Institute of Technology,
under contract with the National Aeronautics and Space Administration.
This research is based in part on data from the Sloan Digital Sky Survey.
Funding for the Sloan Digital Sky Survey IV has been provided by the Alfred P. Sloan Foundation, the U.S. Department of Energy Office of Science, and the Participating Institutions. SDSS-IV acknowledges
support and resources from the Center for High-Performance Computing at
the University of Utah. The SDSS web site is www.sdss.org.
SDSS-IV is managed by the Astrophysical Research Consortium for the 
Participating Institutions of the SDSS Collaboration including the 
Brazilian Participation Group, the Carnegie Institution for Science, 
Carnegie Mellon University, the Chilean Participation Group, the French Participation Group, Harvard-Smithsonian Center for Astrophysics, 
Instituto de Astrof\'isica de Canarias, The Johns Hopkins University, 
Kavli Institute for the Physics and Mathematics of the Universe (IPMU) / 
University of Tokyo, the Korean Participation Group, Lawrence Berkeley National Laboratory, 
Leibniz Institut f\"ur Astrophysik Potsdam (AIP),  
Max-Planck-Institut f\"ur Astronomie (MPIA Heidelberg), 
Max-Planck-Institut f\"ur Astrophysik (MPA Garching), 
Max-Planck-Institut f\"ur Extraterrestrische Physik (MPE), 
National Astronomical Observatories of China, New Mexico State University, 
New York University, University of Notre Dame, 
Observat\'ario Nacional / MCTI, The Ohio State University, 
Pennsylvania State University, Shanghai Astronomical Observatory, 
United Kingdom Participation Group,
Universidad Nacional Aut\'onoma de M\'exico, University of Arizona, 
University of Colorado Boulder, University of Oxford, University of Portsmouth, 
University of Utah, University of Virginia, University of Washington, University of Wisconsin, 
Vanderbilt University, and Yale University.


%

\vspace{5mm}






\vfill
\eject

\appendix

\section{Using an Alternative Specific SFR Measure} \label{sec:appendix_sSFR} 

\subsection{The Star-Forming Main Sequence} \label{sec:appendix_sSFR_A} 

\begin{figure}[ht!]
\plotone{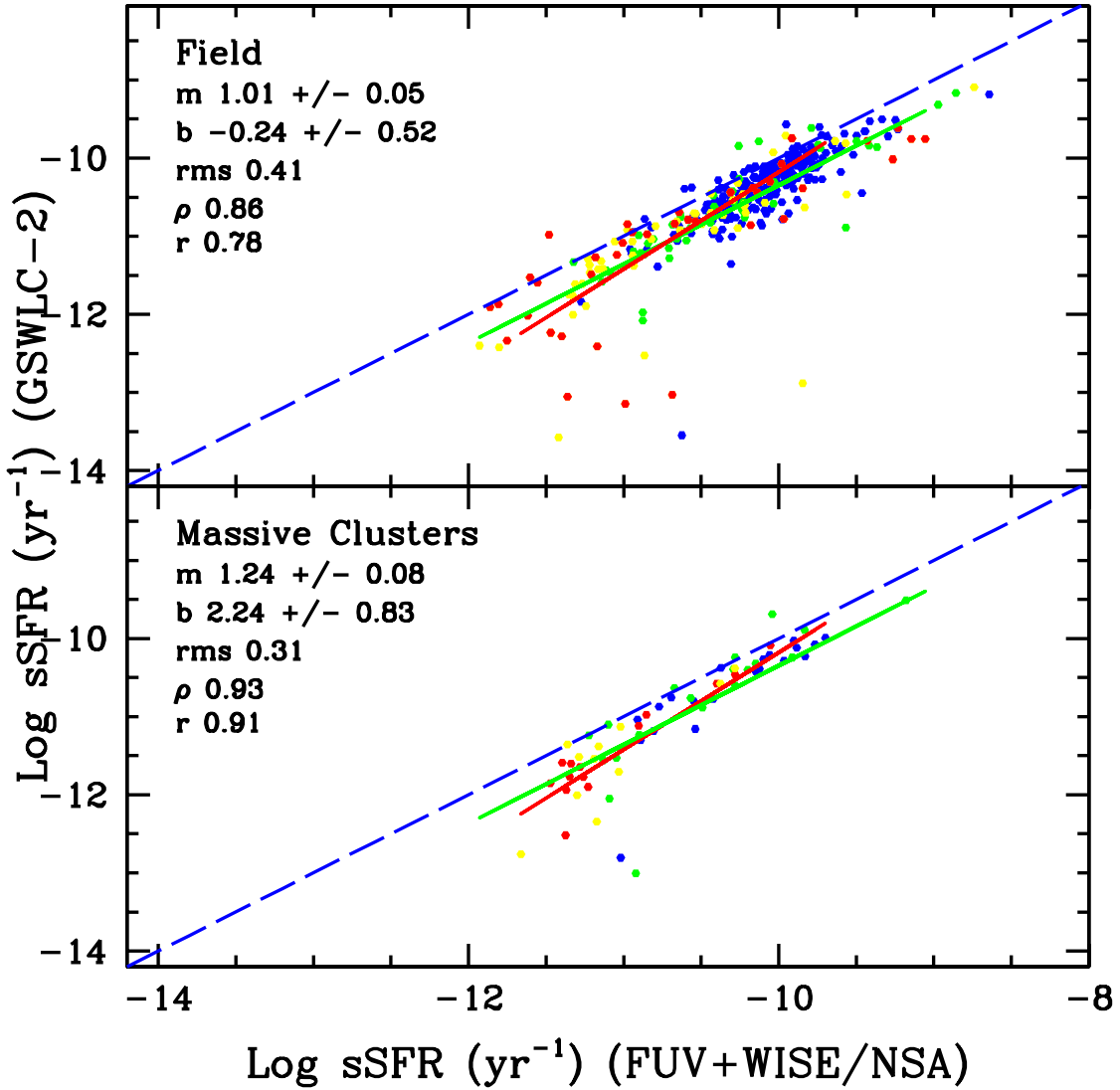}
\caption{
The 
sSFR from the GSWLC-2 vs.\ the sSFR from the NSA FUV fluxes
combined with AllWISE 22 $\mu$m fluxes
using the prescription
from 
\citet{2011ApJ...741..124H}.
Field galaxies are displayed in the top panel, and galaxies in massive
clusters in the bottom panel.
In both plots, the blue dashed line is the one-to-one line,
the green line is the best-fit for the field galaxies, and the red
line is the best-fit for the cluster galaxies.
The slope (m), y-intercept (b), and rms of the best-fit lines 
are printed on the corresponding
plots,
along with the Spearman ($\rho$) and 
Pearson (r) correlation coefficients.
The data points are color-coded based
on concentration (red: C $\ge$ 4.5; yellow: 
4.0 $\le$ C $<$ 4.5; green: 3.5 $\le$ C $<$ 4.0;
blue: C $<$ 3.5).
\label{fig:sSFR_vs_sSFR}}
\end{figure}

Here we derive sSFRs using a different method.
We obtained
22 $\mu$m fluxes for the sample galaxies by cross-correlating
their positions with the AllWISE 
catalog\footnote{https://wise2.ipac.caltech.edu/docs/release/allwise/} 
using a 5$''$ search radius.
We combined these fluxes with 
GALEX FUV fluxes from the NSA, and derived 
SFRs using the prescription in 
\citet{2011ApJ...741..124H}.
We then divided these SFRs by the stellar masses from the 
NSA to obtain a second estimate of sSFR.
In Figure \ref{fig:sSFR_vs_sSFR}, we compare the two
estimates of sSFR, for field (top panel) and cluster galaxies (bottom
panel).  
Galaxies with
log sSFR $<$ -12 in the GSWLC-2 
have higher sSFRs from the FUV+WISE method.  This is associated
with increased
scatter near the bottom of the plot. 
This scatter may be due to contributions to the UV and/or mid-IR fluxes
from an older stellar population, or dust heating
by an older stellar population 
\citep{2016ApJS..227....2S}.
This could lead to an 
over-estimation of the SFRs in some quiescent galaxies.    
Even at high sSFRs, however, there is a offset of the data
from the one-to-one line, with the GSWLC-2 sSFRs being systematically
lower than the FUV+WISE values.   Such an offset between
CIGALE-based SFRs and FUV+IR-based SFRs has been noted
before 
\citep{2018ApJS..234...35Z}.

\begin{figure}[ht!]
\plotone{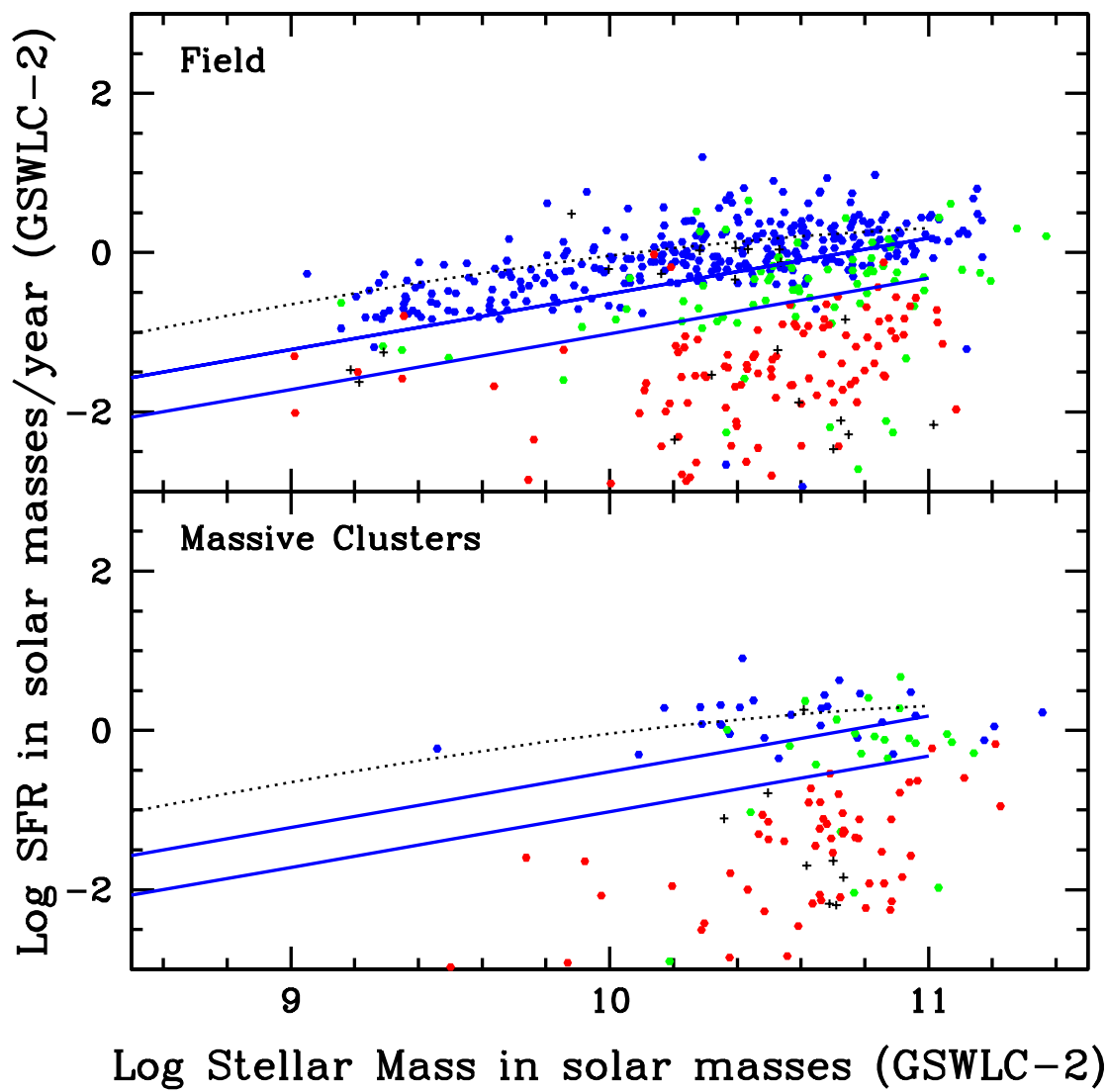}
\caption{
GSWLC-2 SFR vs. stellar mass, for
field galaxies (top panel) and galaxies in massive
clusters (bottom panel).
In both plots, the black dotted line is the star-forming
main sequence as defined by 
\citet{2016MNRAS.462.1749S},
while the upper and lower blue lines, respectively, are the 
\citet{2020MNRAS.491.5406T}
dividing lines between blue cloud
and green valley, and between green valley and red sequence galaxies.
The galaxies are color-coded into blue cloud, green valley, and 
red sequence galaxies according to their color, based on 
the alternative classification scheme of 
\citet{2014A&A...570A..69B},
which relies upon location in the 
NUV - i vs.\ M* plane.
\label{fig:main_seq_GSWLC}}
\end{figure}

\begin{figure}[ht!]
\plotone{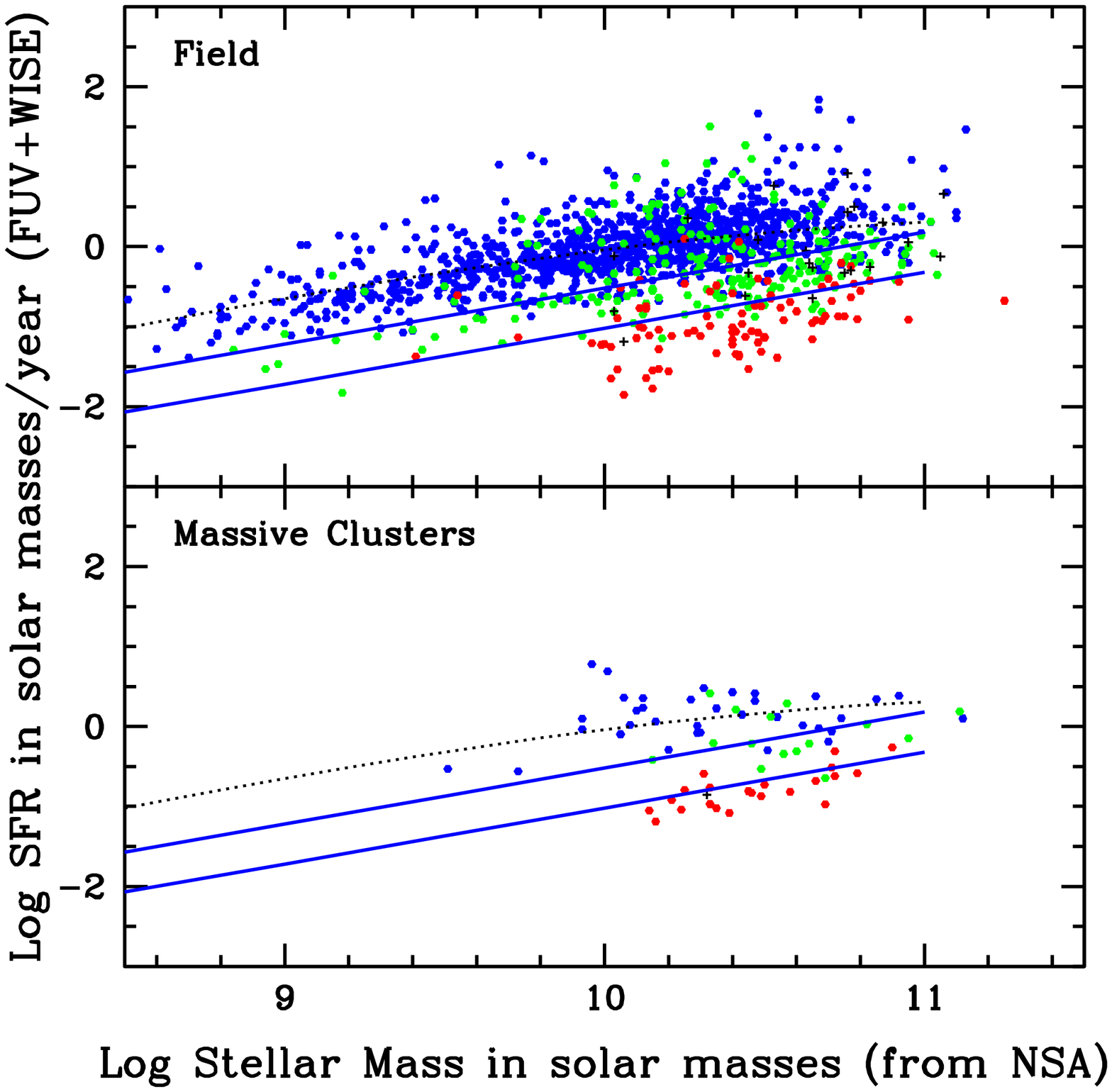}
\caption{
SFRs from FUV+22 $\mu$m vs. stellar mass from the NSA, for
field galaxies (top panel) and galaxies in massive
clusters (bottom panel).
In both plots, the black dotted line is the star-forming
main sequence as defined by 
\citet{2016MNRAS.462.1749S},
while the upper and lower blue lines, respectively, are the 
\citet{2020MNRAS.491.5406T}
dividing lines between blue cloud
and green valley, and between green valley and red sequence galaxies.
The galaxies are color-coded into blue cloud, green valley, and 
red sequence galaxies according to their color, based on 
the alternative classification scheme of 
\citet{2014A&A...570A..69B},
which relies upon location in the 
NUV - i vs.\ M* plane.
\label{fig:main_seq_FUV}}
\end{figure}

To emphasize that many of the galaxies in this sample are
quenched, 
in Figure \ref{fig:main_seq_GSWLC} and \ref{fig:main_seq_FUV} we plot
SFR vs.\ M* for the galaxies
in the field
(top panel) and in massive clusters (bottom panel). 
Figure \ref{fig:main_seq_GSWLC} uses the GSWLC SFRs and M*,
while Figure \ref{fig:main_seq_FUV} uses SFRs from FUV+WISE and M* from the NSA.
As another measure of star-formation activity, we color-code
the galaxies 
in Figures \ref{fig:main_seq_GSWLC} and \ref{fig:main_seq_FUV}
into blue cloud galaxies
(plotted as blue symbols), red sequence galaxies (red symbols),
and green valley galaxies (green symbols), using
the NUV - i vs.\ M* criteria for these three classes from 
\citet{2014A&A...570A..69B}.
Blue cloud galaxies are defined as star-forming galaxies,
while red sequence galaxies are quenched, and green valley are in-between.
In Figures \ref{fig:main_seq_GSWLC} and \ref{fig:main_seq_FUV},
we overlay the star-forming main sequence as defined
by 
\citet{2016MNRAS.462.1749S}
(dotted black curves).
As two blue lines, 
we also plot the dividing
lines between the blue cloud (i.e., main sequence) and the green valley,
and the green valley and the red sequence (quenched galaxies),
as determined by 
\citet{2020MNRAS.491.5406T}.
Many of the field spirals lie in the quenched regime, as do some
of the cluster spirals.
As can be seen from these plots,
there are considerable method-to-method differences
in the definition of quenched galaxies.  However,
all of these methods show that
numerous galaxies in this sample are quenched, 
including both field and 
cluster galaxies.

\subsection{Revisiting Arm Strength
and Bar Strength vs.\ Specific SFR} \label{sec:appendix_sSFR_B} 

\begin{figure}[ht!]
\plotone{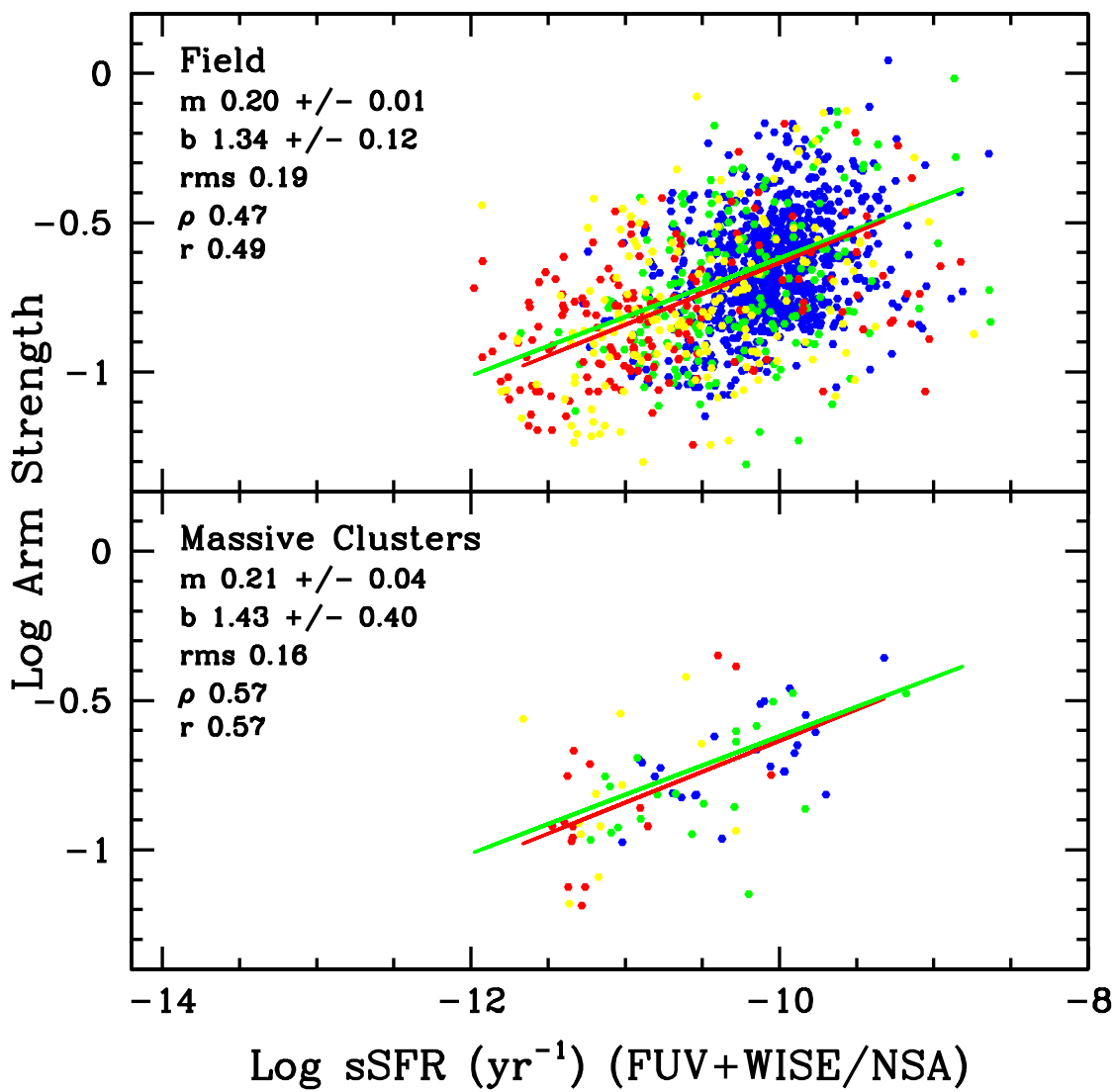}
\caption{ 
The log of the spiral arm strength vs.\ log sSFR 
for 
field galaxies (top panel) and galaxies in massive
clusters (bottom panel).
The sSFRs in these plots come from NSA FUV fluxes
combined with AllWISE 22 $\mu$m fluxes, using the SFR prescription
from 
\citet{2011ApJ...741..124H}.
In both plots, the green and red lines are the best fits for the 
field and cluster galaxies, respectively, with 10 $\le$ log M* $<$ 11.
The slope (m), y-intercept (b), and rms of the best-fit lines 
are printed on the corresponding 
plot,
along with the Spearman ($\rho$) and 
Pearson (r) correlation coefficients.
The data points are color-coded based
on concentration (red: C $\ge$ 4.5; yellow: 
4.0 $\le$ C $<$ 4.5; green: 3.5 $\le$ C $<$ 4.0;
blue: C $<$ 3.5.
\label{fig:sSFR2}}
\end{figure}

In Figure \ref{fig:sSFR2} we plot the arm strengths vs.\ these 
new sSFRs, for field (top panel) and cluster galaxies (bottom panel).
The best-fit line for the field galaxies agrees
within the uncertainties with that of the cluster galaxies.
Note that all of 
the sSFRs derived from the FUV+WISE method have log sSFR $>$ -12
(Figure \ref{fig:sSFR2}),
while the GSWLC-2 sSFRs reach lower values (Figure \ref{fig:sSFR}).

\begin{figure}[ht!]
\plotone{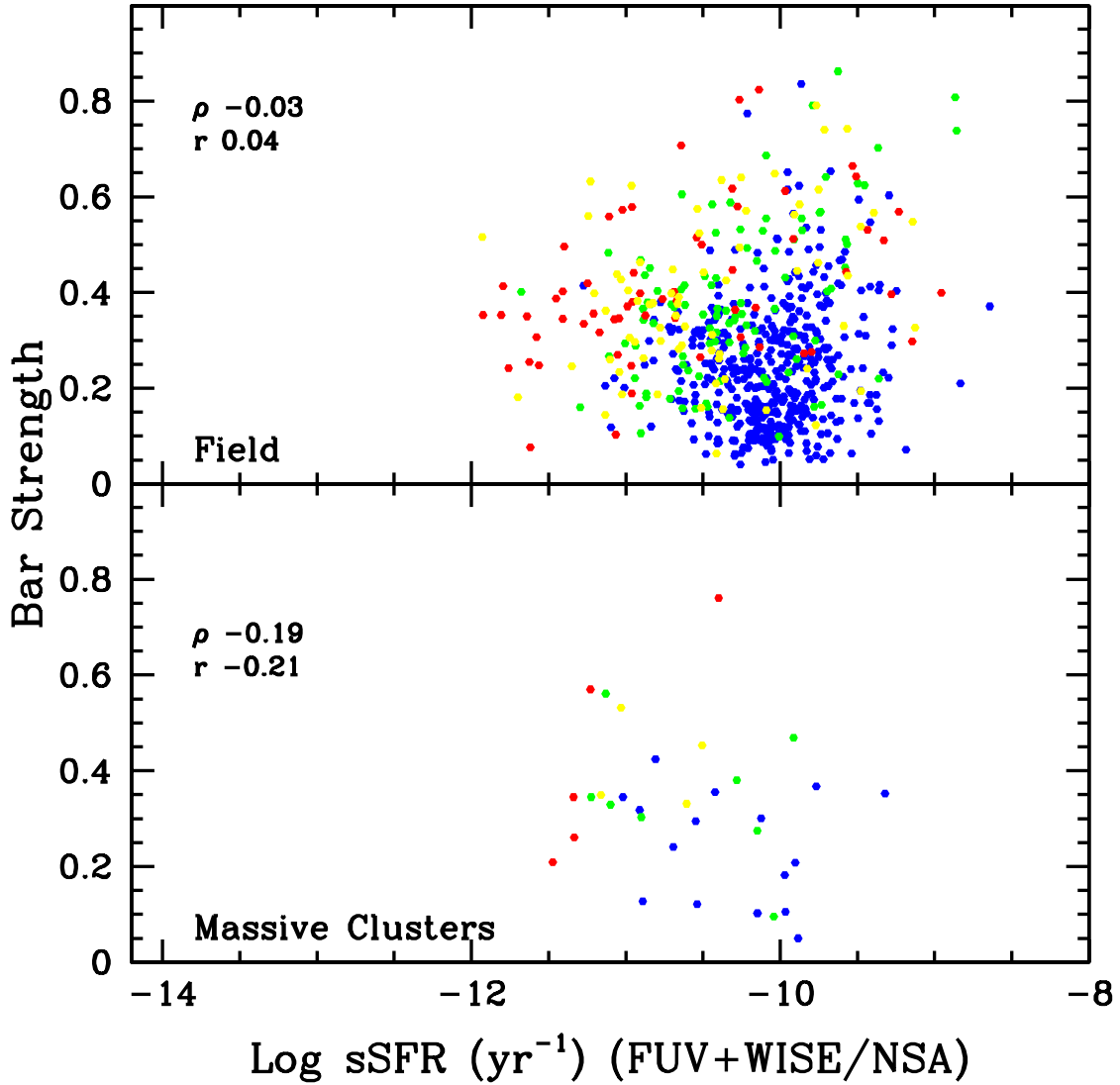}
\caption{
Bar strength
plotted against
the
FUV+WISE/NSA-derived 
sSFR
for
field galaxies (top panel) and galaxies in massive
clusters (bottom panel).
The data points are color-coded based
on concentration (red: C $\ge$ 4.5; yellow: 
4.0 $\le$ C $<$ 4.5; green: 3.5 $\le$ C $<$ 4.0;
blue: C $<$ 3.5).
\label{fig:sSFR_FUV_bar}}
\end{figure}

To test whether the apparent anti-correlation between
sSFR and bar strength for cluster galaxies 
seen in Figure \ref{fig:sSFR_bar}
is real,
we compared bar strength
with our alternative derivation of sSFR in
Figure \ref{fig:sSFR_FUV_bar}.  No correlation is seen for either
the cluster or the field galaxies.

\bibliography{spiral_paper}{}
\bibliographystyle{aasjournal}



\end{document}

%% file: table1_spiral_paper.tex
\begin{deluxetable}{rcccccc}
\tabletypesize \scriptsize
\tablecolumns{7}
\tablewidth{0pc}
\tablecaption{Best-Fit Linear Relations for 10 $\le$ log M* $<$ 11}
\tablehead{   
\colhead{Fig}
&\colhead{Environment} 
&\colhead{Barred/} 
&\colhead{C Range} 
&\colhead{Relation} 
& \colhead{rms} 
& \colhead{Spearman/}
\\ 
\colhead{Num}
&\colhead{}
&\colhead{Unbarred}
& \colhead{}
& \colhead{}
& \colhead{} 
& \colhead{Pearson}
\\ 
\colhead{}
&\colhead{}
&\colhead{}
&\colhead{}
& \colhead{}
& \colhead{} 
& \colhead{Coeff.}
}
\startdata
\ref{fig:C_vs_mass_barred_unbarred}  &  field & both &     &  C = (1.18 $\pm$ 0.09) LOG M*  $-$  (8.44 $\pm$ 0.92) & 0.83 & 0.34/0.32\\
\ref{fig:C_vs_mass_barred_unbarred}  &  cluster & both &     &  C = (0.98 $\pm$ 0.28) LOG M*  $-$  (6.17 $\pm$ 2.96) & 0.68 & 0.28/0.31\\
\ref{fig:C_vs_mass_barred_unbarred}  &  field & unbarred &     &  C = (1.4 $\pm$ 0.12) LOG M*  $-$  (10.65 $\pm$ 1.23) & 0.83 & 0.39/0.38\\
\ref{fig:C_vs_mass_barred_unbarred}  &  cluster & unbarred &     &  C = (1.39 $\pm$ 0.34) LOG M*  $-$  (10.36 $\pm$ 3.57) & 0.59 & 0.45/0.45\\
\ref{fig:bar_strength_vs_C}  &  field & barred &     &  bar strength = (0.07 $\pm$ 0.01) C  +  (0.04 $\pm$ 0.02) & 0.14 & 0.44/0.41\\
\ref{fig:bar_strength_vs_C}  &  cluster & barred &     &  bar strength = (0.08 $\pm$ 0.02) C  +  (0.04 $\pm$ 0.09) & 0.12 & 0.34/0.43\\
\ref{fig:bar_strength_vs_C}  &  field & barred &     &  bar strength vs.\ LOG M* & 0.15 & 0.17/0.13\\
\ref{fig:bar_strength_vs_C}  &  cluster & barred &     &  bar strength = (0.06 $\pm$ 0.08) LOG M*  $-$  (0.24 $\pm$ 0.86) & 0.13 & 0.31/0.30\\
\ref{fig:s_vs_C_barred_unbarred}  &  field & both &     &  s = (-0.041 $\pm$ 0.004) C  +  (0.36 $\pm$ 0.01) & 0.11 & -0.43/-0.30\\
\ref{fig:s_vs_C_barred_unbarred}  &  cluster & both &     &  s = (-0.03 $\pm$ 0.01) C  +  (0.29 $\pm$ 0.04) & 0.08 & -0.33/-0.26\\
\ref{fig:s_vs_C_barred_unbarred}  &  field & unbarred &     &  s = (-0.046 $\pm$ 0.003) C  +  (0.35 $\pm$ 0.01) & 0.08 & -0.49/-0.45\\
\ref{fig:s_vs_C_barred_unbarred}  &  cluster & unbarred &     &  s vs.\ C & 0.07 & -0.29/-0.26\\
 ...  &  field  &  barred  &     & s vs.\ C  & 0.13  &  -0.29/-0.17 \\
 ...  &  cluster &  barred  &     &  s vs.\ C  & 0.08  &  -0.25/-0.14\\
\ref{fig:s_vs_mass}  &  field & both &     &  s vs.\ LOG M* & 0.12 & -0.05/-0.02\\
\ref{fig:s_vs_mass}  &  cluster & both &     &  s vs.\ LOG M* & 0.08 & 0.00/-0.02\\
\ref{fig:s_vs_mass}  &  field & unbarred &     &  s vs.\ LOG M* & 0.09 & -0.18/-0.12\\
\ref{fig:s_vs_mass}  &  cluster & unbarred &     &  s vs.\ LOG M* & 0.08 & 0.01/0.02\\
\ref{fig:s_vs_bar}  &  field & barred & 3 $\le$ C $<$ 4  &  s = (0.33 $\pm$ 0.05) bar strength  +  (0.17 $\pm$ 0.02) & 0.12 & 0.30/0.36\\
\ref{fig:s_vs_bar}  &  cluster & barred & 3 $\le$ C $<$ 4  &  s vs.\ bar strength & 0.06 & 0.14/0.26\\
\ref{fig:s_vs_bar}  &  field & barred & 4 $\le$ C $<$ 5  &  s = (0.5 $\pm$ 0.05) bar strength  +  (0.03 $\pm$ 0.02) & 0.11 & 0.54/0.58\\
\ref{fig:s_vs_bar}  &  cluster & barred & 4 $\le$ C $<$ 5  &  s = (0.31 $\pm$ 0.14) bar strength  +  (0.06 $\pm$ 0.06) & 0.07 & 0.57/0.44\\
\ref{fig:f3_vs_C}  &  field & both &     &  f3 = (-0.052 $\pm$ 0.004) C  +  (0.6 $\pm$ 0.01) & 0.13 & -0.33/-0.33\\
\ref{fig:f3_vs_C}  &  cluster & both &     &  f3 = (-0.08 $\pm$ 0.02) C  +  (0.68 $\pm$ 0.07) & 0.13 & -0.30/-0.38\\
\ref{fig:f3_vs_C}  &  field & unbarred &     &  f3 = (-0.044 $\pm$ 0.005) C  +  (0.58 $\pm$ 0.02) & 0.12 & -0.30/-0.31\\
\ref{fig:f3_vs_C}  &  cluster & unbarred &     &  f3 vs.\ C & 0.13 & -0.17/-0.27\\
...  &  field  & barred  &     &   f3 = (-0.06 $\pm$ 0.01) C  +  (0.63 $\pm$ 0.02)  &  0.14  & -0.36/-0.36 \\
...  &  cluster  &  barred  &     & f3 = (-0.1 $\pm$ 0.03) C  +  (0.77 $\pm$ 0.1)  &  0.13  & -0.46/-0.48  \\
...  &  field & both &     &  f3 vs.\ LOG M* & 0.14 & -0.03/-0.03\\
...  &  cluster & both &     &  f3 vs.\ LOG M* & 0.14 & -0.05/-0.07\\
...  &  field & unbarred &     &  f3 vs.\ LOG M* & 0.13 & -0.02/-0.02\\
...  &  cluster & unbarred &     &  f3 vs.\ LOG M* & 0.13 & -0.08/-0.07\\
...  &  field & both &     &  f3 vs.\ s & 0.14 & 0.04/-0.05\\
...  &  cluster & both &     &  f3 vs.\ s & 0.14 & -0.05/-0.04\\
...  &  field & unbarred &     &  f3 vs.\ s & 0.13 & 0.22/0.14\\
...  &  cluster & unbarred &     &  f3 vs.\ s & 0.13 & -0.08/-0.13\\
...  &  field & barred  &     &  f3 vs.\ s &  0.15  &  -0.13/-0.17 \\
...  &  cluster & barred  &     &  f3 vs.\ s & 0.15  & -0.02/0.01 \\
...  &  field & barred & 3 $\le$ C $<$ 4  &  f3 vs.\ bar & 0.14 & -0.28/-0.29\\
...  &  cluster & barred & 3 $\le$ C $<$ 4  &  f3 vs.\ bar & 0.14 & 0.03/0.07\\
...  &  field & barred & 4 $\le$ C $<$ 5  &  f3 vs.\ bar & 0.13 & -0.27/-0.26\\
...  &  cluster & barred & 4 $\le$ C $<$ 5  &  f3 vs.\ bar & 0.12 & 0.05/0.09\\
\ref{fig:phi_vs_C_and_M}  &  field & both &     &  $\phi$ = (-4.4 $\pm$ 0.3) C  +  (34 $\pm$ 1) & 5.8 & -0.55/-0.52\\
\ref{fig:phi_vs_C_and_M}  &  cluster & both &     &  $\phi$ = (-5.4 $\pm$ 1.2) C  +  (38 $\pm$ 5) & 6.5 & -0.49/-0.54\\
...  &  field & unbarred &     &  $\phi$ = (-4.3 $\pm$ 0.4) C  +  (33 $\pm$ 1) & 6.2 & -0.50/-0.49\\
...  &  cluster & unbarred &     &  $\phi$ = (-5.2 $\pm$ 2.4) C  +  (38 $\pm$ 10) & 7.4 & -0.32/-0.42\\
\ref{fig:phi_vs_C_and_M}  &  field & both &     &  $\phi$ = (-8.7 $\pm$ 0.9) LOG M*  +  (109 $\pm$ 9) & 6.4 & -0.31/-0.32\\
\ref{fig:phi_vs_C_and_M}  &  cluster & both &     &  $\phi$ = (-18.2 $\pm$ 4.4) LOG M*  +  (209 $\pm$ 46) & 6.9 & -0.49/-0.52\\
...  &  field & unbarred &     &  $\phi$ = (-10.6 $\pm$ 1.3) LOG M*  +  (129 $\pm$ 13) & 6.5 & -0.36/-0.38\\
...  &  cluster & unbarred &     &  $\phi$ = (-18.2 $\pm$ 6.4) LOG M*  +  (210 $\pm$ 67) & 7.4 & -0.46/-0.51\\
...  &  field & barred & 3 $\le$ C $<$ 4  &  $\phi$ = (-13.4 $\pm$ 2.9) bar  +  (23 $\pm$ 1) & 6.0 & -0.33/-0.31\\
...  &  cluster & barred & 3 $\le$ C $<$ 4  &  $\phi$ = (-28.2 $\pm$ 23.6) bar  +  (27 $\pm$ 8) & 6.1 & -0.31/-0.39\\
...  &  field & barred & 4 $\le$ C $<$ 5  &  $\phi$ vs.\ bar & 4.8 & -0.11/-0.11\\
...  &  cluster & barred & 4 $\le$ C $<$ 5  &  $\phi$ vs.\ bar & 3.5 & -0.09/-0.21\\
\enddata
\tablenotetext{}{
Fits that do not involve $\phi$ explicitly include
galaxies without
pitch angle measurements.}
\end{deluxetable}

%% file: table2_spiral_paper.tex
\begin{deluxetable}{ccccccccccc}
\tablecolumns{11}
\tablewidth{0pc}
\tablecaption{Kolmogorov-Smirnov/Anderson-Darling Tests: Arm Strengths of Galaxies in Massive Clusters vs. the Field\label{tab:table_s}}
\tablehead{   
\colhead{C}   
& \colhead{barred}
& \colhead{number }    
& \colhead{number } 
& \colhead{KS/AD } 
& \colhead{field} 
& \colhead{field }
& \colhead{field }
& \colhead{cluster} 
& \colhead{cluster }
& \colhead{cluster }
\\ 
\colhead{Range}
& \colhead{or}
& \colhead{field}    
& \colhead{cluster} 
& \colhead{prob} 
& \colhead{median } 
& \colhead{1st}
& \colhead{3rd}
& \colhead{median } 
& \colhead{1st}
& \colhead{3rd}
\\ 
\colhead{} 
& \colhead{unbarred}
& \colhead{ }    
& \colhead{ } 
& \colhead{} 
& \colhead{} 
& \colhead{quartile}
& \colhead{quartile}
& \colhead{} 
& \colhead{quartile}
& \colhead{quartile}
\\ 
}
\startdata
\multicolumn{11}{c}{All Redshifts z $<$ 0.03} \\
\hline
\multicolumn{11}{c}{Both Barred/Unbarred Galaxies } \\
2.8 to 3.2 & both & 236 & 11 & 0.53/$>$0.25 & 0.23 & 0.18 & 0.3 & 0.22 & 0.15 & 0.28 \\
3.2 to 3.6 & both & 199 & 9 & 0.05/0.07 & 0.2 & 0.16 & 0.29 & 0.15 & 0.15 & 0.19 \\
3.6 to 4.0 & both & 173 & 22 & 0.05/0.07 & 0.17 & 0.13 & 0.27 & 0.14 & 0.11 & 0.18 \\
4.0 to 4.4 & both & 242 & 26 & 0.89/$>$0.25 & 0.14 & 0.1 & 0.19 & 0.13 & 0.09 & 0.24 \\
4.4 to 4.8 & both & 224 & 27 & 0.29/$>$0.25 & 0.12 & 0.1 & 0.18 & 0.12 & 0.1 & 0.15 \\
\hline
\multicolumn{11}{c}{Barred Galaxies Only} \\
2.8 to 3.2 & barred & 125 & 5 & 0.71/$>$0.25 & 0.24 & 0.19 & 0.32 & 0.22 & 0.18 & 0.24 \\
3.2 to 3.6 & barred & 103 & 5 & 0.01/0.01 & 0.27 & 0.19 & 0.36 & 0.15 & 0.15 & 0.18 \\
3.6 to 4.0 & barred & 96 & 10 & 0.21/0.14 & 0.22 & 0.15 & 0.35 & 0.17 & 0.13 & 0.26 \\
4.0 to 4.4 & barred & 111 & 9 & 0.81/$>$0.25 & 0.18 & 0.14 & 0.27 & 0.23 & 0.16 & 0.29 \\
4.4 to 4.8 & barred & 96 & 9 & 0.13/0.17 & 0.17 & 0.11 & 0.25 & 0.12 & 0.1 & 0.16 \\
\hline
\multicolumn{11}{c}{Unbarred Galaxies Only} \\
2.8 to 3.2 & unbarred & 111 & 6 & 0.48/$>$0.25 & 0.23 & 0.17 & 0.28 & 0.19 & 0.14 & 0.33 \\
3.2 to 3.6 & unbarred & 96 & 4 & 0.51/$>$0.25 & 0.17 & 0.12 & 0.22 & 0.17 & 0.13 & 0.23 \\
3.6 to 4.0 & unbarred & 77 & 12 & 0.52/$>$0.25 & 0.14 & 0.11 & 0.19 & 0.13 & 0.11 & 0.15 \\
4.0 to 4.4 & unbarred & 131 & 17 & 0.85/$>$0.25 & 0.12 & 0.09 & 0.15 & 0.11 & 0.08 & 0.15 \\
4.4 to 4.8 & unbarred & 128 & 18 & 0.67/$>$0.25 & 0.12 & 0.09 & 0.15 & 0.12 & 0.11 & 0.14 \\
\hline
\multicolumn{11}{c}{Only Galaxies with 0.019 $\le$ z $<$ 0.03} \\
\hline
\multicolumn{11}{c}{Both Barred/Unbarred Galaxies} \\
2.8 to 3.2 & both & 142 & 11 & 0.63/$>$0.25 & 0.22 & 0.18 & 0.3 & 0.22 & 0.15 & 0.28 \\
3.2 to 3.6 & both & 118 & 9 & 0.1/0.07 & 0.19 & 0.16 & 0.3 & 0.15 & 0.15 & 0.19 \\
3.6 to 4.0 & both & 107 & 21 & 0.06/0.07 & 0.17 & 0.13 & 0.28 & 0.14 & 0.11 & 0.18 \\
4.0 to 4.4 & both & 134 & 26 & 0.85/$>$0.25 & 0.14 & 0.1 & 0.19 & 0.13 & 0.09 & 0.24 \\
4.4 to 4.8 & both & 135 & 26 & 0.28/$>$0.25 & 0.13 & 0.1 & 0.18 & 0.12 & 0.11 & 0.16 \\
\hline
\multicolumn{11}{c}{Barred Galaxies Only} \\
2.8 to 3.2 & barred & 77 & 5 & 0.73/$>$0.25 & 0.24 & 0.19 & 0.32 & 0.22 & 0.18 & 0.24 \\
3.2 to 3.6 & barred & 55 & 5 & 0.02/0.01 & 0.27 & 0.17 & 0.36 & 0.15 & 0.15 & 0.18 \\
3.6 to 4.0 & barred & 60 & 10 & 0.19/0.14 & 0.22 & 0.15 & 0.37 & 0.17 & 0.13 & 0.26 \\
4.0 to 4.4 & barred & 62 & 9 & 0.67/$>$0.25 & 0.16 & 0.14 & 0.28 & 0.23 & 0.16 & 0.29 \\
4.4 to 4.8 & barred & 54 & 8 & 0.19/0.17 & 0.17 & 0.12 & 0.23 & 0.14 & 0.11 & 0.18 \\
\hline
\multicolumn{11}{c}{Unbarred Galaxies Only} \\
2.8 to 3.2 & unbarred & 65 & 6 & 0.5/$>$0.25 & 0.22 & 0.17 & 0.26 & 0.19 & 0.14 & 0.33 \\
3.2 to 3.6 & unbarred & 63 & 4 & 0.76/$>$0.25 & 0.17 & 0.13 & 0.22 & 0.17 & 0.13 & 0.23 \\
3.6 to 4.0 & unbarred & 47 & 11 & 0.68/$>$0.25 & 0.14 & 0.11 & 0.19 & 0.12 & 0.11 & 0.15 \\
4.0 to 4.4 & unbarred & 72 & 17 & 0.79/$>$0.25 & 0.12 & 0.09 & 0.16 & 0.11 & 0.08 & 0.15 \\
4.4 to 4.8 & unbarred & 81 & 18 & 0.91/$>$0.25 & 0.12 & 0.1 & 0.16 & 0.12 & 0.11 & 0.14 \\
\enddata
\end{deluxetable}

%% file: table3_spiral_paper.tex
\begin{deluxetable}{cccccccccc}
\tablecolumns{10}
\tablewidth{0pc}
\tablecaption{Kolmogorov-Smirnov/Anderson-Darling Tests: Arm Strengths of Barred vs. Unbarred Field Galaxies\label{tab:table_s_barred_vs_unbarred}}
\tablehead{   
\colhead{C}   
& \colhead{number }    
& \colhead{number } 
& \colhead{KS/AD } 
& \colhead{barred}
& \colhead{barred }
& \colhead{barred }
& \colhead{unbarred} 
& \colhead{unbarred }
& \colhead{unbarred }
\\ 
\colhead{Range}   
& \colhead{barred}    
& \colhead{unbarred} 
& \colhead{prob} 
& \colhead{median } 
& \colhead{1st }
& \colhead{3rd }
& \colhead{median } 
& \colhead{1st }
& \colhead{3rd }
\\ 
\colhead{} 
& \colhead{} 
& \colhead{} 
& \colhead{} 
& \colhead{} 
& \colhead{quartile}
& \colhead{quartile} 
& \colhead{}
& \colhead{quartile}
& \colhead{quartile}
\\ 
}
\startdata
\multicolumn{10}{c}{All Redshifts z $<$ 0.03} \\
2.8 to 3.2 & 125 & 111 & 0.07/0.07 & 0.24 & 0.19 & 0.32 & 0.23 & 0.17 & 0.28 \\
3.2 to 3.6 & 103 & 96 & 2 $\times$ 10$^{-9}$/$<$0.001 & 0.27 & 0.19 & 0.36 & 0.17 & 0.12 & 0.22 \\
3.6 to 4.0 & 96 & 77 & 4 $\times$ 10$^{-7}$/$<$0.001 & 0.22 & 0.15 & 0.35 & 0.14 & 0.11 & 0.19 \\
4.0 to 4.4 & 111 & 131 & 1 $\times$ 10$^{-9}$/$<$0.001 & 0.18 & 0.14 & 0.27 & 0.12 & 0.09 & 0.15 \\
4.4 to 4.8 & 96 & 128 & 8 $\times$ 10$^{-6}$/$<$0.001 & 0.17 & 0.11 & 0.25 & 0.12 & 0.09 & 0.15 \\
\hline
\multicolumn{10}{c}{Only Galaxies with 0.019 $\le$ z $<$ 0.03} \\
2.8 to 3.2 & 77 & 65 & 0.02/0.07 & 0.24 & 0.19 & 0.32 & 0.22 & 0.17 & 0.26 \\
3.2 to 3.6 & 55 & 63 & 7 $\times$ 10$^{-5}$/$<$0.001 & 0.27 & 0.17 & 0.36 & 0.17 & 0.13 & 0.22 \\
3.6 to 4.0 & 60 & 47 & 2 $\times$ 10$^{-5}$/$<$0.001 & 0.22 & 0.15 & 0.37 & 0.14 & 0.11 & 0.19 \\
4.0 to 4.4 & 62 & 72 & 3 $\times$ 10$^{-5}$/$<$0.001 & 0.16 & 0.14 & 0.28 & 0.12 & 0.09 & 0.16 \\
4.4 to 4.8 & 54 & 81 & 8 $\times$ 10$^{-4}$/$<$0.001 & 0.17 & 0.12 & 0.23 & 0.12 & 0.1 & 0.16 \\
\enddata
\end{deluxetable}